\pdfoutput=1

\RequirePackage{fix-cm}
\documentclass[smallextended]{svjour3}       
\smartqed  

\usepackage{graphicx}
\usepackage{latexsym}
\usepackage{natbib}
\usepackage{amssymb}
\usepackage{color}
\usepackage{amsmath}
\newcommand{\diff}[2]{\frac{\partial (#1)}{\partial #2}} 
\newcommand{\eabc}{ {\epsilon^{\alpha\beta\gamma}} }
\newcommand{\bx}{ {\bf x} }
\newcommand{\half}{{1\over2}}
\newcommand{\del}{\nabla}

\begin{document}

\title{Common Envelope Evolution: Where we stand and how we can move
forward}

\titlerunning{Common Envelope Evolution}        

\author{N. Ivanova \and
S. Justham \and
X.~Chen \and
O.~De~Marco \and
C.L.~Fryer \and 
E.~Gaburov \and
H.~Ge \and
E.~Glebbeek \and
Z.~Han \and
X.-D.~Li \and
G.~Lu \and
T.~Marsh \and
Ph.~Podsiadlowski\and
A.~Potter \and
N.~Soker \and 
R.~Taam \and  
T.M.~Tauris \and
E.P.J.~van den Heuvel \and
R.~F.~Webbink
}

\authorrunning{Ivanova et al.} 

\institute{N. Ivanova \at
              Physics Department, University of Alberta, Edmonton, AB T6G 3E1, Canada \\
              Tel.: +1-780-248-1899\\
              \email{nata.ivanova@ualberta.ca}           
           \and
           S. Justham \at
              National Astronomical Observatories, The Chinese Academy of Sciences, Beijing, China \\
              Kavli Institute for Astronomy and Astrophysics, Peking University, Beijing 100871, China \\
              \email{sjustham@bao.ac.cn}
              \and
              X. Chen \at
              National Astronomical Observatories/Yunnan Observatory, The Chinese Academy of Sciences, Kunming 650011, China \\
              Key Laboratory for the Structure and Evolution of Celestial Objects, Chinese Academy of Sciences, Kunming 650011, China\\
              \and
             O. De Marco \at
             Department of Physics \& Astronomy, Macquarie University, Sydney, NSW 2109, Australia \\ 
             \and C.L.~Fryer \at
    Computational Science Division,  Los Alamos National Laboratory, CCS-2, MS D409, Los Alamos, NM 87545, USA\\
\and 
E.~Gaburov \at 
  Center for Interdisciplinary Exploration and Research in Astrophysics (CIERA) \& Department of Physics and Astronomy, Northwestern University, 2145 Sheridan Rd, Evanston, IL 60208, USA \\
\and
H.~Ge \at
              National Astronomical Observatories/Yunnan Observatory, The Chinese Academy of Sciences, Kunming 650011, China \\
              Key Laboratory for the Structure and Evolution of Celestial Objects, Chinese Academy of Sciences, Kunming 650011, China\\
\and
E.~Glebbeek \at
Department of Physics \& Astronomy, McMaster 
  University, 1280 Main Street West, Hamilton, Ontario L8S 4M1,
  Canada \\
   Department of Astrophysics/IMAPP, Radboud University, Nijmegen,  P.O. Box 9010, The Netherlands \\
\and
Z.~Han \at
              National Astronomical Observatories/Yunnan Observatory, The Chinese Academy of Sciences, Kunming 650011, China \\
              Key Laboratory for the Structure and Evolution of Celestial Objects, Chinese Academy of Sciences, Kunming 650011, China\\
\and
X.-D.~Li \at
  Department of Astronomy and Key Laboratory of Modern Astronomy and Astrophysics, Nanjing University,
  Nanjing 210093, China \\
\and
G.~Lu \at
  National Astronomical Observatories / Urumqi
  Observatory, the Chinese Academy of Sciences, Urumqi, 830011, China \\
   School of Physics, Xinjiang University, Urumqi, 830046, China \\
\and
T.~Marsh \at
  Department of Physics, University of Warwick,
  Coventry, CV4 7AL, UK\\
\and
Ph.~Podsiadlowski \at
  Sub-Department of Astronomy, Oxford University, Oxford, OX1 3RH, UK\\
\and
A.~Potter \at 
  Institute of Astronomy, University of Cambridge, The Observatories, Madingley Road, Cambridge CB3 0HA, UK\\
\and
N.~Soker \at 
  Department of Physics, Technion, Israel Institute of Technology, Haifa 32000, Israel\\
\and 
R.~Taam \at
 Center for Interdisciplinary Exploration and Research in Astrophysics (CIERA) \& Department of Physics and Astronomy, Northwestern University, 2145 Sheridan Rd, Evanston, IL 60208, USA \\
  Academia Sinica Institute of Astrophysics and Astronomy-TIARA, P.O. Box 23-141, Taipei, 10617 Taiwan \\
\and  
T.M.~Tauris \at
  Argelander-Insitut f\"{u}r Astronomie, Universit\"{a}t Bonn, Auf dem H\"{u}gel 71, 53121 Bonn, Germany \\
  Max-Planck-Institut f\"{u}r Radioastronomie, Auf dem H\"{u}gel 69, 53121 Bonn, Germany \\
\and
E.P.J.~van den Heuvel \at
 Astronomical Institute `Anton Pannekoek', P.O.Box 94249, 1090GE Amsterdam, The Netherlands \\
\and
 R.~F.~Webbink  \at
    Department of Astronomy, University of Illinois, 1002 W. Green St., Urbana, IL 61801, USA \\   
\and      
            }

\date{Received: date / Accepted: date}

\maketitle

\begin{abstract}
This work aims to present our current best physical understanding of common-envelope
evolution (CEE). We highlight areas of consensus and disagreement, and
stress ideas which should point the way forward for progress in this
important but long-standing and largely unconquered problem. Unusually
for CEE-related work, we mostly try to avoid relying on results from population
synthesis or observations, in order to avoid potentially being misled
by previous misunderstandings.  As far as possible we debate all the
relevant issues starting from physics alone, all the way from the evolution of the binary
system immediately before CEE begins to the processes which might
occur just after the ejection of the envelope. In particular, we
include extensive discussion about the energy
sources and sinks operating in CEE, and hence examine the foundations of the standard energy
formalism.  Special attention is also given to comparing the results of
hydrodynamic simulations from different groups and to discussing the
potential effect of initial conditions on the differences in the
outcomes. We compare current numerical techniques for the problem
of CEE and also whether more appropriate tools could and should be
produced (including new formulations of computational hydrodynamics,
and attempts to include 3D processes within 1D codes).  
Finally we explore new ways to link CEE with observations. We compare
previous simulations of CEE to the recent outburst from V1309 Sco, and
discuss to what extent post-common-envelope binaries and nebulae can provide
information, e.g.\ from binary eccentricities, which is not currently being fully
exploited.
\keywords{Close binaries \and Stellar structure, interiors, evolution \and Hydrodynamics}
\end{abstract}

\tableofcontents

\section{Introduction: The importance of common-envelope evolution}

\label{sec:intro}

Common-envelope evolution (CEE) is the name given to a short-lived phase in the
life of a binary star during which two stars orbit inside a single, shared envelope.
CEE is believed to be a vital process in the evolution of a
large number and wide diversity of binary stars. This almost certainly 
includes the progenitors of Type Ia supernovae, X-ray binaries and
double neutron stars. Hence understanding the outcome of CEE is
required in order to understand the production of the most important 
cosmological standard candles, the nearest known black holes and the 
most promising stellar-mass gravitational-wave sources.

The reason for the importance of CEE is relatively simple to explain, especially for
compact binaries. The stars which produced the compact component of many interesting
systems must once have been orders-of-magnitude larger than would fit within the
present-day system. CEE is currently accepted as allowing the formation
of these systems. The standard reference for CEE is \citet{Pa76}.
This cites private communication with Ostriker, along with
Ron Webbink's PhD thesis, for the origin of this idea (see also \citealt{vdH1976}).
After the ejection of the
common envelope (CE), the remains of the binary stars can then be left in the tight orbits we observe.

However, once a CE phase begins, envelope ejection is not inevitable. When CEE leads to
envelope ejection (and a tighter binary) and when it leads to a merger
is one of the questions which we can still not answer from our own
theoretical understanding: all we have been able to do with comparative certainty is appeal to
the existence of apparently post-CE binaries. Work which discussed the physical situation involved in 
CEE was published before 1976 \citep{B-K+S1971,SparksStecher1974,Refsdal+1974}, and interest 
in such cases helped to inspire the realisation that CEE might be a formation mechanism for close binaries.
Nonetheless, today's theoretical picture of the endpoint of CEE -- and the consequent utility 
of CEE for producing observed systems -- is more based on evolutionary necessity than physical calculation.
In the absence of a complete  physical solution, simplified treatments containing free parameters,  
have been adopted (see \S~\ref{sec:history} for an
  introduction to the history of this process, and for details of the
  recipes see \S~\ref{sec:energy} and \S~\ref{sec:angular_momentum}).
The free parameters in these simplified treatments
are sometimes tuned to match observations, and sometimes values are
assumed in order to make predictions. That is problematic since there
is little reason to believe that these parameters should take a global
value; the time-scales and energy sources and sinks could (and
probably do) vary considerably between situations. 

In general it is also not sufficient to state that CE must act in a certain way in order to produce the observed systems, since perhaps alternative formation channels are available.  For example, population synthesis codes are able to reproduce the observed population of black-hole low-mass X-ray binaries if they set CE ejection efficiencies to high enough values, but the best physical constraints we have seem to preclude the formation of one subset of them \citep[][and references therein]{PRH2003}.  Taking this formation restriction seriously, rather than assuming CEE is somehow efficient enough, led to the proposal of new formation mechanisms which might also help explain, e.g., the strange abundances of the donors in these systems \citep{Justham+2006,ChenLi2006,Ivanova2006,podsi10}.  

Since CEE remains central to our understanding of the formation of many types of system, it is uncomfortable that in many cases we are still fitting parameters with few physical constraints.  CEE is one of the most important unsolved problems in stellar evolution, and is arguably the most significant and least-well-constrained major process in binary evolution \citep[for alternative reviews see][]{Taam00,Web08,Taam10}.

\subsection{A crucial astrophysical process}

Because CEE is important in the formation of a wide variety of
systems, a discussion of the astrophysical importance of CEE in the
context of compact binaries could easily be lengthy; we will give a
very incomplete survey. 

As with most astrophysical processes, we cannot wait long
  enough to watch the formation of many systems by CEE. Nor can we
  normally infer the precise prior history of individual systems.  
  So in order to make quantitative tests of our formation
  theories we model entire populations of objects and
  then compare the properties of those synthesised populations to
  reality. The tools which allow us to do this are called
  \emph{population synthesis} codes. To distinguish this type of
  population synthesis from those used in other areas of astrophysics,
  the more specific term \emph{binary population synthesis} (BPS) is
  often used. Such calculations turn statistical descriptions of stellar initial conditions
  -- such as the initial mass function (IMF) and binary separation
  distribution -- into predictions for, e.g., the formation rates for
  different type of stellar exotica, or the expected present-day
  distribution for the masses and orbital-periods of the type of
  compact binary under investigation.  To do this, BPS simulates the
  evolution of many different binary systems. Obtaining meaningful results
  for rare classes of system or event (such as X-ray binaries or type
  Ia supernovae) may require calculating the evolution of hundreds of
  millions of individual binary systems; hence BPS codes necessarily include 
  simplified and parametrised descriptions of evolutionary processes such
  as CEE. Sometimes BPS is used to try to determine which values of CEE parameters best
  reproduce reality, although the many uncertainties and
  nonlinearities involved in binary evolution mean that this must be
  done cautiously.  Nonetheless, BPS certainly enables us to
  see how our poor understanding of CEE converts to uncertainties in
  predictions, as we now illustrate.

One currently important example is how uncertainties in the outcome of
CEE carry through into large uncertainties in theoretical predictions
for compact-object merger rates, as have been used to help
  justify observational facilities such as LIGO. Gravitational-wave observatories clearly have an interest in the expected merger rates of compact objects in the local universe, in order to try to predict the rate of events they should detect.  Such mergers, when resulting from primordial binaries, are typically expected to involve at least one CE phase in their production. Some of the merger event rate could also be produced following dynamical interactions in dense stellar systems (e.g.\ globular clusters).

Taking results from population synthesis calculations, \citet{Abadie+2010}, quote `realistic' rates for mergers of a NS with a stellar-mass BH which span more than two orders of magnitude (the full range quoted in their table 7 covers four orders of magnitude).  Whilst there are certainly other significant unknowns, almost the full range of uncertainty within the set of rates quoted as realistic can emerge just from altering how one class of systems entering CEE during a particular evolutionary phase is treated \citep{Belczynski+2007}, and the potential occurrence of a special case of CEE can produce one of the higher realistic rates \citep{Dewi+2006}.   The BPS rates for the merger of two stellar-mass BHs quoted in \citet{Abadie+2010} are even more uncertain (a range of more than three orders of magnitude for the field-binary models considered realistic), and again changes in just how CEE is treated could encompass most of that range of rates \citep[e.g.\ for][with otherwise identical assumptions, the presence or absence of a single CE channel can affect the BH-BH merger rate by a factor of 500]{Belczynski+2007}.

\begin{centering}
\begin{figure}
\begin{centering}
\includegraphics[angle=90,height=0.85\textheight]{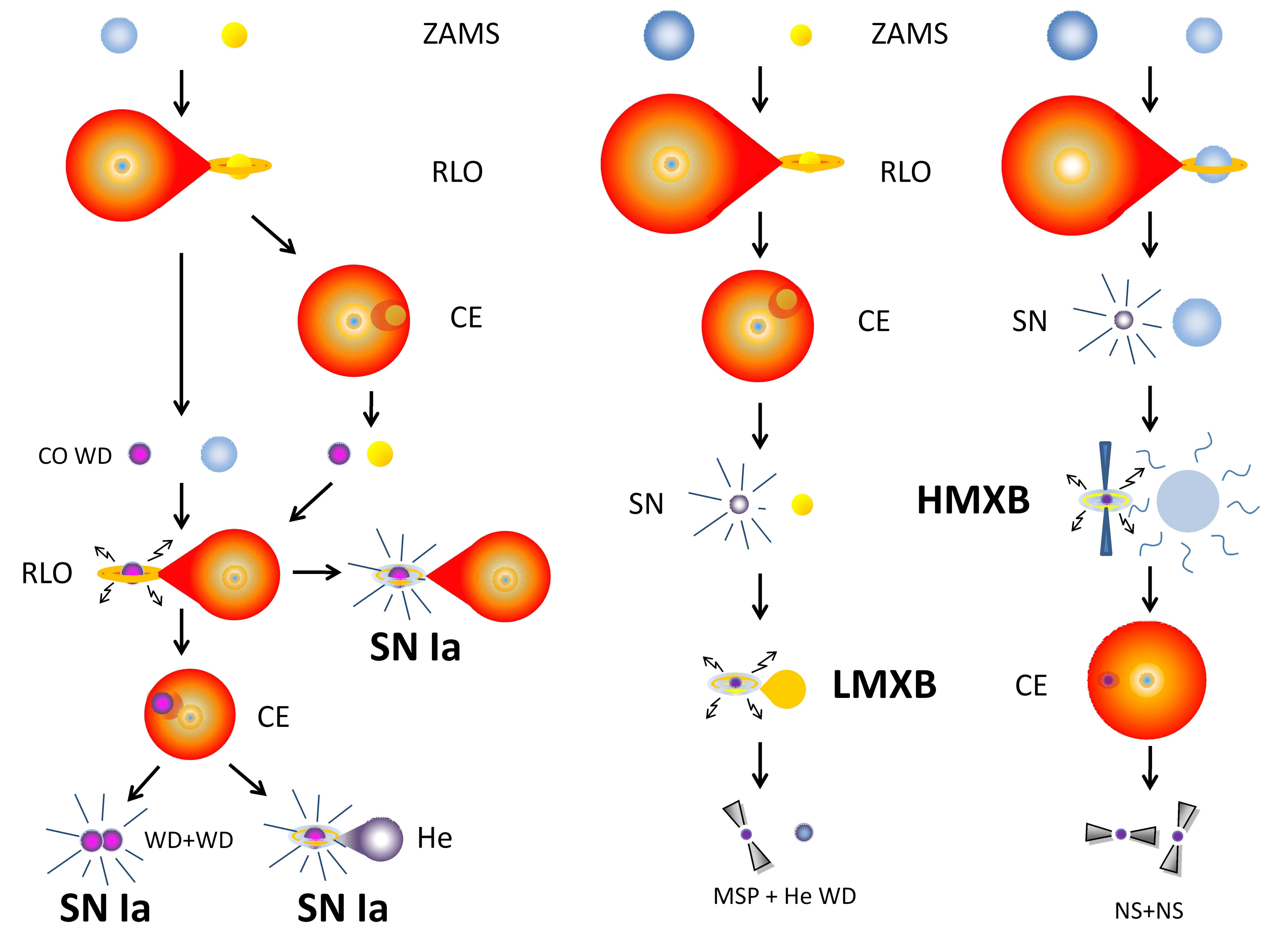}
\caption{Examples of evolutionary channels where CEE plays a crucial
  role in the formation of the final system. The leftmost column presents
  a variety of ways to form potential SN Ia progenitors, including
  double-degenerate mergers and accretion onto a CO WD from a
  non-degenerate companion. 
  The middle and rightmost columns illustrate the
  formation of systems containing neutron stars: one
  route by which a binary millisecond pulsar may form, and one way to 
  produce a double pulsar (formation of which could also involve an additional CE
  phase before the first SN). Other variations of these 
  channels exist. Abbreviations: 
ZAMS -- zero age main sequence, RLO -- Roche lobe overflow, CE -- common envelope, CO WD -- carbon-oxygen white dwarf, He -- He star, HMXB -- high-mass X-ray binary, LMXB -- low-mass X-ray binary, MSP -- millisecond pulsar, NS -- neutron star, SN -- supernova.}
\label{fig_channels}
\end{centering}
\end{figure}
\end{centering}

Another type of compact-object mergers -- of carbon-oxygen
  white dwarfs (CO WDs) -- is potentially also responsible for type Ia
supernovae (SN Ia). Indeed, the paper regularly cited for introducing the
energy parameterisation of CEE \citep[][see \S
\ref{sec:energy}]{Webbink84} was aiming to study WD-WD mergers,
including them as potential progenitors of SN Ia \citep[see also][]{IbenTutukov1984}. 
If these double-degenerate mergers need to be of roughly
Chandrasekhar-mass or more in order to lead to an explosion then the
individual CO WDs need to be
relatively massive. In turn, this suggests that the core evolution of the
stars which produced the CO WDs was not truncated very early; hence
the initial binary separation needs to have been wide enough to allow
at least the primary to evolve into a relatively massive CO WD.  At
the point when both WDs have been formed, the orbital separation needs
to be small enough for gravitational radiation to be able to lead to a
merger within the age of the universe. This is another classic case
where CEE is required to turn a long-period binary into a short-period
one.  Figure \ref{fig_channels} schematically illustrates
potential evolutionary scenarios leading to such a WD-WD merger. There
are two distinct possibilities for the character and outcome of the first mass transfer
episode. Probably currently physically preferable is that the first
mass-transfer episode is stable; in this scenario then such double-degenerate SN Ia
progenitors only require one CE phase in their production. However, the
dominant formation channel which emerges from many BPS predictions
typically involves an unstable first episode of mass transfer followed
by CEE. If this second option does dominate then CEE would be involved
twice in forming potential double-degenerate SN Ia.
In this case then the energy transfer during CEE must be extremely efficient
($\gtrsim50$\%) in order to keep the binary fairly wide after the
first CE phase. Hence population synthesis predictions for the rates
of such mergers tend to adopt very high CE efficiencies, and tend
to be very sensitive to reductions in that efficiency. For example, in the
calculations by \citet{Ruiter+2011}, perfect CE efficiency
(specifically $\alpha_{\rm CE} \lambda$=1, for which see \S \ref{sec:energy})
predicts a Chandrasekhar-mass CO WD merger rate just consistent with
the empirical SN Ia rate.  However, a reduction in overall CE
energetic efficiency by a factor of 8 reduces the predicted rate of SN Ia from the
CO WD merger channel by more than three orders of magnitude at 100
Myr after the starburst and makes the merger rate almost completely
negligible from $\approx$3 Gyr after the starburst; the overall
predicted SN Ia rate here falls far below the observed rate.  
Unfortunately we cannot firmly state whether the first mass-transfer
  phase leads to CEE or not, since we lack a sufficiently detailed
  knowledge of mass transfer stability. However, this specific example has been the
  subject of considerable debate (see \S \ref{sec:angular_momentum}). 
  Understanding the general stability of mass transfer is a
  problem strongly related to CEE itself and will also be discussed later
  (\S \ref{sec:onset})

 Instead of, or in addition to, WD-WD mergers
then SN Ia might be produced by accretion onto a CO
WD in single-degenerate systems \citep[see, e.g.,][]{WhelanIben1973}.
These systems also involve CEE in their formation, so an improved understanding of CEE
should help us to understand their production. However, the predicted 
formation rates of  SN Ia through single-degenerate progenitors tend
to be less strongly dependent on CE efficiency than predictions for the
double-degenerate systems. 
Indeed, in the models of \citealt{Ruiter+2011} then the calculations which assume a
lower CE efficiency lead to an increase in the single-degenerate SN Ia
rate at some epochs (see also, e.g., \citealt{HanPhP2004}, where
the highest assumed CE efficiency produces the lowest overall SN Ia rate for
each otherwise equivalent set of models).  If population calculations are to help
determine which channels actually produce SN Ia then tighter physical constraints
on CE ejection, along with a better understanding of when mass
transfer leads to CEE, would be very helpful.  We note in passing
that birthrates of particular classes of system are not necessarily
monotonically dependent on CE ejection efficiency \citep[see, e.g.,
table 2 of][]{Willems+2005}.

The formation of both classes of gamma-ray bursts (GRBs) probably also involves CEE. Some GRBs are believed to result from compact object mergers (as above); these are associated with the observed set of short-duration, harder-spectrum bursts.  The typically longer-duration observational subclass of GRBs, whose parent populations are strongly linked with recent star formation, are also likely to have CEE in their formation channels \citep{Fryer+1999}.  These are thought to arise from a special-case of core-collapse in massive stars; it is believed that the cores should be rotating rapidly enough to cause a massive accretion disc to form as the core collapses.  In addition, the progenitor star is expected to have lost its envelope, both on theoretical grounds (to enable the jet of the GRB to escape) and on observational grounds (when these GRBs have been linked with a supernova, that supernova has been of a stripped-envelope star, typically a broad-lined type Ic supernova). Stripping the envelope and spinning-up the core can be achieved by several channels involving CEE (for a review of this and of alternative possibilities, see \citealt{Fryer+2007}; also \citealt{podsi10}). One notable recently-observed GRB has been explained using a specific CEE-based model (see \citealt{Thone+2011}; for the underlying model see, e.g., \citealt{FryerWoosley1998}).

The physics of CEE also has the potential to revolutionise our understanding of the evolution of `single' stars.  A star might have its evolution altered by CE-type inspiral of a planet or brown dwarf. In particular, it has been proposed that planets might eject the envelopes of some red giants (\citealt{Soker1998}; see also \citealt{NelemansTauris1998,SokerHarpaz2000,SokerHadar2001,MarSok11}). Planet-driven envelope ejection might potentially explain the formation of single low-mass white dwarfs (as proposed by \citealt{NelemansTauris1998}; see also the discussion and comparison of alternatives in \citealt{Justham+2010}). This possibility is now being driven by observation as well as theory. \citet{2006Natur.442..543M} have observed a low-mass white dwarf ( $\approx 0.39~M_{\odot}$) with a close brown-dwarf ($0.053~M _{\odot}$) companion \citep[see also, e.g.,][]{Setiawan+2010}.  We should perhaps consider the long-term evolution of every `single' star with planets as effectively that of a binary (or multiple) system with an extreme mass ratio.

\subsection{An extraordinary physical problem}

Despite the importance of CEE, it is essentially unsolved.  The situation is extremely challenging for both computation and analytic treatment; from beginning to end the problem involves a complex mix of physical processes operating over a huge range of scales.  A relatively common problem would be one in which a neutron star (NS) spirals into the envelope of a giant. Simulations of such a CE event might need to cover a range in timescale of $\sim 10^{10}$ (i.e. from perhaps 1s, which is already three orders of magnitude longer than the dynamical timescale of the NS, to $\sim 1000$yr, the thermal time of the envelope and plausible duration of the CE phase; note that this ignores the duration of the onset of CEE. An interesting range in scale could be $\sim 10^{8}$ (i.e. from $\approx 10$ km, the size of the NS, to $\approx 1000 R_{\odot}$), and even more if the details of the accretion onto the NS are important (as it might be; see \S \ref{sec:hypercritical}), or if shocks within the envelope need to be resolved more accurately than this allows.  There is no prospect of simulations with anything like a resolution of $(10^{8})^{3}$ in the relevant future, nor ones which continue for $10^{10}$ timesteps.  Even for less extreme examples, in which the inspiralling secondary is not a compact object, comprehensive models are still beyond the reach of our ability. Calculations trying to capture the most important aspects of CEE have been attempted for many years \citep{Taam+1978,mmh79}, but even today's sophisticated simulations necessarily ignore some almost certainly significant physics  (see \S \ref{meth_comp} and \ref{sec:methods}).

\subsection{A little history: how we arrived at the current situation}
\label{sec:history}

Whilst the physical complexity and numerical demands of CEE still leave us
with a very incomplete understanding of how it proceeds, it was
recognized early on that very general considerations of energy and
angular momentum conservation might provide useful constraints on the
outcomes. We note that those early thoughts were not vastly less
physically sophisticated than the pictures used today.
These fundamental constraints would then enable population synthesis
studies. The aim was to model the evolution of an ensemble of hypothetical
binaries in order to unravel the evolutionary channels that lead to the wide
variety of highly-evolved binaries actually observed, and also perhaps to
predict families of evolved binaries yet to be discovered or
recognized.

The earliest treatment of CEE to be widely employed was one assuming
that the energy needed to eject the common envelope was derived
entirely from orbital energy dissipation \citep{vdH1976}.  The current 
rationale for neglecting other possible sources and sinks of energy is
discussed below in \S \ref{sec:energy}, although even at the very
beginning it was recognised that there were several possible complications. 
\cite{Pa76} identified frictional drag as driving transfer of
both orbital energy (as heat) and angular momentum from the binary orbit to
the common envelope, and realised that
a combination of angular momentum and energy conservation
would drive the envelope expansion. Moreover, \cite{Pa76} also 
discusses the fact that the expanding envelope could be expected to
radiate energy away at an increasing rate, and that the consequences of
such effects for the overall scenario are hard to quantify. 

In the energy formalism that was adopted, the energy
budget for the binary is fixed at the onset of mass transfer and
the post-common-envelope system is constrained to have
an orbital energy which is negative enough to provide the energy necessary for
envelope ejection.  In reality,
common-envelope ejection cannot be completely efficient (since, for
example, the ejecta carry away some terminal kinetic energy), and so
an efficiency parameter, $\alpha_{\rm CE}$, was introduced to
characterize the fraction of dissipated orbital energy actually used
to eject the common envelope \citep{Livio88}.

When it comes to quantifying the different terms appearing in the
energy budget, elementary orbital mechanics tells us unambiguously
that the total orbital energy (potential plus kinetic) of a binary with
separation $a$  is $E_{\rm orb} = -Gm_{1} m_{2}/2a$.  Evaluation of the
envelope binding energy, $E_{\rm bind}$, is a more problematic affair
(see \S \ref{sec:energy}).  \cite{Webbink84} introduced a simple
parameterization, $E_{\rm bind} = Gm_{1} m_{\rm 1,env}/R_{1}$  based on
evaluation of the gravitational potential energy plus internal energy
of envelopes of a handful of models of giant branch stars he had on
hand; ionization/dissociation energy was neglected.  Unfortunately,
his paper failed to stipulate which energy terms were included or
excluded in the approximation for $E_{\rm bind}$, but the simple
expression introduced there was clearly intended only to provide an
order-of-magnitude estimate of $E_{\rm bind}$.

More realistic evaluation of $E_{\rm bind}$ depends on the detailed
structure of the donor envelope.  To that end, an additional factor,
$\lambda$, was introduced \citep{deKool1990} to allow for differences in
envelope structure:
\begin{equation}
E_{\rm bind} = G \frac{m_{1} m_{\rm 1,env}}{\lambda R_{1}}.
\end{equation}
As conceived, $\lambda$ depends on the structure of the donor star,
although in practice it is sometimes treated as a free parameter.

The introduction of the $\lambda$ parameter should have improved
matters quantitatively.   It would have been desirable to define this
factor in the inverse, i.e. $E_{\rm bind} = \lambda Gm_{1} m_{1,\rm
  env}/R_{1}$, thereby avoiding nasty singularities when $E_{\rm bind}$
changes sign (because of the recombination term), but the convention
is now irredeemably established.   Unfortunately, when
\cite{deKool1990} introduced $\lambda$, he included only the
gravitational term, and this seems to have led to the notion that the
internal energy was somehow separable from it.  Of course
the Virial Theorem tells us that these terms are strongly related,
though not necessarily in a simple way.  So when
\cite{DewiTauris2001} turned to this issue, they introduced, in
addition to $\lambda_{\rm g}$ parameterizing the gravitational
potential energy, a second $\lambda_{\rm b}$ to parameterize the sum
of gravitational and internal energy, and the issue immediately arose
over which, if either, parameterization should be used.  Further, as
it is noted now, the formal values of $\lambda$ depend strongly on where
one places the mass cut for the ejected envelope.

While $\lambda$ was invented to improve and simplify calculations, particularly for 
population synthesis, it is now clear that not only is having a fixed
value for all possible systems wrong, but it is also still not 
certain how to calculate $\lambda$ for any given star,
however well-known that star's structure is.
We will return to this in more detail in \S \ref{sec:energy}.
 
The next formalism to be invented was based on conservation of angular momentum.
The historical necessity for this alternative, known as the $\gamma$-formalism  \citep{Nelemans00},
was to find at least some explanation for formation of the known double-white dwarf (DWD) binaries. 
There it seemed that the standard energy formalism failed, as it
could only explain the observed systems if energy is
\emph{generated} during CEE, i.e. $\alpha_{\rm CE}>1$. (More precisely
stated, an unknown source of energy appeared to be needed to 
replace the expected role of the orbital energy
source, since the orbital energy actually acts as a further energy \emph{sink} for these
systems.) 
Apparent violation of energy conservation law is rather stressful for a physicist,
so a less obviously troublesome conservation law was called upon to help. 
Again, as no self-consistent numerical simulations 
could have been performed at the time, the angular momentum budget  
had to be parametrized and then its free parameter has been fine-tuned using the
observations of several known-to-the-date DWD systems.
This did not \emph{resolve} the apparent energy generation problem, only hid it. 
Nonetheless, it opened a discussion about the possibility to eject an
envelope by some other mechanism other than a standard common envelope event.
We will consider this formalism in more detail in \S~\ref{sec:angular_momentum}.
Note that the current explanation for the increase of the
  binary separation during this first mass-transfer phase is that it
  is quasi-conservative, such that the mass transfer is driven by nuclear
  energy input and thermal expansion. So there is no longer any
  apparent need to resort to unexplained energy generation.

In nature, during a real CEE, both fundamental conservation laws must
-- of course -- be obeyed. However, neither of these two {\it
  simplified} formalisms were designed to simultaneously 
obey both conservation laws.
It has to be understood that these approximate methods were
invented mainly because of our inability -- which continues to the
present day --  to self-consistently model a complete CEE event. 
However, after many years of use in population synthesis, 
the severe limitations of these educated guesses
seem to have sometimes been forgotten.
These expressions can -- all too easily -- be
used to make apparently predictive statements which may have limited
justification. 

To summarise, an energy formalism first emerged based on the
  argument that common-envelope ejection must be a dynamical
  process.  If its duration were as long as a thermal time-scale, the
  input from available energy sources could be lost to radiation, 
  but also other additional energy sources might well play a role. 
  A complete combination of all possible sinks and sources acting 
  on different timescales would lead to a very complex and difficult picture. In this work we
will examine the physics underlying CEE, and see to what extent we
can hope to move beyond these uncertain simplifications.

\subsection{This work}

This work aims to take a physical approach to the problem of
  CEE. It considers CEE from first principles, trying not to let preconceptions and potentially-misinterpreted observations or population synthesis calculations mislead us. Hence it does not aim to be a comprehensive review of all possible implications of the common envelope problem, but it does hope to build the state-of-the-art in \emph{understanding} CEE.

The next section (\ref{sec:phases}) gives an overview of a notional CE event, dividing it into phases within which different processes are dominant, and also pointing to relevant sections within the remainder of the text.  \S \ref{sec:energy} then considers at length the overall energy balance within CEE, whilst \S \ref{sec:fate} considers the situation at the end of the CE phase.  \S \ref{sec:angular_momentum} briefly discusses the application of angular momentum conservation to CEE.  In \S \ref{sec:onset} we look at the conditions which produce and precede CEE.   \S \ref{meth_comp} then compares the results from different modern hydrodynamic simulations, and \S \ref{sec:methods} discusses the best present-day simulation tools along with potential future improvements in those methods.  \S \ref{sec:hypercritical} discusses the possibility of hypercritical accretion.  \S \ref{sec:observations}  considers what we can learn from observations of post-CE systems; there we also compare observations of a recent transient event, which may well have been produced by CEE, to the expectations produced by CEE simulations. The conclusions, \S \ref{sec:finale}, include a list of some promising directions for possible progress.

\section{Main phases}
\label{sec:phases}

\begin{centering}
\begin{figure}
\begin{centering}
\includegraphics[height=.5\textheight]{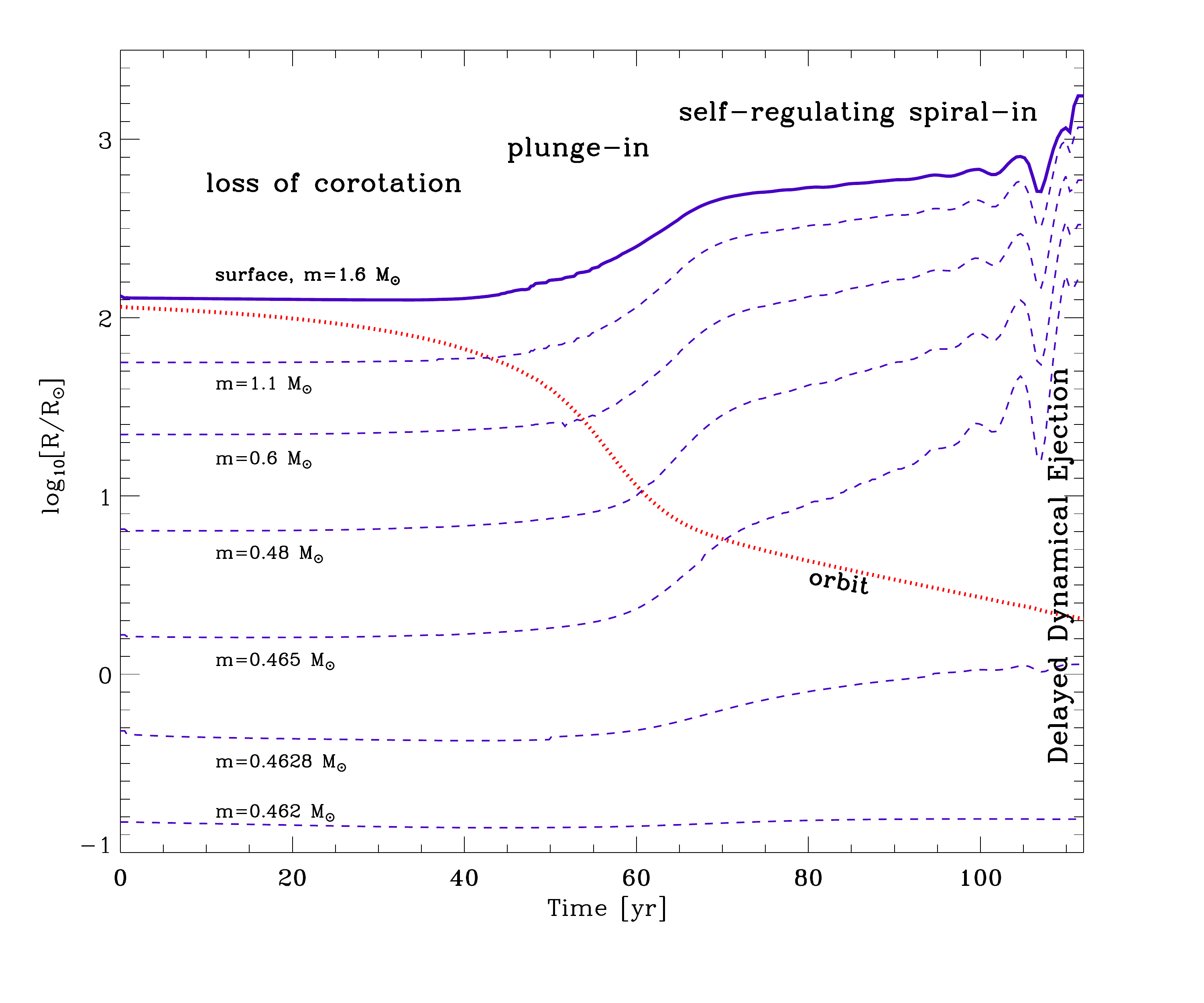}
\caption{The main potential phases of a CE event prior to the envelope ejection or the merger. 
This example is for a 1.6 $M_\odot$ red giant and a 0.3$M_\odot$ WD, using data from one-dimensional hydrodynamical simulations in \cite{Iva02}. Not all phases are expected to happen during all CE events. The dashed lines represent locations at fixed mass coordinates, and the dotted line shows the location of the inspiralling secondary.}
\label{fig_phases}
\end{centering}
\end{figure}
\end{centering}

It is convenient to break down the progression of an idealised CE event into several distinct phases, where each phase operates on its own timescale \citep[][see also Fig.~\ref{fig_phases}]{PhP2001}:

\begin{itemize}

\item {\bf I: Loss of corotation}

During this stage a stable and probably non-eccentric binary, where the rotation of the donor is also likely to be synchronized with the orbit, is transformed into its complete antithesis -- a spiralling-in binary.

The start of the spiral-in could be caused by, e.g.: 

{
\begin{enumerate}
\item {a dynamically unstable (runaway) mass transfer. This happens if the donor, either due to its evolution or due to its immediate reaction upon mass loss, {\it expands} relative to its Roche lobe (for more details see \S~\ref{sec:onset})}.
\item {an instability such as the Darwin instability
    \citep[][]{dar79}, or a secular tidal instability \citep{Hut80,
      Lai93,Egg01}. The Darwin instability occurs when the spin
    angular momentum of the system is more than a third of its orbital
    angular momentum (see also \S~\ref{sec:onset_tides}).}
\item {the reaction of the accretor leads to matter filling the binary orbit. For example, if mass transfer proceeds at too great a rate to be accreted by the compact companion, but the system is also unable to quickly expel the mass, then a common-envelope is naturally formed. Potential cases include an envelope temporarily trapped around a neutron-star being fed at super-Eddington rates \citep{Beg79, Hou91, Kin99}, or reincarnation of an accreting white dwarf which tries to form a red giant \citep{Nom79, Nom07}; or perhaps even in nova systems when the expansion of the nova shell engulfs the companion.}
\end{enumerate}
}

The loss of corotation itself occurs on a dynamical timescale. Prior to that moment, however, the stellar structure is strongly affected by the mass-transfer history before the dynamical instability sets in. This preparatory stage could last hundreds of years, from {\it dozens of dynamical timescales to a thermal time-scale} \citep[see \S~\ref{sec:onset} and ][]{podsi02}. 

\item {\bf II: Plunge-in and its termination}

A rapid spiral-in, during which the orbital energy is deposited in the envelope, drives its expansion and {\it may} lead to its dynamical ejection right away, or to a rapid merger of both stars. This stage is purely {\it dynamical} and is the best studied stage to-date. Typical hydrodynamical simulations for CEE ending with a merger or with a binary formation are shown on Figs.\ref{fig_jamie} and \ref{fig_passy}, and for more technical details see discussion in \S~\ref{meth_comp}.    
 
\begin{centering}
\begin{figure}
\begin{centering}
\includegraphics[height=6.25in]{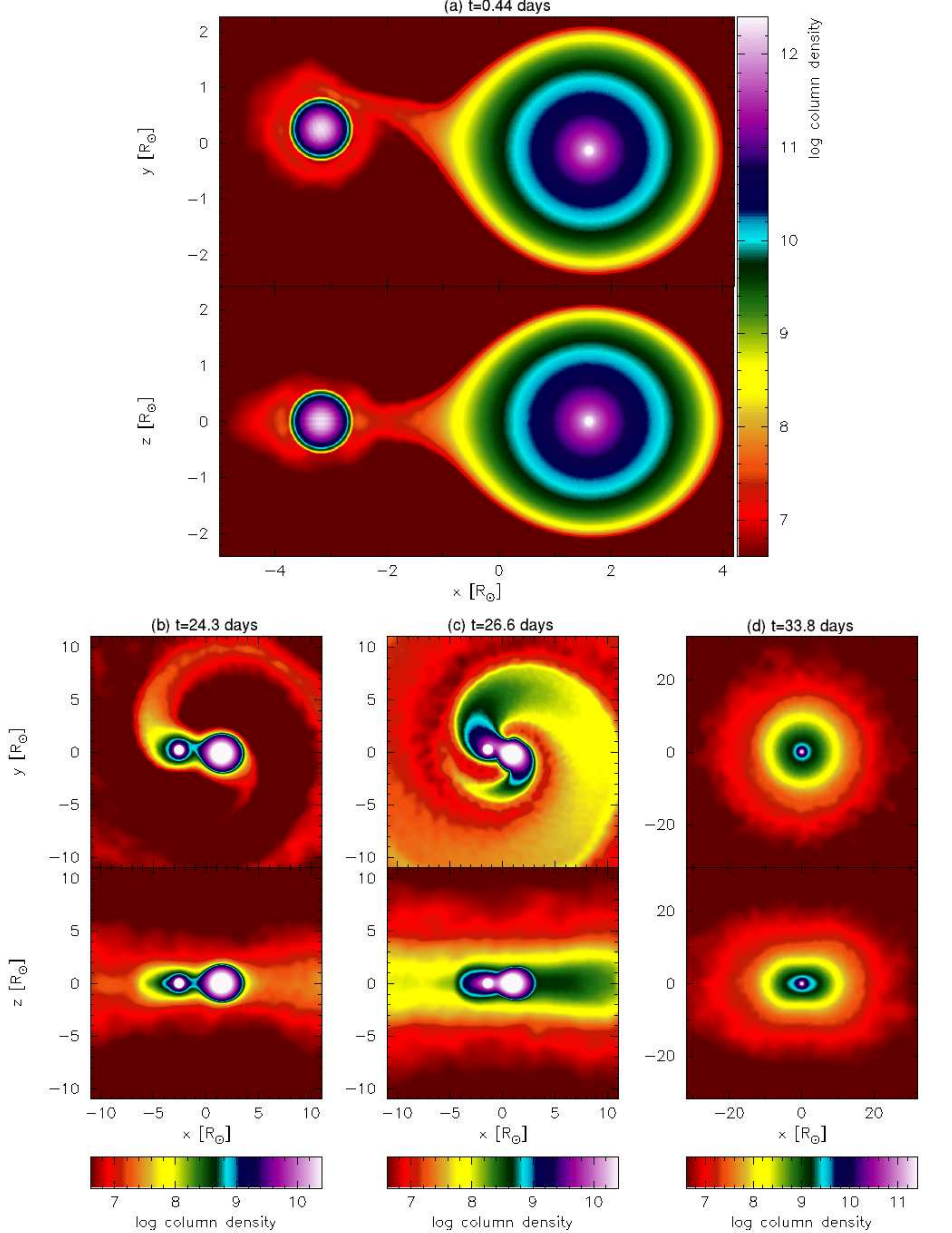}
\caption{Common envelope event with a $1.2M_\odot$ early giant and $0.6 M_\odot$ MS star, resulting in a merger of two stars. 
Simulation performed for this review by J. Lombardi and R. Scruggs, simulated with
$2.2\times 10^5$ SPH particles.  For more technical details on the
code, see \cite{Gaburov10} and \cite{Lombardi11}.  Vizualization (images and on-line video) 
are generated using SPLASH \citep{Price07}.}
\label{fig_jamie}
\end{centering}
\end{figure}
\end{centering}

\begin{centering}
\begin{figure}
\begin{centering}
\includegraphics[height=0.9\textheight]{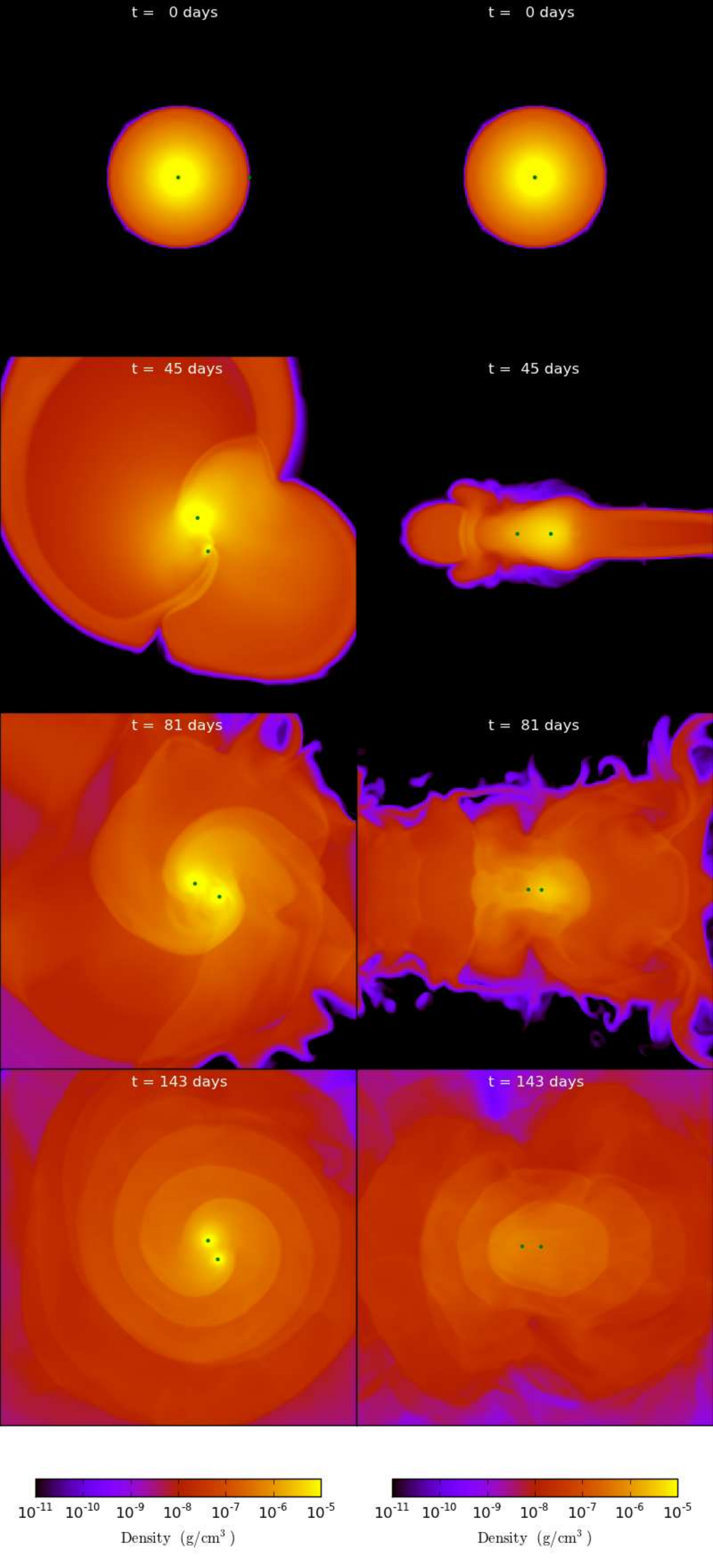}
\caption{Common envelope event with $0.88M_\odot$ giant and $0.6 M_\odot$ MS star, likely leading to the formation of a close binary. Shown are density slices in the orbital plane (left) and in the perpendicular plane (right) at different times, each panel is $430\ R_\odot$ on a side. Simulations were carried out with the gird-based code ENZO \citep{2005ApJS..160....1O}, and a resolution of 256$^3$ cells. The image was created for this review by J.-C.Passy. For more details on simulations see \citep{Passy11} and \S~\ref{meth_comp}.}
\label{fig_passy}
\end{centering}
\end{figure}
\end{centering}

\item {\bf III:  {Self-regulating} spiral-in}

The envelope may expand enough that the spiral-in slows down. In this way a self-regulating state can be formed, in which frictional luminosity released by the spiral-in is transported to the surface where it is radiated away \citep{mmh79}.  This is expected to happen, for example, in some cases if the rate of spiral-in is determined by the local density in the region of the secondary: too little instantaneous heating means that the local density rises, increasing the rate of spiral-in and therefore heating (and vice versa).  This phase is non-dynamical and operates on the {\it thermal time-scale} of the envelope.  How this difference in time-scale affects the energetics of CEE is discussed in \S \ref{sec:energy}.   Recent hydrodynamic simulations of phase II have found non-local energy dissipation \citep{Ric08,Passy11}; if those long-range effects continue to dominate beyond the initial dynamical spiral-in then it is less clear whether a self-regulating state is likely to form. 

\item {\bf IV: Termination of the self-regulating phase}

The self-regulated spiral-in ends with the ejection of the envelope (e.g., via delayed dynamical ejection, \citealt{Iva02}, \citealt{han02}), or when either of the secondary or core of the primary overfills its Roche lobe. The second case can result in a (slow) merger \citep{Iva02, Iva03}, but also provides a further route for envelope ejection \citep{ips02,podsi10}. This phase takes {\it several dynamical time-scales}.

In principle, a self-regulated spiral-in (`phase III') could also be followed by another dynamical plunge (`phase II') if the mechanism maintaining self-regulation somehow ends. That plunge could in turn be followed by another self-regulated phase. It is not clear how unlikely such a repeat is to happen in reality, but there seems to be no first-principles physical reason why the sequence of phases could not be I--II--III--II--III--[...]--IV  in some cases.   

\item{\bf V: Post-CE evolution}

The final properties of the post-CE system are not necessarily set until some time after envelope ejection. For example, the eccentricity of a surviving binary can be changed by any remaining circumstellar matter, which might well include a circumbinary disk. Thermal evolution of the remnant cores might drive further mass transfer, and winds from the remnant cores could widen the system. (For more details see \S~\ref{sec:observations}.)

\end{itemize}

\section{The energy budget during CEE}
\label{sec:energy}

The standard way to predict the fate of a common-envelope
phase is known as the {\bf energy formalism}
\citep{vdH1976,Webbink84,Livio88,IbenLivio93}, in which the energy difference between the
orbital energies before and after the event is compared with the
energy required to disperse the envelope to infinity, $E_{\rm bind}$. 
\begin{equation}
E_{\rm bind} = \Delta E_{\rm orb} = E_{\rm orb,i} - E_{\rm orb,f} = -\frac{ G m_1 m_2} {2 a_{\rm i}} + \frac{ G m_{1\rm,c} m_2} {2 a_{\rm f}} 
\label{enform}
\end{equation} Here  $a_{\rm i}$ and $ a_{\rm f}$ are the initial and
final binary separations, $m_1$ and $m_2$ are the initial star masses
and $ m_{1\rm,c}$ is the final mass of the star that lost its
envelope $m_{1,\rm env}$. As not all the available orbital energy can be used to
drive the envelope ejection, the concept of \emph{common-envelope
efficiency} is introduced, which is parametrized as $\alpha_{\rm
CE}$. This is the fraction of the available orbital energy which is
usefully used in ejecting the envelope.

We could alternatively state the energy budget for CEE by
writing that the combined total energy of the
immediate products of CEE cannot be greater than the total
energy of the system at the onset of CEE.
This statement plus a few approximations leads to 
Eq.~\ref{enform}. We also need to decide
which physical contributions should be counted
in this energy budget, but if they are physically complete
then $\alpha_{\rm CE}$ should never need to exceed unity.

\begin{figure}
\begin{center}
\includegraphics[width=1.\textwidth]{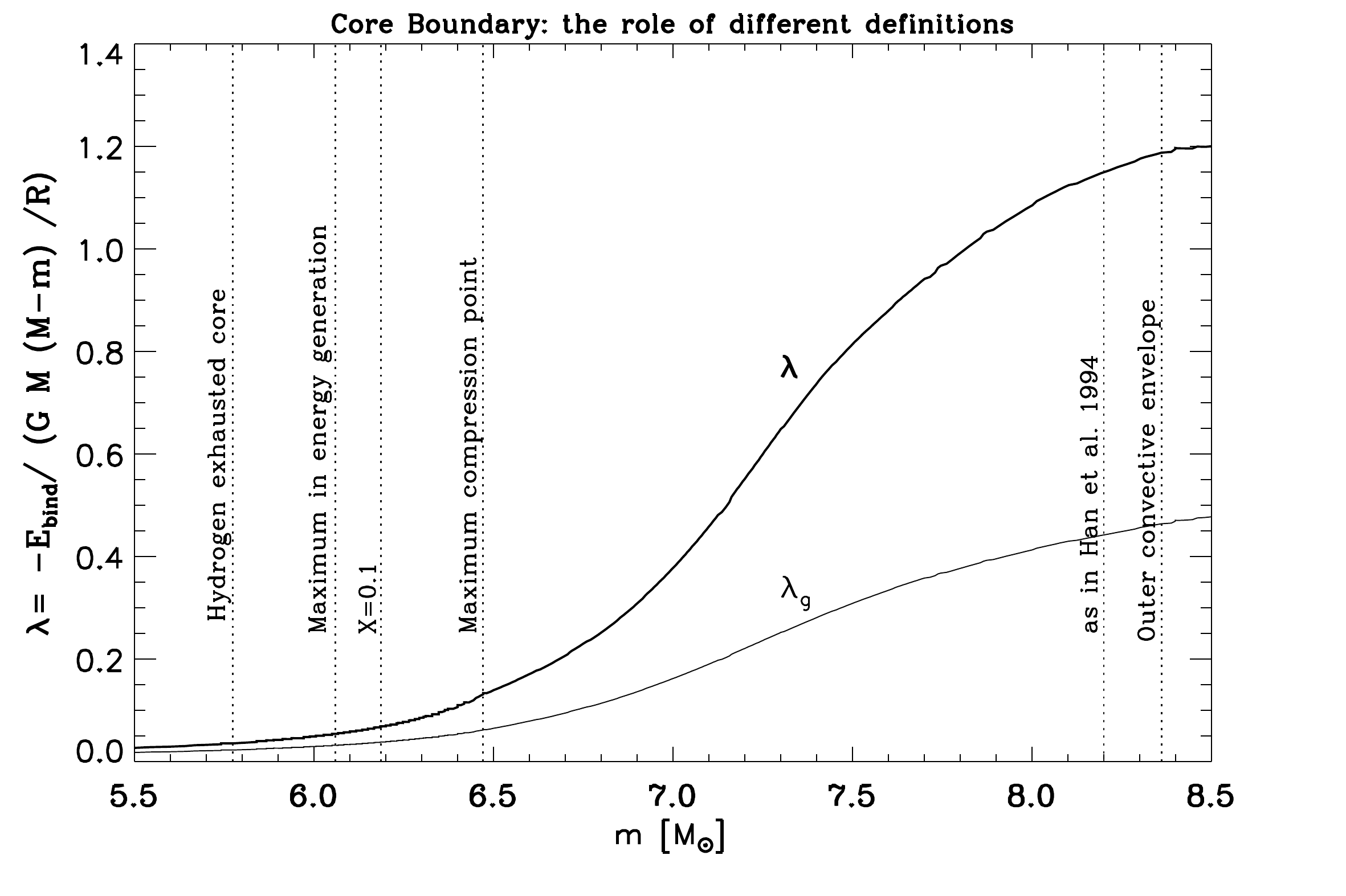}
\caption{$\lambda$ as a function of mass shown on example of 20 $M_\odot$ star when it has $R=750 R_\odot$ ($Z=0.02$, overshooting $0.2$ of the pressure scale
and no wind loss). For comparison shown $\lambda_{\rm g}$ when only gravitational binading energy is taken into account (thin solid line) and when internal energy is taken into account as well (thick solid line). Dotted lines correspond to several possible core definitions, as discussed in \S4.1.}
\label{lam_core}
\end{center}
\end{figure}

There are subtly different ways of writing the energy formalism.
However, all implictly assume that the ejected material departs with
\emph{precisely} the local escape velocity, i.e.  $\alpha_{\rm CE} = 1$ does not
only imply perfect energy transfer, but also \emph{perfect
  fine-tuning}.  Since kinetic energy scales as the square of
velocity, matter would need to escape within a factor of $\approx$1.4 of the
escape velocity for $\alpha_{\rm CE} > 0.5$ to be allowed.

A significant technical improvement in the application of this formalism was the
inclusion of a second parameter, $\lambda$, to account for the particular
structure of each star in calculating $E_{\rm bind}$ for that star
\citep{deKool1990,Dew00,DewiTauris2001}. Following this
  addition, the most commonly used form for the energy formalism in
  population studies is now: 
\begin{equation}
\frac{m_1 m_{\rm 1, env}}{\lambda R_1} = \alpha_{\rm CE} \left ( -\frac{ G m_1 m_2} {2 a_{\rm i}} + \frac{ G m_{1\rm,c} m_2} {2 a_{\rm f}}  \right )
\end{equation}
\noindent This expression allows the two free parameters
to be simply joined into a single unknown, $\alpha_{\rm CE}\lambda$, and this
convenient combination can be commonly seen in population synthesis
papers. Of course, using a global value for the product $\alpha_{\rm
  CE}\lambda$ does lose the advantage gained when using $\lambda$ to
describe the individual binding energy of specific stars. 

We note that different
definitions of $\lambda$ exist in the literature, depending on whether
the authors include only the contribution from gravitational binding
energy or also the internal energy of the star (see Fig.~\ref{lam_core}).
The value of
$\lambda$ can change greatly between stars, so using a global value in
calculations is unsatisfactory. An important physical question
associated with this is how to determine
the boundary between the remnant core and the ejected envelope, since
$\lambda$ can be extremely sensitive to that location \citep{Tau01};
this is discussed in \S \ref{sec:fate}.

Note that the envelope does not just need to become unbound
from the giant, as it must also be lost from the binary.
Eq.~\ref{enform}, even when using detailed binding-energy calculations
for the giant star, neglects this. (One way of thinking about this is
that the zero of potential
energy for the envelope is redefined between the initial and final
states.) The appropriate correction would
usually be small, but it is often forgotten. 

When calculating  $E_{\rm bind}$, it is vital to know whether to include
\emph{only} the gravitational terms. \citet{Webbink84} performed a
full integration over both the
gravitational binding energy and the thermal energy of the gas, since
they are inextricably linked (but did not include recombination 
energy, for which see \S \ref{sec:recombination}), but early
parametrisations only included the gravitational terms. Physically it might 
be preferable for, e.g.,
the thermal energy of the gas to be thought of as a potential source of energy
rather than as something which reduces the magnitude of the binding
energy; in either case we need to think about how internal energy
might be converted to mechanical work if it is to help eject the envelope.
This depends partly on the timescales over
which the CE event happens, as we will discuss below. 
Likewise, those timescales help to control whether
other energy sources can contribute to the ejection besides the
orbital energy reservoir. 

\subsection{Applicability of the energy formalism: timescales and energy conservation}
\label{sec:energycons}

It is crucial to realise that the standard energy formalism (as in
Eq.~\ref{enform}) was introduced to explain a common envelope event as an event
taking place on a dynamical timescale. The formalism also presumes that
only the energy stored in the binary orbit, or in the initial internal energy of the common envelope,
could play a role in the envelope ejection.   If the energy formalism is mis-applied (for
example, to quasi-conservative -- thermal or nuclear time scale --
mass transfer) then
artifacts like an apparent efficiency greater than unity
($\alpha_{\rm CE}\gg 1$, i.e. non-conservation of energy) could easily
take place. This would clearly be misleading and unphysical,
  but the situation could arise since this approximation neglects some
  potentially-important energy sources and sinks.  
 Among the likely sinks are radiative losses from the common envelope and
 energy stored in microscopic or macroscopic degrees of freedom
 (i.e.\ internal energy of the matter and terminal kinetic energy of
 the ejecta). 
 Prospective sources are nuclear energy input -- either from burning at
 the base of the common envelope or from burning ignited at the
 surface of the accretor -- and accretion energy from matter retained
 by the companion star. 
 Note that, although mass transfer involves the liberation of gravitational
 potential energy to heat the accreted envelope, this exchange of
 gravitational potential energy for thermal energy neither introduces
 new energy sources nor new energy sinks.

The longer the CE phase lasts, the more opportunity there is for
deviation from the energetically closed system described above.
For example if the event takes place on a thermal timescale
or longer, then energy lost in radiation from the
envelope's photosphere might have to be taken into account. For static
equilibrium models we might feel justified in assuming that this loss is balanced by 
heating from the stellar core, but this is unlikely to remain true as the
star's structure alters during the CE event. Either the radiation from the surface or heating from the core might
be larger in different CE events. Predicting future
radiative losses in general would be challenging if not impossible.

Similarly, predicting the details of changes in the nuclear energy
sources during CEE is not straightforward, since their output might increase (see
\S \ref{sec:nuclear_energy}) or fade away due to adiabatic expansion of the core in response
to mass loss. Qualitatively, however, it seems reasonable to expect that if the donor star is in thermal
equilibrium at the onset of mass transfer, 
then radiative losses initially balance nuclear energy input. 
Then radiative losses seem likely to grow relative to input from nuclear
sources. This is because we anticipate that the 
emitting area will probably increase whilst the nuclear sources, if
anything, seem most likely to decline in output, since
the internal decompression attending mass
loss will tend to quench nuclear burning.

Qualitatively it is also possible to argue that accretion
  during CEE is not commonly significant for non-degenerate companion.
The common envelope itself typically possesses much higher specific
entropy than the surface of the accretor, with the consequence that
matter accreted by the companion star reaches pressure equilibrium at
the surface of that star with much higher temperature, and vastly
lower density, than the accretor’s initial surface layer.  A
temperature inversion or roughly isothermal layer is expected to
bridge this entropy jump with the result that, over the duration of
the CEE (which is much shorter than the thermal time scale of the accretor), the
accretor is thermally isolated from the common envelope, while the
common envelope itself becomes increasingly tenuous.  If this picture
is correct then one would expect
very little net accretion onto a non-degenerate companion star \citep{1988covp.conf..403W,1991ApJ...370..709H}.
For degenerate companions, in this same context, the ignition of nuclear
burning at the surface of the accretor might be inhibited by the very high
entropy of accreted material -- which would be extremely buoyant, and
difficult to compress to ignition conditions -- although detailed simulations 
of the process should be performed (see also \S \ref{sec:acc_en} and
\S \ref{sec:hypercritical}).

It should be clear that it is very difficult to make any general
statements once the common envelope ejection is non-dynamical. 
Once the spiral-in or the envelope
preliminary expansion takes place on a timescale longer than
dynamical, energy conservation in the simple original form above \emph{is not
expected to work}.

\subsection{Relating loss of orbital energy to heat input and outflow of the envelope}

The energy from orbital decay is often assumed to thermalise
locally, typically by viscous dissipation in the region of the in-spiralling secondary.
However, hydrodynamic simulations \citep{Ric2011} form large-scale spiral waves, with tidal
arms trailing the orbit of the binary.  
Spiral shocks transfer angular momentum to the matter in the envelope. 
Furthermore, some of the energy in those spiral shocks will be dissipated as
heat a long way from the secondary. 

It also seems possible that some matter is flung out as a result
of these spiral waves, i.e.\  orbital kinetic
energy is directly transferred to the kinetic energy of the envelope. 
If the spiral-in ends during the dynamical plunge-in, without entering a self-regulating phase, 
then a significant fraction of the orbital energy transferred to the
ejected envelope might not have been thermalised.  

Avoiding a thermal intermediate stage would have the clear
advantage that the energy input is less likely to be radiated away,
but might reduce the chance that any other heat source could help with
that part of the ejection. If we could decide to what extent the
envelope is ejected directly (by kinetic energy imparted from spiral shocks) or indirectly (by heating and a
pressure gradient) -- an apparently simple distinction -- then it
might help us conclude how much the suggestions in the following
subsection are likely to be helpful. 

The results of \citet{Ric2011} are discussed in more detail in \S
\ref{meth_comp}. Here we note that $\approx$ 25\% of the envelope is
ejected during their dynamical plunge-in calculations. The
distribution of entropy production within the envelope
may well be different during any subsequent self-regulated
spiral-in. The dominant driving mechanism for further envelope loss might
therefore also change.

\subsection{Is orbital energy the only relevant source of energy?}

Section \ref{sec:energycons} hopefully made it clear that there could
easily be scope for additional sources of energy to participate in CE
ejection. In the following we discuss several possibilities. The first
is widely accepted, though physically unproven to help, but the others
are less normally included.

\subsubsection{Internal Energies}

It has become standard practice to include the internal energy of the
envelope in CE binding energy calculations. It is arguably physically clearer to think of the
internal energy reservoir as another energy source, and we shall do so
here, but it is also
natural to modify the definition of $E_{\rm bind}$ such that it becomes the sum of the potential energy and
internal energy of the envelope.  This has typically been calculated using detailed
stellar models via:

\begin{equation}
E_{\rm bind} = - \int_{\rm core}^{\rm surface} \left ( \Psi(m) + \epsilon (m) \right ) dm  
\label{ebind}
\end{equation} 
Here $\Psi(m)=-Gm/r$ is the gravitational potential
and $\epsilon$ is the specific internal energy.  If integrated over
the whole star, Eq. \ref{ebind} gives the total energy of the star. 
However, when applied only to a part of the star, it is no longer
formally valid, in part due to how gravity is taken into account.

This contribution of internal energies was first explicitly applied by \citet{Han+1994}, and can make a very large
difference to the energetic ease of envelope ejection during some phases of
stellar evolution.  Some authors only allow a fraction of the available
internal energy reservoir to contribute to the ejection, in which case a second efficiency parameter,
$\alpha_{\rm th}$ is used to denote the fraction of the internal energy
which is available to help eject the envelope. 

Eq. \ref{ebind} neglects the response of the core, which we discuss
further in \S\ref{sec:fate}. Here we note that, if the core expands
during mass loss, this could do mechanical work on the envelope. So
the binding energy should formally be calculated as the difference
between the initial ($E_{i}$) and final ($E_{f}$) total energies of the star:
\begin{equation}
\label{ebindWebbink}
\begin{split}
E_{\rm bind} = E_{i} - E_{f} = - \int_{\rm centre}^{\rm surface,i}
\left ( \Psi_{i}(m) + \epsilon_{i}(m)\right )  dm  \\
+ \int_{\rm centre}^{\rm surface,f} \left ( \Psi_{f}(m) + \epsilon_{f}(m) \right ) dm
\end{split}
\end{equation} 
where the integrals are now through the whole star, not
just the envelope \citep[][]{Ge10,del10}. For stars with degenerate
cores it seems unlikely that this correction is large, but it has not yet
been definitively shown to be unimportant.

It is not guaranteed that the internal energy should make a
significant contribution. The simplest physical version of this change seems to presume that
a significant part of the envelope expansion is subsonic, i.e. that pressure equilibrium can
be maintained. Otherwise the envelope's gas would seem unable to
transfer its internal energy into envelope expulsion via thermal pressure.

Furthermore, some stars appear marginally unbound when their
internal energy is included in the binding energy calculation, yet
they retain their envelopes.
Evidently, a net excess of internal energy over gravitational binding energy is not a
sufficient condition to unbind the envelope, even when this situation is maintained over many dynamical timescales.
Of course it is easy to speculate that 
the CE event might somehow trigger the release of this energy.\footnote{
At least some population synthesis calculations \citep[see, e.g.,][]{Han+2003} have found better agreement
with observations for particular classes of system by including the internal energy reservoir.}
Arguments have been made that positive internal energy is the
condition which determines spontaneous envelope ejection for single stars, and
that this helps to match the intial-final mass relation
\citep{Han+1994,Meng+2008}. If this is the case, then at metallicities
$\gtrapprox 0.02$, stars with initial mass $\rm \lessapprox 1.0
M_{\odot}$ do not ignite helium \citep{Meng+2008}.

\subsubsection{Internal energy, thermal energy and recombination energy}
\label{sec:recombination}

It seems worth exploring the details of the `internal energy' term
included in Eq.~\ref{ebind}. In particular, we wish to highlight that 
the contributions used  separate into two distinct groups. 

The natural components of internal energy are the thermal terms
familiar from kinetic theory, which we collectively label $U_{\rm th}$. These measure the energy of the matter
relative to the state where stationary (cold) electrons and ions are separated to
infinity, i.e. the natural zero-energy state.  This combines the
internal kinetic energy of the particles and the energy stored in radiation.
Per unit volume, we write:
\begin{equation}
\frac{U_{\rm th}}{V} = a T^{4} + \sum_{particles} \sum_{d.o.f.} \frac{k_{\rm B}T}{2}
\end{equation}
where the summations are over the particles (including molecules) present, and their
available degrees of freedom. (We have not written down the corrections to the electron
energies due to Coulomb interactions and degeneracy, which are not
likely to be significant in stellar envelopes.)

The second set of contributions arise because we expect that more
energy than $U_{\rm th}$ is available to be released from the matter
in the envelope during envelope ejection. The plasma can recombine
and some atoms will form molecules; those processes will
release binding energy. This extra store of available energy is typically
referred to as \emph{recombination energy}, $\Delta E_{\rm recomb}$.
It can be calculated by adding the appropriate ionisation and
dissociation potentials for each ion and atom present, though it is
usual to neglect dissociation of any other molecule than $\rm
H_{2}$. We note that recombination energy was suggested much earlier to
be a potential driving mechanism for the ejection of ordinary
planetary 
nebulae \citep{Lucy1967,Roxburgh1967,Paczynski1968}.

These two, very different, components have been mixed into `internal
energy' when discussing envelope ejection and stellar binding energies
\citep[see, e.g.,][]{Han+1994, han02}.  One of the reasons why this
might be physically confusing is that recombination energy does not contribute to 
the standard internal energy which enters the virial theorem. 
This is also one of the reasons why recombination energy is
potentially helpful in CE ejection. For a stellar envelope
which is dominated by gas pressure such that the gravitational binding
energy is $U_{\rm th}/2$ then, if $\Delta E_{\rm recomb} = U_{\rm
  th}$, the star's envelope would be formally unbound even before CEE.

Their relative magnitude can be crudely estimated by comparing the value of
$k_{\rm B}$  (i.e. ${\rm 8.6 \times 10^{-5}~eV~K^{-1}}$) with the
ionisation potentials of hydrogen and helium (79.1 eV/ion for He, 13.6
eV/ion for H). Assuming a 10:1 ratio of hydrogen to helium (by number)
gives an average of $\approx$ 20 eV available per ion, in which case energy stored in
thermal terms dominates energy stored in the ionisation state of
the plasma for temperatures above $\sim 2 \times 10^{5}$ K.  

So there seems very likely to be a strong contrast in where the energy release from these
two components will happen. The thermal
terms, with specific energy $\sim 3/2 k_{\rm B}T$ per particle in most giant envelopes, 
will store and release energy at high temperatures, i.e.\ deep within the
star. The release of binding energy during recombination and
molecule formation will take place at relatively low temperatures. 

The fact that the gravitational potential well is deepest far from
the possible recombination zones seems worth pursuing. This might help 
explain how internal energy can help CE ejection, even though stars which
are marginally unbound after calculating the integral in
Eq.~\ref{ebind} (when including recombination terms) are stable
to perturbations.  When the CE spiral-in has made the 
envelope expand and cool enough then 
recombination would be triggered, perhaps giving the final push
to make a loose envelope unbound. 

On the other hand, it is also possible that
recombination energy is liberated so close to the surface that it is
is more easily convected to the surface and radiated away. 
The helium recombination zones in red giants are typically well
below the photosphere (at optical depths $\gg$ 100), so if the
giant structure is roughly preserved during CEE then we do expect
the energy from recombination to be thermalised. Even if the envelope 
above the recombination zone became optically thin in the continuum,
line-driven expansion might still be favoured by remaining optically
thick in the recombination lines. However, there is
very little mass above those recombination zones, and the recombination
zones themselves tend to help drive convection.  

The distinction between the recombination and $kT$ components 
is not normally made.  It may be that using a single $\alpha_{\rm th}$ parameter for all
internal energy contributions is
currently sufficient for use in population synthesis, and we should
certainly be
careful about introducing yet another fitting parameter. 
Nonetheless, if we aim to
understand the physics underlying CEE then in future work it seems
sensible to aim to deal separately with the thermal and recombination terms.

\subsubsection{Tidal heating}

Tidal heating is sometimes discussed as an additional effect which
might help the envelope ejection, and sometimes presumed
to work more efficiently than orbital energy taken into account in
the energy formalism. This deserves a special note of clarification.
Tidal heating is clearly \emph{not} an energy source but
rather a transfer \emph{mechanism}, taking energy out
of the binary orbit and stellar spin. 

The orbital energy reservoir is no larger than
if tidal heating is ignored, and that contribution has already been
taken into account in the energy budget even in the original energy
formalism. In this respect then tidal heating obeys exactly the same
law of energy conservation as would dynamical spiral-in. 

In principle there might be a small correction, due to the energy stored in the
stellar spin, whilst corotation is enforced. Energy stored in spins is
usually ignored in the energy balance equation. Yet it only seems likely
to be at all helpful if the giant is rotating faster than corotation, and is spun
\emph{down} as tides take effect. This is the opposite of the
strongly expected situation.  Indeed, taking into account spin
energy in the overall energy budget seems most likely to make the situation
\emph{worse}: some of the available orbital energy will go into enforcing corotation.

Moreover, the tidal heating timescale seems likely to be longer than
that of the  dynamical spiral-in.  In which case, the star can lose
more of this orbital energy via radiation from the surface layers than
if tidal heating was ignored. So potentially tidal heating can
\emph{decrease} the efficiency if energy conservation is applied
using Eq.~(\ref{enform}). 

So, for several reasons, invoking tidal heating should 
not increase the amount of energy available to eject the envelope.
It should not result in $\alpha_{\rm ce} > 1$.

\subsubsection{Nuclear energy}
\label{sec:nuclear_energy}

Another energy source that could play a role in the envelope ejection
is nuclear fusion  \citep{Iva02,Iva03}.  If one considers a binary
that is doomed to merge, but does not yet merge 
during the dynamical plunge-in phase, then during the self-regulating
spiral-in phase a non-compact companion (e.g., a main sequence star)
will, at some point, start to overfill its Roche lobe.  This can be
considered to be the end of the normal spiral-in.  
Due to continued frictional drag from the envelope on the mass-losing companion,
the orbit continues to shrink, forcing the mass
transfer to continue and even to increase.  A stream of hydrogen-rich material
can then penetrate deep into the giant's core, reaching even the
He burning shell and leading to its complete explosion \citep{ips02},
since the released nuclear energy during explosive hydrogen burning
could exceed the binding energy of the He shell (in massive
stars this can be a few times $10^{51}$ erg).  The rest of the CE is 
much less tightly bound and is also ejected during the
same explosion. This leaves behind a compact binary consisting of
the core of the giant and whatever remains of the low-mass companion
after the mass transfer.  The companion is not expected to remain Roche-lobe
filling immediately after the explosion.

Such  {\it explosive CE ejection}  could both help a less massive
companion to survive the CE  (this makes the formation of low-mass
black-hole X-ray binaries more plausible). It also seems to naturally produce a fast-rotating core
which has been stripped of both hydrogen and helium \citep{podsi10}. 
The remnant star could then produce both a long-duration
$\gamma$-ray burst and a type Ic SN, helping to explain their observational connection.

\subsubsection{Accretion energy}
\label{sec:acc_en}

Another potential source of energy is the luminosity of accretion onto
the secondary during the common envelope phase \citep[see, e.g.,][]{Iva02,VossTauris2003}.  
The Eddington luminosity would release $\sim 5\times10^{45}$ ergs per year per
$1~M_\odot$ of the accretor. In which case, if a slow spiral-in lasts from 100 to 1000
years, the energy released through accretion could become comparable
to the energy release from the binary orbit via tidal interaction and viscous
friction \citep[for the comparison of contributions in the case of
different masses for a donor and a giant, see ][]{Iva02}.  
In most cases, standard methods predict that the available accretion
rate for an inspiralling companion exceeds its Eddington-limited accretion rate. 
However, hydrodynamical simulations found that whilst the spiral-in is still dynamical,
the commonly-used Bondi-Hoyle-Lyttleton prescription for estimating the
accretion rate onto the companion significantly overestimates the true
rate \citep{Ric2011}, in which case the contribution of accretion to
the energy budget could easily be negligible (see also \S \ref{sec:hypercritical}).

The balance between orbital energy release and accretion luminosity
should change at different stages of the CE process. 
When a compact object is orbiting inside the outer regions of the
envelope of the giant (where the binding energy per unit mass is low
and the spiraling-in timescale is long) then it seems easiest for
accretion energy release to dominate orbital energy deposition. 
A special case of accretion energy release would occur if an
inspiralling compact object orbits deeply enough to cause
the core to overfill its Roche lobe \citep{Sok2004}. 
This might cause a brief, powerful release of accretion
energy to help envelope ejection.  If that process occurrs, it might
disfavour the formation of Thorne-\.{Z}ytkow objects \citep{ThorneZytkow1975,ThorneZytkow1977}. 

Accretion energy release might be able to help envelope ejection in
ways other than via heating. Kinetic outflows -- jets -- might be driven by
accretion onto an inspiralling compact companion. \cite{Sok2004}
argued that this should be the expected outcome for an inspiralling WD or NS.
Many parameters are poorly determined for this entire process, but
Soker argues that the jets can blow hot bubbles within the envelope, 
causing some mass loss and potentially slowing the spiral-in.

\subsection{Does enthalpy help to unbind the envelope?}
\label{sec:enthalpy}

Above we have given some possible extensions to the canonical
energy formalism. In particular, we have explored a set of potential
additional energy sources which might help unbind the
envelope. However, it has recently been proposed by \cite{Ivach11} that the standard
framework is seriously physically incomplete if the CE ejection
happens during the self-regulating phase. 

In particular, \cite{Ivach11}  argued that the condition to start outflows is similar to the energy requirement
in Eq.~\ref{ebind}, but with an additional $P/\rho$ term,
familiar from the Bernouilli equation:

\begin{equation}
E_{\rm flow}= - \int_{\rm core}^{\rm surface}\left (\Psi(m) +  \epsilon(m) + \frac{P(m)}{\rho(m)} \right ) dm \ .
\label{bernsimple}
\end{equation}

\noindent Since $P/\rho$ is non-negative, the condition to
start outflows during slow spiral-in occurs before the envelope's
total energy become positive. As a result, this ``enthalpy'' formalism
helps to explain how low-mass companions can unbind stellar envelopes 
without requiring an apparent $\alpha_{\rm CE} > 1$.  
Although this consideration may change the requirements for the energy
budget, we emphasize that this was derived without reference to the
total energy budget for envelope ejection, and it arises from a condition that
separates stable envelopes from envelopes that are unstable with
respect to the generation of stationary outflows.

This would be a radical change in the standard picture of CE
energetics; understanding this question is clearly important.
An energetic debate over whether the arguments in \cite{Ivach11} are correct is
still continuing, and we outline two opposing points of view below;
there are others.

\subsubsection{Against: energy redistribution during dynamical envelope ejection}
\label{en_dyn}

The $P/\rho$ contribution in the Bernouilli equation expresses the
fact that the pressure gradient helps to accelerate the envelope outwards.

Hence the gas expelled from the outer regions carries more kinetic energy than what
would be calculated without the work of the pressure included. But
this energy comes at the expense of the energy of the inner regions of
the envelope. So the $P/\rho$ term is important, but this
only redistributes energy rather than being a new, previously forgotten,
energy source.

This can be demonstrated by a simple case. Consider a gas of adiabatic
index $\gamma$ with a uniform initial pressure $P_0$ and initial
density $\rho_0$, occupying a cylindrical pipe in the region $x_l < x
< x_r$ (where $x_{l}$ is left and $x_{r}$ right, corresponding to the
inner and outer edges of the envelope). At t=0 the valve at $x_r$ is opened. 

This classic problem is solved in $\S 99$ of \cite{LanLif}. 
The velocity of the gas at the right (outer) edge reaches a value of $v_r=2 C_s/(\gamma-1)$, where $C_s=\gamma P_0/\rho_0$
is the initial sound speed. Its specific kinetic energy $2 C^2_s/(\gamma-1)^2$ [e.g., $(9/2) C^2_s$ for $\gamma=5/3$],
is much larger than the initial specific internal energy $C^2_s
/\gamma(\gamma-1)$ [e.g., $(9/10) C^2_s$ for $\gamma=5/3$]. 
This `extra' energy comes at the expense of the energy of gas elements
further to the left (i.e.\ further inside).
A rarefaction wave propagates to the left and reduces the
internal energy of the gas there. The further to the left a mass segment is, the lower its velocity is.

The same qualitative flow structure holds for the ejected CE. The
pressure gradient accelerate the outer parts of the envelope at the
expense of the inner parts. The energy is unevenly distributed: the
outer parts escape with a speed much above the escape velocity, but
the very inner parts might not reach the escape velocity. They will
fall back, unless extra energy is deposited to the still-bound
envelope segments.

This uneven energy distribution is clearly shown for a case where
the energy is deposited over a short time in the inner part of the
envelope \citep{Kashi2011}. The inner parts of the envelope expand at
velocities below the escape velocity. They fall back to the binary
system. If they contain sufficient angular momentum, a circumbinary
disk might be formed. Note, however, that this may no longer be valid
if the orbiting companion continues to add energy at the base of the
envelope, or if heat can flow outwards from the core on a short enough
timescale.

To maintain a negative pressure gradient (that accelerates outward) in the inner regions during the
ejection process, the bottom of the envelope must gain sufficient heat
from the core (which requires a sufficiently long timescale for
ejection), or by continued energy input from the binary (the
conditions on which are unclear). 
However, in the simple case where the envelope is energetically isolated after
the start of envelope ejection then the $P/\rho$ term only
redistributes energy within the envelope.

\subsubsection{For: Outflows during self-regulating spiral-in}
\label{en_ther}

The arguments above assume that the ejection timescale is short, but the derivation of
Eq.~\ref{bernsimple} implicitly required that ejection happens on a
thermal timescale.   The arguments which lead to the use of Eq.~\ref{bernsimple} rather
than Eq.~\ref{ebind} were based on considering stellar stability
criteria.  The original assumption for the energy formalism is
that the energy required to eject the envelope equals $E_{\rm
  bind}$.  This is based on either of two assumptions: that an envelope is dispersed once its
total energy $W_{\rm env}>0$, or that an envelope with $W_{\rm env}>0$ is unstable.
The connection between $W$ and $E_{\rm bind}$ presumes that $E_{\rm bind}$
is in fact $W_{\rm env}$. But \cite{Ivach11} argued that those assumptions are not
foolproof, as both a star with $W>0$ can be \emph{kinetically
  stable}  \citep{BKZ67}, and a star's stability condition against
adiabatic perturbations is not the same as having $W>0$.

\cite{Ivach11} instead considered quasi-steady surface outflows, which 
would develop on the same timescale as it takes for the envelope to redistribute heat
released during the spiral-in, i.e. the thermal timescale of the
envelope.  These outflows could only take place if slow
spiral-in occurred, not during a dynamical plunge-in phase.
It is important to realise that such {\it steady} flows do not behave
the same way as the {\it non-stationary} flows described in \S~\ref{en_dyn}.
Since the base of the envelope could have time to take energy
from the core, the final total energy requirement for envelope ejection
might be more than that given by Eq.~\ref{bernsimple}. 
However, the energy which might be released
by the reaction of the core cannot easily be evaluated at the
start of the CE phase; full mass-loss calculations would be needed.

\subsubsection{Summary}

Whether enthalpy helps with CE ejection
may therefore be determined by the timescale over which the ejection occurs.

\emph{Both} arguments above might be correct in different binary systems. If the envelope can be
ejected during the dynamical plunge-in, then the envelope may act as a closed energetic
system (depending on the timescale of ejection compared to the
timescale of energy input from the binary orbit). 
But if that rapid ejection doesn't happen, and the spiral-in reaches the
self-regulating phase, then it may becomes possible for quasi-steady
outflows to develop on the thermal timescale of the envelope, and also
for further heat input to come from the core or from the binary orbit. 
In cases where the $P/\rho$ term
only acts to redistribute energy within the ejected envelope then it
might make the overall ejection more \emph{difficult}, in other cases
it might be helpful. A priori it is not clear which situation is more
likely to be common.

Although it is still unclear to what extent enthalpy helps with CE
ejection, both sides of the debate above suggest that the
$P/\rho$ term might be vital in determining the point which defines the depth from which the envelope
is ejected, i.e. the bifurcation point which separates the material
which remains bound from the material which escapes.
How to physically determine this location will be addressed
in the next section.

\section{The end of the CE phase \& the fate of the remnant}
\label{sec:fate}

The previous section discussed the widely-used energy formalism and
variations upon it. The main question which the energy formalism is trying to answer
is {\it where does the spiral-in stop?} That is: can we take the
initial conditions of a CE phase and predict the outcome? If a merger
is avoided, what does the remaining binary look like? 

It is not sufficient to conclude that there is, in principle, sufficient
useful energy available to eject the envelope.  Perhaps spiral-in does
stop as soon as sufficient gravitational potential energy has been
released to unbind the envelope, but this standard assumption is at
best crude. Physically, it might be that (almost all) the envelope is
ejected, but the spiral-in of the companion still continues until it merges with
the core. (A post-CE binary might alternatively merge during thermal
relaxation following envelope ejection.)

Even more fundamentally, it is not trivial to define the boundary
between the `core' and the `envelope'. Nor is it clear how close that boundary is
to the bifurcation point which separates the material which is ejected from
that which stays bound.  So far we have treated these points as if
they were well-known, but they are not. These locations are 
needed in order to calculate $E_{\rm bind}$ correctly, and different
definitions can lead to large differences in CE outcome \citep{Tau01}.

\subsection{Locating the bifurcation point}
\label{sec:thebifurcation}

A number of possible criteria can be found in the literature which 
aim to define the boundary between 
the remaining core and the ejected
envelope. Some are related to plausible definitions of the core
mass, some attempt to predict a natural bifurcation point on other
grounds. 

Obvious possibilities are the minimum possible core mass (the
hydrogen-exhausted core) and the maximum possible core mass 
(the transition between the radiative zone
of the H-burning shell and the bottom of the outer convective
envelope). Equivalent descriptions of the latter point include:
 (i) where the entropy profile has a transition between the
increasing and flat parts \citep{Tau01}; (ii) where the effective
polytropic index is discontinuous \citep{Hje87}. The proposed
conditions can be grouped into three main categories as follows:

\begin{enumerate}
\item connected to the nuclear energy generation: 
\begin{itemize}
\item at the maximum nuclear energy generation within the H shell  \citep{Tau01}
\item at the maximum nuclear energy generation plus a condition on the mass of the remaining envelope, which itself is a function of the evolutionary status of the donor \citep[for low mass red giants and asymptotic giant branch stars,][]{Mar11}
\item where the nuclear energy generation falls below some threshold  \citep{Mar11}
\end{itemize}
\item based on the chemical composition:
\begin{itemize}  
\item the central mass which contains less than 10\% hydrogen \citep{Dew00}
\item core is everything below the location where $X=15\%$ \citep{Xu10}
\end{itemize}
\item connected with thermodynamic quantities:
\begin{itemize}
\item where $\partial^2 \log \rho /\partial^2 m =0 $ within the H-burning shell \citep{Bis98}
\item where the function of the binding energy $y=\sinh^{-1}(E_{\rm bind})$ has the transition  
between a sharp increase and a fairly slow increase in the outer
envelope \citep{Han+1994}
\item where the value of $P/\rho$ is at its maximum within the H-burning shell.
  This could be described as the point of maximum compression, or
  maximum sonic velocity \citep{Iva11}.
\item by using the entropy profile to predict the surface luminosity
  of any possible remnant. Comparison of this predicted luminosity with the
  current nuclear luminosity might suggest whether that potential
  remnant would expand or contract (on a thermal timescale) after
  being exposed. 
\end{itemize}
\end{enumerate}

Not all the definitions are applicable to every star: some only work
for low-mass giants or asymptotic giant branch stars, and some conditions can not be
found or determined uniquely in all the stars  (e.g., the
condition $\partial^2 \log \rho /\partial^2 m = 0 $ does not always
give a unique answer for massive stars). In Fig.~\ref{lam_core} we demonstrate 
how different definitions of the bifurcation point can work.

As the binding energy within the hydrogen shell greatly exceeds the
binding energy of the outer convective envelope, different core
definitions for the same star could lead to final binary separations
different by factors of up to 100 \citep{Tau01,Iva11,Iva11a}; 
for the star illustrated in Fig.~\ref{lam_core} the
  different core definitions predict 
  envelope binding energies which vary by a factor of 34.  It is
therefore of paramount importance to find the bifurcation (core boundary) point
as accurately as possible.  For that, understanding the physical reasons
behind the existence of such a point is very important.

Most of the core definitions are simply ad hoc and do not carry
much meaning except that they could be used as fixed comparison
points between different population studies.  However, some (e.g.,
the thickness of the remaining envelope) are based on a known feature of low-mass
giants (those which have degenerate cores): the ability to re-expand
their envelope back to a giant structure if the remaining envelope mass
exceeds some (small) value \citep{Dei70}.  In general, the envelope
mass that still re-expands needs to be found for every core mass, but
when adopting a criterion it is usually approximated as some small fixed mass.

The definition in which the core is determined by the compression point,
$m_{\rm cp}$, is in some sense a generalization of the case of
low-mass giants described above.  Here the bifurcation is also based on opposite reactions of
different parts of the H shell to the very rapid mass loss; both
immediately after the envelope ejection and subsequently on a
thermal timescale.  However, this type of
divergence point for giant stars exists in all giants, including
massive ones.  In the general case, it can be said that if mass is
removed to below the divergence point then the remnant contracts on its thermal
timescale. On the other hand, if mass remains above the divergence point then the star expands during its
thermal readjustment. During that thermal reexpansion the remnant
could either develop an outer convective envelope or experience
strong thermal pulses. This divergence point does not reliably
coincide with any of the other proposed bifurcation points described above.

Additional characteristics of that bifurcation point 
$m_{\rm cp}$ have been found \citep[for more details see][]{Iva11, Iva11a}, where
the most important is that the energy expense required to shed the
envelope down to $m_{\rm cp}$ is minimal, if both the
expansion during CEE and thermal readjustment after CE ejection are
considered. This is related to the question of whether the enthalpy
formalism for the energy balance should be applied (see
\ref{sec:enthalpy} above), but it seems that $m_{\rm cp}$ should be
the natural bifurcation point whether the ejection is on a thermal
timescale or dynamical. Hence it seems plausible that $m_{\rm cp}$ could be the long-searched-for
and physically motivated point which defines where the spiral-in
stops.

\subsection{Interaction with a post-CE disk}

If not all of the envelope is ejected then, due to
angular momentum conservation and further interaction of the fallback gas with
the binary system, a circumbinary disk may
well be formed \citep{Kashi2011,Mar11}. Various numerical
simulations have also suggested that a substantial fraction of the envelope might stay
bound \citep[e.g.][]{San98,Lombardi06,Passy11}.  That circumbinary disk is
expected to have a thick structure
\citep[e.g.][]{Sok2004,Sok92,San98} and its interaction with the 
binary system may further reduce the orbital separation
\citep{Kashi2011}.  In the context of the energy formalism it would
effectively mean that $\alpha_{\rm CE} \ll  1$ and so in many cases this could lead to a
merging immediately after the dynamical phase of the CE. 
\citet{Kashi2011} find that this effective value of $\alpha_{\rm CE} < 1$ \citep[see
also][]{Iva11} can explain the recent findings of \cite{Mar11}: they also
found that the value of $\alpha_{\rm CE}$ they deduce from
observations is much smaller than what their numerical simulations of
the CE phase give \citep{Mar08,Mar09,Mar11}.

\section{The angular momentum budget}
\label{sec:angular_momentum}

We have previously considered energy conservation as a constraint on
CEE. The total angular momentum of the system should also be
conserved, but as of yet this law has been less widely applied
when studying CEE. 

From first principles it seems surprising that angular momentum would
be the dominant factor in determining the final state of any CEE in which the 
binary separation ($a$) is significantly reduced. This is because most of the 
transfer of angular momentum is expected to happen at
wide separations ($J \propto \sqrt{a}$). In contrast, most of the gravitational energy release
($\propto 1/a$) should occur later in CEE, i.e.\ when the post-CE
separation is being finalised.  Nonetheless, the
physical necessity of angular momentum conservation may be
particularly useful in
understanding the early stages of CEE, and CE phases where $a$ does not
significantly decrease.

\subsection{The plunge}

The need for a dynamical plunge-in at the very beginning of
CEE can be qualitatively understood by considering when the
orbital energy and angular momentum have to be shed from a
spiraling-in binary. If we write the orbital energy $E$ of a binary 
in terms of its eccentricity $e$, 
angular momentum $J$, total mass $M$, and reduced mass $\mu$,
\begin{equation}
E = - \frac{G^2 M^2 \mu^3 (1-e^2)}{2J^2},
\end{equation}
we see that in the limit that $dE \approx 0$ (with $M$ and $\mu$ assumed constant),
\begin{equation}
\frac{de^2}{1-e^2} \approx -2 \frac{dJ}{J}.
\end{equation}
This suggests that, in the regime when the orbital energy $E$ is
almost constant (i.e., at the start of the spiral-in), the
binary's eccentricity grows roughly as fast as angular momentum $J$
is transferred to the envelope.

The orbit is not expected to circularize until $a/R \lesssim r_{\rm
  g}^{4/3}$, where $R$ is the radius of the common envelope, and $r_g$ 
its dimensionless radius of gyration (then $I=r_{\rm g}^2 M R^2$ is the moment of intertia) 
and for giants $r_{\rm
  g}^2\approx 0.1$.

\subsection{The  $\gamma$-formalism}

Considering conservation of angular momentum might
avoid some of the problems with trying to apply energy
conservation that we outlined in \S \ref{sec:energy}. This gives
physical motivation for trying an alternative parametrisation.  In
this subsection we begin by considering such a parametrisation: 
the $\gamma$-formalism, in
which angular momentum is considered to be the deterministic
quantity \citep{Nelemans00,Nelemans05}. The governing equation is:
\begin{equation}
\frac{\Delta J_{\rm lost}}{J_{\rm i}}= \frac{J_{\rm i} - J_{\rm f}}{J_{\rm i}} = \gamma \frac{m_{1,\rm e}} {m_1+m_2} 
\label{gammaform}
\end{equation} 
where $J_{\rm i}$ and $J_{\rm f}$ are the orbital angular
momenta of the initial and the final binaries, and $m_{1, \rm e} =m_{1}
- m_{1, \rm c}$ is the mass of the ejected envelope.  

This has come to be widely used in BPS studies as an alternative
to the standard energy-based methods for predicting the outcome of
general CE events. However,  the $\gamma$-formalism was first proposed
in an attempt to explain a narrower set of systems 
for which the standard
energy prescription for CEE appeared to be particularly problematic. For clarity we first 
address the more restricted original set of cases, and in the next subsection (\S
\ref{sec:inspiralling-gamma}) we consider the potential broader application
of angular-momentum based parametrisations like the
$\gamma$-formalism to predicting the outcomes of canonical CEE. 

\subsubsection{The origins of the $\gamma$-formalism}

The $\gamma$-formalism was developed in order
to explain some particular DWD systems \citep{Nelemans00,Nelemans05}. 
These were thought to have formed
following two CE episodes, during the first of which the orbital
period might have \emph{increased}. This requirement arose because the 
older WD in those DWD systems has a smaller mass, and since radius and core mass are
coupled in low-mass giants, the orbital separation at the onset of the
second mass-transfer episode had to be wider than at the onset of the
first one. The energy formalism would not naturally describe a
CE phase which \emph{widened} the binary orbit as appeared to have
happened for these systems.
Moreover, since that relationship between giant radius
and core mass allows the properties of the pre-CE systems
to be reconstructed, the $\gamma$-value for each unstable
mass-transfer episode can also be inferred \citep{Nelemans00}. 
Intriguingly, that reconstruction method found
that all those observed DWD systems could be explained by very similar
values of $\gamma$. This led to interest in whether the narrow range
of inferred $\gamma$-values was related to a deeper meaning. In
addition, it was also suggested that Eq.~\ref{gammaform} provides a better
tool than the energy formalism for predicting the post-CE properties
of a wide range of binaries -- using a single value of $\gamma$
for all occurrences -- including those where
the method originally used to reconstruct the pre-CE properties of the
DWD systems would not work \citep{Nelemans05}.

\subsubsection{The sensitivity of the outcome to $\gamma$}
\label{sec:gammasensitivity}

The narrow range of $\gamma$ produced by reconstruction techniques
did suggest that the method could be very valuable. The ability to
predict the outcomes of CEE for several disparate classes of system
using one value of $\gamma$ would make it a powerful tool, and
understanding the origin of a universal $\gamma$-value might help
to illuminate the physics taking place during CEE.
However, \cite{Web08} 
explained that this may be understood as an intrinsic property of the
formalism itself rather than giving deeper insight into CEE. 
In this way both the success in the initial fitting and later
problems in the application of the  $\gamma$-formalism
for other types of binaries may arise from the 
mathematical consideration of how the $\gamma$-formalism (as described by
Eq.~\ref{gammaform}) performs a transformation of an inital binary
into a post-CE binary.  Specifically, a small range of $\gamma$ is
capable of leading to a very wide range of outcomes: the mapping from
the initial to final separation is very sensitive to $\gamma$ 
(see \citealt{Web08} and \citealt{woods10}, and for
a more formal mathematical explanation see \citealt{WoodsChile2011}). This
naturally leads to reconstruction methods inferring a small range
of $\gamma$, as if the process itself is divergent, the inverse process is
convergent. In this case a wide range of outcomes (observed binary separations) 
was connected to a narrow range of inputs ($\gamma$ values). 
This sensitivity to small changes in $\gamma$ suggests that 
 Eq.~\ref{gammaform} should be at least reformulated. 

\subsubsection{The physical basis of the $\gamma$-formalism}

Eq.~\ref{gammaform} is a fitting mechanism for the outcome of a mysterious period of
canonically-unstable RLOF. Beyond this, there is no clear physical picture of the underlying
processes which the $\gamma$-formalism represents. One
could interpret Eq.~\ref{gammaform} as describing the angular momentum
which is carried away by each particle of mass ejected from the
system. However, the prescription only gives the overall angular-momentum loss at
the end of the mass-ejection phase; it contains no assumptions about the
specific angular momentum loss at each instant.  Nonetheless, as this
review aims to understand the physics of CEE, we direct the
interested reader to \S \ref{sec:comparisonAM} and
\ref{sec:1Dsimulations}, which discuss angular momentum
transport and loss in simulations of CEE. 
Hopefully our understanding
of the physics of CEE will soon improve enough to allow predictions
to simultaneously take advantage of both energy and angular-momentum
conservation.

Since the systems to which the $\gamma$-formalism was first applied
are precisely the ones with which the energy
formalism struggled, it is clear that the $\gamma$-formalism does
not automatically describe a CEE phase where a limit from simple energy conservation is
expected.
Replacing the energy formalism with a
different parametrisation does not solve the apparent 
\emph{physical} problem if energy generation during CEE is
required to form a particular post-CE system.
Some systems predicted by the $\gamma$-formalism
can be described as having apparently violated energy conservation
during their formation if only orbital and thermal energies are available. 
This suggests that the timescale of the ejection in the $\gamma$-formalism is longer than 
the thermal timescale of the stellar envelope, which makes it more plausible 
that an additional energy source -- such as the star's own fusion energy -- could be used (see \S\ref{sec:energycons}).
We stress that introducing Eq.~\ref{gammaform} does not solve the underlying issue
in the formation of the DWD binaries (i.e.~exactly what happened during the first MT episode)
to which it was first applied,
even if it was an effective parametrisation of the outcome.

\subsubsection{Resolving the problem of DWD formation with mass-transfer stability?}

The proposal of the $\gamma$-formalism highlighted a set of
mass transfer episodes which apparently led to orbital expansion via
CEE. 
However, an expanding orbit -- as required to explain the DWD systems which the
$\gamma$-formalism was first used to parametrise -- can be a
consequence of \emph{stable} mass transfer. 
The first Roche-lobe overflow episode in the formation of those DWD
systems was not necessarily unstable, and hence did not necessarily
lead to CEE (as recently argued by \citealt{woods11}). 
In this case the progenitor systems that form DWDs are different to those that form DWDs
via CEE and the $\gamma$-formalism.\footnote{The mass-transfer phases calculated in
  \cite{woods11} were previously thought to be dynamically unstable
  for two reasons. Firstly, no realistic mass-transfer calculations
  were performed and only simplified radius-exponents in the adiabatic
  approximation were used to evaluate the stability 
  (see \citealt{woodsmt11} for why the adiabatic approximation is
  imperfect). Secondly the mass-transfer was considered to be fully
  conservative even though the transfer rate may exceed the Eddington
  limit of the accretor.}
This means that use of the $\gamma$-formalism does not substitute for
following the MT episode in detail. Further study is required to finally
determine which systems form DWDs, but the existing work related to the
formation of these systems should certainly remind us that understanding
mass-transfer stability is as important as understanding CEE
itself. 

\subsection{Angular-momentum based parametrisations of classical,
  inspiralling CEE}
\label{sec:inspiralling-gamma}

As noted previously, angular momentum is probably the most natural
conserved quantity to consider when the binary separation does not significantly
decrease. That condition does apply to the systems for which the
$\gamma$-formalism was developed.  
However, the majority of canonical CEE cases involve a
major spiral-in; indeed, that reduction in separation was the serious problem which CEE was
invented to solve. Nonetheless, since angular-momentum conservation is
physically true it seems worth considering whether a formalism
similar to the $\gamma$-formalism could be used for all CEE events. 

Moreover, numerous population synthesis studies have already 
adopted the $\gamma$-formalism as an alternative way of predicting the outcome of
general CE phases (normally only to compare with the standard $\alpha$--$\lambda$ prescription). 
Hence it is important to consider whether such
use is likely to lead to undesirable outcomes. A simple test finds that blanket
use of the standard $\gamma$-prescription with
a single value of $\gamma$ and a typical initial binary population
leads to apparent energy input in
a large fraction -- roughly half -- of the CE events that avoid merger.\footnote{ Using
  {\sc{StarTrack}},   we find that $\approx$1/2 of surviving post-CE binaries end with
  apparent energy input for $\gamma$ of 1.5 or 1.75, and over 1/3 of
  them when $\gamma=2$. 
We considered the population of pre-CE binaries at the time 
of their \emph{first} dynamically-unstable RLOF, then determined the
outcome predicted by Eq.~\ref{gammaform} for each system. If the CE
event does not lead to a merger, 
we analyzed which of the post-CE binaries have more orbital energy
than they did before the onset of CEE.
The initial population took  primary stars
  from 1--100 $M_{\odot}$ following a Kroupa IMF, with flat
  distributions for both the initial mass ratio and the logarithm of the
  initial orbital period, i.e.~typical assumptions.}
If the $\gamma$-formalism -- or a similar angular-momentum-based
prescription -- becomes the standard way to predict
outcomes of CEE in population synthesis then this high fraction of
events which require unexplained energy input should recieve greater attention. 
We emphasize that any purely angular-momentum-based prescription is a
fundamentally different way of treating CEE from one where the outcome
is guaranteed to be limited by the available orbital energy.
The $\gamma$-formalism isn't \emph{only} an alternative choice of
parametrisation; it is also a qualitatively different picture.

\subsubsection{Angular-momentum-based fitting for CE with significant spiral-in}
\label{sec:restrictedgamma}

Motivated by the above, we now analytically examine the behaviour of an
angular-momentum based prescription for situations with significant
spiral-in. We do this by taking the $\gamma$-formalism and adding 
the additional condition that the orbital energy decreases during CEE.  

If we assume circular, Keplerian post-CE orbits then for each such CE event we can define two limiting values of $\gamma$: 
\begin{itemize}
\item{$\gamma_{\rm E}$} The value of $\gamma$ for which
  Eq.~\ref{gammaform} predicts that the post-CE orbital energy will be
  higher than the pre-CE orbital energy.
\item{$\gamma_{\rm M}$} The value of $\gamma$ for which
  Eq.~\ref{gammaform} predicts that the system will merge. We
  define this cautiously, such that for $\gamma_{\rm M}$ \emph{all} of the orbital angular
  momentum of the system is carried away by the envelope ejection;  mergers could
  happen for less extreme values than $\gamma_{\rm M}$.
\end{itemize}
From the above definitions can be derived a relation between
$\gamma_{\rm E}$ and $\gamma_{\rm M}$, specifically:
\begin{equation}
\label{eq:gammaEgammaM}
\frac{\gamma_{\rm E}}{\gamma_{\rm M}} = 1 -
\sqrt{\left(\frac{M_{\rm c}}{M_{1}}\right)^{3}
  \left(\frac{M_{1}+M_{2}}{M_{\rm c}+M_{2}} \right)}
\end{equation}
where $M_{1}$ and $M_{2}$ are the pre-CE masses of the components and
$M_{\rm c}$ is the core mass of the primary star (i.e.\ $M_{\rm
  c}=M_{1}-M_{\rm ej}$, where $M_{\rm ej}$ is the mass ejected during CEE).
The above can be rewritten as:
\begin{equation}
\label{eq:gammaEgammaMalt}
\frac{\gamma_{\rm E}}{\gamma_{\rm M}} = 1 -
  \left( \left(1-x\right)^{3/2}
  \left(1-\frac{x}{k}\right)^{-1/2} \right)
\end{equation}
where $x=M_{\rm ej}/M_{1}$, i.e.\ the fractional mass of the pre-CE
primary star which is ejected during the CE event, $k=1+(1/q)$ and $q=M_{1}/M_{2}$ is the mass ratio
of the system prior to CEE. Note that $k$ is a weak function of $q$
for the range of likely cases at the onset of CEE (i.e. from $q \gg 1$
to $q \approx 1$, which correspond to $k \approx 1$ and $k \approx 2$). For
those two limits:
\begin{itemize}
\item When $k \approx 1$ (i.e.\ $q \gg 1$) then Eq.~\ref{eq:gammaEgammaMalt}
simply reduces to $(\gamma_{\rm  E}/\gamma_{\rm M}) 
\approx x$.  
\item When $q=1$ then the binomial expansion, truncated after the
  first two powers of $x$, gives
$(\gamma_{\rm
  E}/\gamma_{\rm M}) \approx (5/4)x-(3/32)x^{2}$.  
\end{itemize}

\noindent The above indicates that the
range of  $\gamma$ between $\gamma_{\rm E}$ and $\gamma_{\rm M}$ is
dominated by the fractional mass ejection from the primary. 
Cases where the surviving core mass is a small fraction of the
donor mass -- in which case $x$ is a large fraction of unity -- are expected to be common.
It should be clear that in such systems the range of $\gamma$ between
$\gamma_{\rm E}$ and $\gamma_{\rm M}$ is a small fraction of
$\gamma_{\rm M}$. This is related to the more general
sensitivity of the formalism discussed in \S\ref{sec:gammasensitivity}.

Appendix B uses model stellar structures to numerically demonstrate
the limited range between $\gamma_{E}$ and $\gamma_{M}$ for some
unexceptional cases.
Importantly, the range of values between $\gamma_{E}$ and
$\gamma_{M}$ is expected to differ from system to system (see also
\citealt{Web08}, \citealt{woods10}, and \citealt{WoodsChile2011}).
Hence a single global value of $\gamma$ seems unlikely to be effective
in describing all CE phases for which there is significant spiral-in and
significant mass ejection.

\subsubsection{Wider application of the $\gamma$-formalism} 

As noted previously, the specific successes of the $\gamma$-formalism have led to some
population synthesis studies adopting it as a general alternative to
the energy formalism.  Some comparisons of the observed populations of
post-CE binaries with population
synthesis models have found inconsistencies with the
$\gamma$-formalism when applying it to general CE events 
\citep[see, e.g.,][]{Dav10,Zor10}.   
However, we have argued elsewhere against drawing over-strong conclusions
about the process of CEE from population synthesis alone, and the same
principle should apply here. 

Nonetheless, use of the
$\gamma$-formalism to make predictions for systems other than those
for which it was calibrated should be done cautiously. This is
especially true for systems which undergo serious spiral-in, or for
which the final orbital energy is lower than the initial; then the
outcomes predicted will be highly sensitive to the chosen value of
$\gamma$, as explained above. In particular, it seems very unlikely
that a single value of $\gamma$ could apply to all
CE phases which occur in the Universe. 
Avoiding unexplained energy input in a significant
fraction of post-CE binaries requires fine-tuning of $\gamma$ for
particular cases.  We have made clear that the parameters in the
energy formalism are also likely to be different for different
systems, hence such variation in $\gamma$ is not a fundamental argument against the
use of a parametrisation based on angular momentum, but it is a strong
practical warning to those who make and interpret BPS models (especially
combined with the high sensitivity of the current $\gamma$-formalism to
changes in $\gamma$).   

Overall, the classes of CEE to which the $\gamma$-formalism might
currently be well-suited are almost certainly limited. Population synthesis
modellers who intend to employ the
$\gamma$-formalism should consider this point.
The recent work by \cite{toonen12} adopts a set of
restrictions which may be useful guidelines: they do not apply the
$\gamma$-prescription for a second episode of dynamically 
unstable mass transfer, nor when the companion star is a compact remnant, 
nor when the dynamical instability is due to a tidal instability.

\section{The onset of the common-envelope phase}
\label{sec:onset}

The onset of the common-envelope phase is not immediate.  This process
involves both the time during which unstable  RLOF is turning into the 
common-envelope phase and also the recent pre-RLOF evolution of the donor.

\subsection{Enhanced mass loss before RLOF}

The donor might lose a significant amount of its mass during the approach to RLOF,
i.e. before the actual RLOF starts.  
Mass lost through a tidally-enhanced wind was proposed by
\citet{Tout88} to explain the observed mass-ratio inversion in 
some RS CVn binaries.
In addition, very massive stars that
approach the Humphreys-Davidson limit could be subject to enhanced
winds, and even spontaneous envelope loss
\citep[e.g.,][]{1991A&A...252..159V,Egg02}. AGB superwinds might also be enhanced or
triggered by the
presence of a close companion (e.g., Chen et
al. 2011, in prep).  One of the driving mechanisms for AGB superwinds is
likely connected to pulsations, and such pulsations can be either
amplified (e.g., due to tidal interactions)  when the star is close to
its Roche Lobe (RL), or strong pulsations can start earlier than it
would be in a case of a single star.   A similar effect might happen for stars
which are close to other pulsation instabilities, such as the Cepheid
instability \citep{Egg02}.  
Another potential driving mechanism for enhanced winds can be
connected to the rotational velocity of the star. The rate-of-rotation
of the donor is likely to increase prior to RLOF due to synchronization
of the stellar spin with the orbit via tidal interactions \citep{Bear2010}.
So it seems possible that during this pre-RLOF stage a star might lose mass at the same
rate as AGB superwinds, $10^{-4} M_\odot{\rm yr}^{-1}$ (or at an even
greater rate for massive stars). This mass loss occurs without loss of
orbital energy (i.e. without reducing the semi-major axis of the orbit
before the onset of the main CE phase).  Note, however, that wind loss
will tend to widen the binary, which may lead to avoidance of CEE; it
should certainly increase the stability of RLOF against CEE.

The obvious consequences of any enhanced mass loss prior the onset
of CE, compared to evolution as a single star of the same mass, is
that both the mass of the envelope and its binding energy can be
decreased due to matter re-distribution.  This may lead to an {\it
apparent} increase of $\alpha_{\rm CE}$ for the overall sequence of
events.  
The binding energy decrease
at the tip of the AGB is even argued to lead to a state when the envelope
becomes almost unbound or blown away by a superwind, and a binary
may even completely avoid the formation of a common envelope \citep{Chen2011}.
Even after the secondary enters the
giant envelope, the rotational velocity could be high enough to keep inducing an
enhanced mass-loss rate. Significantly more systems might survive the
CE phase if these preceding spin-up and mass-loss phases were taken into account.

\subsection{Duration of tidal interactions before dynamical instability}
\label{sec:onset_tides}

Generally we can estimate that tidal interaction becomes significant when the orbital separation is
two-three times larger than the giant radius: see, e.g., \citealt{PZ93} for {\it immediate} tidal interactions.
Due to {\it continous} tidal interaction during the time a donor evolves near the tip of the RGB or AGB, $\tau_{\rm ev}$, 
a giant donor is argued to be tidally spun-up even at larger separations \citep{Sok96}: 
\begin{equation}
a_{\rm max} \simeq 5 R_g
\left ( \frac{\tau_{\rm ev}} {10^6 {\rm yr} } \right )^{1/8}
\left ( \frac{M_2}{0.1 M_\odot } \right )^{1/8}
F(L_{\rm g},R_{\rm g},M_{\rm env}),
\label{eq:amax}
\end{equation}
where
$F(L_{\rm g},R_{\rm g},M_{\rm env})$ is a slowly varying function of the primary's luminosity, radius, and envelope mass, respectively.
Note that there are a number of uncertainties and simplifications
within that expression. It is based on Zahn's impressive
theory of tidal spin interactions \citep{zahn77,Zahn89}, but the
numerical factor should be treated with caution, as it is often found
that this must be tuned to match observed binary systems. 
However, the important point
here is qualitative:  $a_{\rm max}$ increases with the mass of the companion.

In order to avoid the onset of CEE after synchronization has been achieved, the system
needs to remain stable against the Darwin instability \citep{dar79}. 
Qualitatively, this instability is a consequence of the
  fact that removing angular momentum from the binary orbit causes the
  orbital period to decrease, i.e. spin faster. 
  Hence, in a tidally-locked binary, if the giant extracts angular
  momentum from the orbit (e.g.\ by expanding and thereby changing its
  moment of inertia) then tidal locking forces it to
  extract \emph{additional} angular momentum from the orbit in order
  to stay synchronised (since the orbital period will itself have been decreased by the star's
  initial demands). It should be clear that if the
  moment of inertia of the binary orbit is far larger than the moment
  of inertia of the individual stars then this exchange of angular momentum will
  not destabilise the system.
  However, in some cases there are no stable solutions, i.e.\ if
  the attempts by the orbit to supply the spin angular momentum
  demanded by the star are unable to lead to equilibrium.
  When such a runaway occurs then the stars merge (i.e., in this case,
  enter CEE). 
Quantitatively, the condition to avoid that instability -- assuming
that the system is tidally-locked -- is that the orbital moment 
of inertia $I_{\rm orb}$ be more than 3 times $I_{\rm g}$ the giant's moment of inertia
$I_{\rm orb} >  3 I_{\rm g}  = 3 r_{\rm gyr}^2 M_{\rm env} R_g^2$
\citep{Hut80}. Here $r_{\rm gyr}$ 
is the gyration radius of the giant and is usually about 0.1. 
We have also made the usual simplifying assumption that the moment of
  inertia of the giant is much larger than that of the other star. 
A more massive companion makes the system more stable with respect to Darwin instability.
As the giant's radius grows, the binary system becomes less stable;
for a discussion of the competition between orbital separation increase due
to mass loss and orbit decrease due to tidal interaction, and the possible
onset of the Darwin instability, see \cite{Bear2010}.

It follows, that more massive secondaries could be more efficient in bringing the
giant envelope to synchronization before entering the CE phase. They are also more efficient in maintaining
this stage for a long time. During that time the giant loses more of its envelope in the wind.
When the CE finally occurs, as the wind carries angular momentum and/or the giant expands, there is
less mass in the envelope. More massive secondaries would therefore
tend to have less mass to expel during the CE phase, and so would end
the CE phase with a wider orbital separation. In the energy
formalism a larger final orbital separation indicates a larger value of $\alpha_{\rm CE}$.
Because of this, massive secondaries could be expected to appear to cause larger values of $\alpha_{\rm CE}$.
However, more massive companions would be expected to have larger post-CE separations even with the same $\alpha_{\rm CE}$, simply since they carry more pre-CE orbital energy. So the post-CE observational signature of these tidal interactions is not unique.

Nonetheless, this prediction based on pre-RLOF tidal interactions is in fact 
\emph{contrary} to that deduced from observations by \cite{Mar11}, who
find that there is a possible negative correlation between the mass ratio of the two stars
and the value of $\alpha_{\rm CE}$. Namely, for larger $M_2/M_g$ the
average value of $\alpha_{\rm CE}$ is smaller, as also found by \cite{Dav11} 
(although \citealt{Zor10} don't find indications for a dependence of $\alpha$
on the mass of the companion;
also, there are enormous observational selection effects favouring short orbital periods).
It can be noted that the final separations 
from observations are {\it all} low, irrespective of the mass ratio. 
De Marco argues that, without needing to make complex reconstructions, this already tells you 
that the low mass systems have a larger $\alpha$ in the energy formalism.

Soker argues that a possible reconciliation of the apparent contradiction between the
finding of \cite{Mar11} and his estimate may 
come from the distribution of initial binary parameters (e.g., more
massive secondaries could reside closer to their parent star and so
they enter the CE phase at earlier epoch). However, 
so far there is no observational evidence for such distributions.
Alternatively, the difference in effective $\alpha_{\rm CE}$ could arise because
the difference in mass affects the physics of the CE
ejection. For example perhaps CEE involving more massive
companions occurs on a shorter timescale; that could affect the 
energy redistribution within the envelope to make complete envelope 
ejection more difficult (see the discussion in section \ref{sec:enthalpy}).
And, finally, it may simply mean that tidally induced synchronization before RLOF 
does not play a significant role in the outcome of CEE.

\subsection{RLOF and the development of dynamical instability}

Once a model donor star overfills its Roche lobe, a theoretical criterion is
usually applied to try to determine whether the mass transfer is dynamically
unstable. If the RLOF is dynamically unstable, it is usually expected
to lead to CEE. However, some special-case systems do exist which we would expect to have
experienced dynamically unstable mass transfer seem to have avoided
CEE \citep[see the discussion in ][]{PhP+1992}.

The standard analysis of the stability of mass transfer compares the differential reaction of
the Roche lobe to mass transfer to the reaction of the donor to
mass loss on different timescales \citep[see][]{Hje87}.
For the purpose of this analysis, the donor has often been treated as a
composite polytrope \citep{Hje87,Sob97}. The donor's reaction is mainly a
function of whether the envelope is convective or radiative. However,
it is wrong to forget that the existence of a core can make a substantial 
difference to mass transfer stability (\citealt{Hje87,Sob97}; unfortunately this is not an unusual
misconception -- see, e.g., the discussion in \citealt{PhP2001} and references therein).
As an example, a commonly used critical mass ratio ($q_{\rm crit}$)
for the stability of RLOF from convective donors
\emph{with isentropic envelopes} is $q_{\rm crit} \simeq 2/3$; this
value is only relevant for \emph{fully} convective stars, as 
stars with cores are more stable \citep{Hje87,Sob97}.

A polytropic equation of state has also been used to derive an
analytic solution for the mass transfer rate during the lead up to
runaway \citep{WI87, Web2010}. This phase is difficult to treat
self-consistently with a full stellar evolution code so, although the
assumptions used are highly idealised, this solution may be of use in setting
up the initial conditions of hydrodynamic simulations of CEE.

Recent progress has been made in studying the problem of mass transfer
stability using the adiabatic approximation but
using realistic stellar structures rather than polytropic stellar models.
\citep{Ge10}.  These studies have preliminary shown significant
differences to the old criteria for when the instability occurs, as
well as considerable changes for the same star at different points along the giant
branch. The more detailed models show greater stability, with $q_{\rm
crit}$ as large as 10 for some of the stars (Ge et al.\ 2011, in prep).

Nonetheless, such work carries the main disadvantage of old studies:
the adiabatic approach literally means that the reaction of the star is
studied by keeping the entropy profile (at each mass coordinate)
fixed. The thermal adjustment time of the outer layers of the star is
so short that, even when the mass transfer is taking place on
timescales shorter than the global thermal timescale of the star, the
entropy profile within the star can deviate considerably during mass
transfer from the fixed profile used in adiabatic codes 
\citep{PhP+2002, woodsmt11,PH11}. In particular, the superadiabatic
spike near the surface of the star is not lost in the way that the
adiabatic approximation predicts; some of the strong
expansion predicted in adiabatic codes is suppressed by retaining this spike.

A further stabilising effect present in reality but absent in
adiabatic codes is the finite time taken for the development of
the dynamical instability after the start of mass transfer \citep[see the
discussion in][]{han02}.  The critical mass ratio also depends on
how conservative the mass transfer is, where less conservative mass 
transfer leads to more stability and higher $q_{\rm crit}$ \citep[see,
e.g.,][]{PhP+1992,Kal96,Soberman+1997,Han+2001,woods11}. The dynamical
stability of RLOF could also be increased by tidal spin-orbit
couplings \citep{TaurisSavonije2001}.

Adiabatic codes are elegant, and provide a clean \& well-defined 
answer about when instability occurs. Adiabatic codes could also  
be modified by adding artificial thermal relaxation, 
essentially placing a superadiabatic blanket on top of an adiabatic
envelope. Indeed \citet{Ge10} found that in this case the reaction of
the star is typically calculated to lie between the predictions from
detailed stellar codes and those produced by adiabatic calculations.
However, any modern detailed
stellar/binary evolutionary code can also provide $q_{\rm crit}$,
without needing to resort to the adiabatic approximation. 
For example, \citet{han02} explicitly calculate 
values of $q_{\rm crit}$ for use in their own population synthesis
calculations; \citet{ChenHan2008} also use a full stellar evolution code to 
investigate $q_{\rm crit}$ in detail.  Of course both approaches are
approximations, and therefore potentially misleading, since neither type of
code is really treating the full three-dimensional problem. It may
even be that the structure in the vicinity of the inner Lagrangian
point is closer to the predictions from adiabatic codes, since there the
superadiabatic layer may not be able to rebuild itself (for studies of
the flow in this region, see \citealt{Paczynski1972} and figure 3.6 of \citealt{Eggleton2006}).

\subsection{3D and hydrodynamic effects}
\label{sec:onset3Dhydro}

Fully understanding the onset of CE might well require the inclusion
of physics beyond standard stellar calculations.
There are two important factors affecting how dynamically unstable the
initial phase will be, according to current studies by means of 3D
hydrodynamical simulations:

\begin{itemize}
\item how strongly the donor is in or out of corotation with the binary
\item what is the value of the total angular momentum
\end{itemize}

These two issues are worthy of further consideration. If the initial
conditions 
for hydrodynamic CE simulations are such
that the donor is not in corotation with the binary, or if the companion
is simply placed at the surface of the donor, then the dynamical
plunge-in phase is being \emph{forced to start artificially quickly}.  In
both of those cases then the system as a whole is missing some of the
angular momentum which it should posess (for companions massive enough that we expect them 
to spin-up the giant's envelope).   Neither of the
approximations reflects the real situation, and the consequences
are not yet well understood.  

Unfortunately such initial conditions have been commonly used 
in published simulations, but the degree of non-corotation varies
from one research group to another. If we compare two cases: one with 95\% of the appropriate orbital
velocity required for corotation in \citet{Ric2011} against 0\% as
in \citet{Passy11}, it seems that the more rapid intial rotation may help to eject
more material to infinity from the system. Conversely, less rotation
could lead to more material being trapped, perhaps in a bound circumbinary disk (this comparison
is considered in more detail in \S \ref{meth_comp}). Determining what the
angular momentum distribution is in a binary system when the donor
overfills its Roche lobe is an important question in properly treating
the initial stages of CEE, and is important input in order to make the
most of computer time.

\subsection{Onset of CE from dynamically stable RLOF}
\label{sec:onsetDynamicallyStable}

It is possible that mass-transferring binaries which do not experience
CEE following a standard dynamical
instability are still dragged into CEE. This might happen because the
accretor cannot accept matter at the rate at which it is being
transferred, and also the system as a whole cannot eject the matter
rapidly enough. In this case a de facto common
envelope could be built up around the stars.   Until relatively
recently it was thought that thermal-timescale mass transfer in X-ray binaries could
lead to CEE in this way. However, it is now acknowledged that Cygnus X-2 passed
through such a thermal-timescale phase and avoided CEE
\citep{King+Ritter1999,PhPRappaport2000,
Tauris+2000,Kolb+2000}.    Another relevant system in this context is
SS433, which seems to be transferring matter at $\gtrsim 10^{-4}
M_{\odot} {\rm yr^{-1}}$ but -- so far -- appears to have avoided CEE \citep{Blundell+2001,PhP2001}.

Double-core evolution is a special case of this
(\citealt{Brown1995}, see also \citealt{Dewi+2006}). In this case the
CE phase ejects the envelopes of \emph{both} stars. 
Unusually, it requires the mass ratio to be close to 1 
(typically within a few percent). If the primary then overfills its
Roche-lobe as a giant, then accretion onto the secondary might 
cause it to expand and also overfill its Roche-lobe. This leads to
a joint CE, in this case formed by matter from both stars, and
inspiral of both cores.

\section{Comparison of state-of-the-art 3D simulations}
\label{meth_comp}

3D hydrodynamic simulations of common-envelope evolution
have been carried out by \cite{Ric2011} (hereafter RT) using the grid-based, adaptive mesh
refinement (AMR) code FLASH  \citep{Fry2000}  and by  \cite{Passy11}
(hereafter  PDM)
using the grid-based code Enzo \citep{2005ApJS..160....1O} in single grid mode and the Lagrangian  code SNSPH \citep{2006ApJ...643..292F}.

The  star simulated  by  PDM  was a  0.88~M$_\odot$,
85~R$_\odot$  giant   with  companions  in  the  mass   range  0.1  to
0.9~M$_\odot$,  and  RT  considered  a 1.05~$M_\odot$,
31.6 $R_\odot$ giant with a 0.6~$M_\odot$ companion.  As the initial
masses are similar, the main difference between the initial conditions
chosen for these simulations were the
initial  conditions for the  rotation, where RT considered a
donor which is  almost in corotation with the  orbital motion (spun up 
to $95\%$ of the orbital angular velocity), whilst PDM took
the case when the giant is not rotating at all (see also \S \ref{sec:onset3Dhydro}).

In PDM, the  grid-based  models with 256$^3$  resolution and  the  Lagrangian  500\,000
particle models reach essentially  the same conclusions,  which gives
some confidence  that   there  are  no   major  numerical  issues   in  the
simulations.  The effective resolution of the RT simulations was 2048$^3$.

\subsection{Final separations and envelope ejection}

The dynamical in-fall phase lasts  of the order of 
50 days in RT and  10 to 100 days in PDM, 
and the   final   separations   are  between   a   few   and
$\sim$30~R$_\odot$. The final orbital separation in RT are few
times smaller than the PDM simulations with the same companion mass.  
However for both groups these final separations are systematically larger than
the observed separations of post-common envelope binaries \citep{Mar11,Zor10}.  One explanation is that the phase
immediately following the dynamical in-spiral phase further alters the
post-common  envelope   binary  separation.  \cite{Kashi2011}
suggested that  even a  small amount  of fall-back  mass can  create a
circumbinary  disk which can then tighten the immediately-post-CE binary orbit through tidal interactions.

In the RT simulations about 25\% of the envelope is ejected. 
The PDM simulations stop at the end of the dynamical spiral-in, at
which point \emph{most of the envelope is still loosely bound} -- only a small fraction  of the stellar
envelope is unbound. If this result is physical then the next phase of
evolution seems likely to be an in-falling envelope that will then
form a disk. The fall-back disk  envisaged by \cite{Kashi2011} is
far less massive than  the  mass of  the  fallback  material in the PDM
simulation (at only 1--10\% of the total envelope mass).   The
difference in the amount of ejected matter may simply be consistent with RT
producing a shorter final binary period. The reason why RT
and PDM systematically disagree about the final orbital period is not clear, though
different initial conditions seem a likely reason.

\subsection{Angular momentum}
\label{sec:comparisonAM}

The main difference in initial conditions between RT and PDM is the
amount of angular momentum in the giant donor envelope before the
spiral-in phase. Since this should affect the speed of the initial
plunge, such a difference could easily lead to differing outcomes,
perhaps playing an important role in determining the ejection efficiency. 
Yet the simulations of \cite{San98} found that primary spin did \emph{not}
substantially alter the results for their heavier primaries, 
though for smaller mass ratios than considered by RT.
Tidal spin-up of the primary should be efficient for larger companion masses, while for
the lower  masses (e.g., $M_{2}  \le 0.05~M_\odot$) it might make little or no difference and the primary
could be spinning slowly. So, even if pre-CE spin is a factor in
ejecting  the  common  envelope,  it  could apply only to some
interactions.  

Nonetheless this pre-inspiral stage should be considered carefully,
as it seems likely to be important in affecting the
simulations.  In PDM  the companion  is always placed on
the primary's surface with a Keplerian velocity, where the in-spiral starts immediately. Clearly
this is unrealistic, since the companion would interact with the giant
tidally and through wind accretion for a reasonably long time before
falling in. What is not clear is how this initial phase influences the
outcome of the interaction. A comparison test ran with the companion placed 5\% farther out or with an orbital
velocity  slightly  larger than  the  Keplerian  value in \cite{Passy11} did not alter the final results,
though it resulted in marginally larger eccentricity.

We note that RT  find that the outflowing matter carries significant angular
momentum.  That is, the highest velocity components are in 
the tangential direction rather than in the radial direction.  
PDM, on the other hand, find the opposite; this difference should be
pursued further. 

\subsection{Variations with initial mass ratio}

\cite{San98} found that the fraction of ejected
mass increases with mass ratio (companion mass to red giant) of the
system. The result found in RT (for systems with mass ratios closer to
unity) are consistent with this trend. 

In PDM, lower mass companions take longer to in-spiral, in particular
initially, and  come to  rest at smaller  orbital separations,  as one
might expect  (although note that these are not the final separation as the envelope is still bound). 
However, not only do the observed post-common envelope systems cluster at
smaller orbital separations, these separations do not appear to be
a  function of  mass  ratio  nor secondary  mass.\footnote{For details
  of the how their sample of post-CE systems was selected see \cite{Mar11}.}  
The  observations therefore suggest that more massive
secondaries (i.e.~ systems with larger mass ratios, $q$) are less efficient at unbinding the
envelope  and so they sink deeper into the envelope despite  having plenty  of orbital  energy to
deliver.  This is  in line  with what  was determined  by \cite{Mar11} 
and independently by  \cite{Dav11}. Alternatively,
more massive  companions might suffer further in-spiral {\it after}
the envelope is ejected due to one or more alternative physical
mechanism(s).

\subsection{Energetics and $\alpha_{\rm CE}$}

RT find that, for their hydrodynamic transfer of orbital
energy to the envelope, $\alpha_{\rm CE}$ is $\sim 25\%$ based on the amount of
matter ejected.   PDM deemed it  inappropriate  to calculate  the values  of
$\alpha_{\rm CE}$ when the envelope has not been ejected. Clearly
further energy sources might help envelope ejection if they were
included (see \S \ref{sec:energy}). One obvious candidate for inclusion in simulations is the 
reservoir of recombination energy; it has
not yet been shown whether that energy release can be efficiently
converted into kinetic energy of the envelope.

\subsection{Eccentricity}

In PDM, the initial eccentricity of the orbit is zero for most of
their simulation. By the end of the dynamical in-fall phase a small eccentricity is driven into the
system  ($e \sim 0.1$). Their  two simulations  where the  companion  was placed
further out or had a larger-than-keplerian orbital velocity, resulting
in  a  mild  initial   eccentricity,  finished  with  slightly  larger
eccentricities  than the  simulations  that were  started in  circular
orbits. Eccentricity measurements on real post-common-envelope systems
have  not  yet  reached  the  level of  precision  to test this.

\subsection{Entering self-regulation?}

A vital question is whether the endpoints of the simulations in both RT and PDM are
simply the start of a longer, self-regulated phase. In the terminology
of \S \ref{sec:phases} is this the end of phase II and the start of
phase III?   Alternatively, has phase II ended with envelope ejection
and no further spiral-in (or a rapid merger)?  It would be unrealistic for calculations like this
to follow phase III. Furthermore, since the in-spiral timescales involved become
so long, \emph{it would be natural for calculations like this to
  look like they are converging on a steady-state in either case}.

It might be that, if 25\% of the envelope is ejected in this
phase (as found by RT), the rest is ejected in a separate later phase of the CE
event, following a period of self-regulated
spiral-in. Asymptotically-slowing calculations are very sensibly stopped so as
not to waste computer time, but we encourage thought as to how to
distinguish whether such simulations are entering a phase of self-regulation.

As the timescale of these simulations starts to approach the thermal
timescale, processes other than pure hydrodynamics begin to become
important. This is the regime in which 1D stellar-evolution type codes seem most useful,
as they can typically include more physics than is present in 3D hydro codes. 
However, we should still be careful to check that assumed symmetries are not
too problematic.

\section{Numerical Methods}
\label{sec:methods}

At present it is not possible to treat the whole common-envelope 
problem with only a single code and a realistic amount of computer
time. The dynamical plunge-in phase could be treated with some hydrodynamic
codes. The pre- and post-plunge-in stages (the onset of
mass transfer and the slow spiral-in) are each likely to occur on a thermal
timescale or longer, and could only be treated with a code that includes
a full equation of state, and both radiative and convective energy transfer.
An appropriate code for these longer phases would be a stellar-evolution code that is adapted
to treat at least some specific features of the common envelope
evolution, although such codes would currently only treat the problem
in 1D and so could miss other key aspects of the situation.

\subsection{Existing 3D hydrodynamic methods and their limits}

A wide variety of numerical tools are available to model the
stages of common envelope evolution which are dominated by
hydrodynamics.  In principle, Lagrangian codes are the most straightforward and
accurate: comparing pressures at the centres of adjacent zones gives
the acceleration on each zone edge. However, Lagrangian grids suffer in
multi-dimensional problems, as the zone edges can become tangled.
Eulerian codes avoid mesh tangling, but the relative motion between the
matter and mesh leads to numerical advection.  A number of advances to
the Eulerian grid-based
technique have increased the power of Eulerian techniques in modeling
common envelope: nested grids, rotating grids, adaptive mesh
refinement (AMR).  Modern computers also allow sufficiently high resolution
smooth particle hydrodynamics (SPH) calculations to study CEE.  
In addition, adaptive Lagrangian-Eulerian (ALE), particle-in-cell
and spectral methods are becoming more common in astrophysics.
All of these computational methods provide a wide range of choices for
modelers of common envelope evolution.  Here we discuss these techniques,
focusing on their application to common envelope model simulations.

An important aspect of numerical modeling is understanding the
strengths and weaknesses of a given technique and how these
strengths and weaknesses will affect the results in a given application.
We present an introductory summary of these below.

ALE, as the name implies, tries provide the best features of both
Lagrangian and Eulerian codes. Usually they behave like Lagrangian
code, with Eulerian-like re-zoning available to avoid mesh
tangling. Unfortunately the increased complexity can produce new
difficulties. ALE codes are strong in problems such as core-collapse
where the stellar core collapses, nearly spherically, several orders
of magnitude in space before turbulence sets in.  In such problems, a
strict Lagrangian code, followed by an Eulerian turbulence calculation
takes advantage of the strengths of the adaptive Lagrangian-Eulerian
technique.  It is not clear that the common-envelope problem has
features where a pure Lagrangian capability will be important and
ALE's strengths may not be well-suited for the common envelope
problem. 

Particle-in-cell codes are generally adopted where detailed
microphysics must be modeled, and we are not at this stage for common
envelope calculations.  Finally, the sensitivity of spectral methods
to boundary conditions make complex problems such as modeling the
common envelope process daunting.  At this point, it is not clear that these 3
``new'' techniques are ideally suited for the problems associated with
CEE.  Instead, we will focus on basic grid-based and
SPH techniques.

{\bf Strengths of Eulerian, Grid-Based Codes}
\begin{itemize}
\item History and Code Base: The long history of grid-based
schemes in computational physics has led to a number of schemes
developed to better model shocks and include additional physics
such as radiation transport.

\item Tracing space, not mass:  Grid-based codes are ideal for
low-mass flows: e.g. winds, mass streams in accreting binaries, and
the low-density cavities that might be formed during CEE. 
\end{itemize}

{\bf Weaknesses of Eulerian, Grid-Based Codes}
\begin{itemize}
\item Advection term:  The advection term in the hydrodynamic equations
of a grid-based code generally does not allow strict momentum or angular
momentum conservation and it leads to numerical diffusion of heat and
materials. Local non-conservation leads to global non-conservation. 
For calculations of the common-envelope problem, 
the lack of angular momentum conservation can alter the final result.

\item Tracing mass:  Grid-based codes are not ideal for tracing
mass, and that makes following the ejecta in a common envelope
calculation difficult.

\item Shock modeling schemes: Although the shock modeling schemes used
 in grid-based codes are ideal for shocks along the grid, they are
 not so accurate off-axis and conserve total energy often at the
 expense of getting erroneous internal energy estimates.

\item The re-zoning in AMR cannot simultaneously conserve energy, density and
pressure gradients and some care must be given to re-zoning algorithms.
\end{itemize}

{\bf Strengths of Smooth Particle Hydrodynamics}
\begin{itemize}
\item Linear and Angular momentum are conserved.  However, strict conservation
is not maintained with gravity implementations.

\item Ideally suited for problems tracing mass, e.g. the ejecta in a
common envelope phase.
\end{itemize}

{\bf Weaknesses of Smooth Particle Hydrodynamics}
\begin{itemize}
\item Low-mass streams are difficult to model.  SPH is not an ideal
tool to model the initial onset of the common envelope
phase \citep[though see, e.g.,][]{Church+2009}.

\item Low-density bubbles or cavities formed inside the (departing) envelope
might also suffer from poor resolution.

\item Most implementations use artificial viscosity to model shocks.
This typically broadens the shock front, preventing crisp shock
models. In addition, the artificial viscosity may over-estimate the amount
of friction in the flow.

\item Setup is generally more difficult.  For example, careful thought and wisdom
  is needed to make the best choice of particle mass for a particular
  problem. 

\item Few off-the-shelf packages are available with which to include additional physics.
\end{itemize}

For further discussion of the practical strengths and weaknesses
of SPH see, e.g., \citet{Price2011}; for more formal reviews see,
e.g., \citet{Rosswog2009} and \citet{Springel2010}. 

In any CE calculation adopting either grid- or particle-based schemes, we must worry about
how the scheme implements gravity.  Typically, SPH schemes use
tree-based gravity schemes, as do many AMR codes 
\citep{BarnesHut1986,WarrenSalmon1993,WarrenSalmon1995}.  Multipole schemes
are also prevalent in grid-based codes.  Each gravity routine carries
with it numerical artifacts and these must be understood.  Tree-based
schemes are accompanied by a multipole acceptibility criterion (MAC) 
and this can be easily tuned to determine the errors in the gravity
routine \citep{SalmonWarren1994}.

Boundary conditions can also pose problems for both grid- and particle-based schemes.

Code comparison can be an extremely powerful tool
to distinguish between the numerical artifacts of different schemes,
as performed for CEE by \cite{Passy11}.

Finally, we stress that that \emph{any numerical scheme must be used with care}.
Understanding the weaknesses of a technique is \emph{critical} to
interpreting the results.

\subsection{A novel generalisation of mesh-less methods}

As with other approaches, new numerical methods can be
developed. In the Appendix we demonstrate this by showing that
Lagrangian  particle-based methods are a subset of more general
mesh-less finite-volume schemes.  The spatially-discrete equations have
the same form and properties as the ones for mesh-based finite volume
numerical schemes, whilst the geometrical quantities (corresponding to
volumes and areas in mesh-based schemes) are expressed as spatial integrals in mesh-less schemes.
As a concrete example we also show that several approximations are needed to obtain the SPH
equations in closed form suitable for numerical integration, and these
approximations introduce certain inaccuracies. The approximation can be
improved with high-order numerical quadratures, but the computational cost and
complexity of these may well be comparable to that of unstructured mesh construction
in mesh-based schemes. This mesh-less generalisation breaks down the
artificial differences between mesh- and particle-based methods, and
hopefully opens the way for codes which have the advantages of both
types.

\subsection{1D simulations: what can be learned?}
\label{sec:1Dsimulations}

Early attempts at simulating the CE phase in one dimension produced
some successes. The simulations of \cite{mmh79} set the timescale
for CE evolution at around $1000$ years. However these simulations
were unable to model higher-dimensional effects such as the
preferential ejection of material in the orbital plane
\citep{bod84} or the spiral shocks and circulation currents
generated by the infalling cores \citep{Taam00}. If these effects are
not included, simulations of CE evolution lead to very different
results and often suggest no mass ejection at all. 

Clearly we would like to be able to run full three-dimensional
high-resolution hydrodynamic simulations of the CE
phase for multiple systems, but unfortunately the computing power 
required to do so on a reasonable timescale is still many years away.  
It would be extremely useful if we could use detailed
three-dimensional models to gain sufficient understanding of the
non-spherical processes so that we could derive a one-dimensional 
parameterization of the missing effects.  One-dimensional
models have the strong advantage that they can be run sufficiently
quickly that the CE phase of a large number of systems can be modelled
at the expense of relatively little computing time. This would allow
us to come up with quantitative prediction for the outcome of a CE
phase for a wide range of systems.

The early one-dimensional simulations of \citet{mmh79} assumed that
the angular momentum in the CE was deposited into the envelope by the
spiraling cores and then redistributed diffusively by convection
leading to a steady state distribution satisfying

\begin{equation}
\frac{\partial}{\partial r}\left({\mu r^4\frac{\partial\Omega}{\partial r}}\right)=0
\end{equation}

\noindent where $\mu$ is the convective diffusion coefficient which
was taken to be uniform. This is a very simple approximation which
could easily be improved upon given our current knowledge. In
particular, we stress that it is \emph{essential} to restore the time-dependence of the angular momentum 
distribution because the evolution of the envelope can occur on a dynamical timescale.

An example of a similar model including some of the missing physical effects is:

\begin{eqnarray}
\diff{r^2\Omega}{t}&+&\frac{1}{r^2}\diff{r^4\Omega
  U}{r}+\frac{1}{r^2}\frac{\partial}{\partial r}\left({\mu(r)
    r^4\frac{\partial\Omega}{\partial r}}\right) \nonumber \\
&+&\frac{1}{r^2}\frac{\partial}{\partial r}\left({\nu(r)
    r^{2}\frac{\partial r^{2}\Omega}{\partial r}}\right)=\dot{J}(r)  
\end{eqnarray}

\noindent where we have used a model similar to
\citet{mmh79} -- based on angular momentum conservation -- but we have
included a number of important terms: 

\begin{itemize}

\item $U$ is a term for advection of angular momentum by
circulation,similar to the Eddington-Sweet circulation expected in
rotating stars \citep{zahn92}. 

\item $\mu$, the standard diffusion coefficient, has been retained, but we
can now reasonably model its spatial variation. It has
been noted that some numerical simulations predict a single convective
cell in the CE \citep{Taam10}. This may require revising the diffusion
coefficient from the one predicted by standard mixing-length
theory. Note that this diffusion coefficient assumes angular momentum is
transported by shear-induced turbulence or some similar process so
that the system tends towards solid body rotation. 

\item $\nu$ is an additional diffusion coefficient. This represents the
alternative possibility that fluid parcels are able to retain their
angular momentum. In this case the system tends towards a state of
uniform specific angular momentum \citep[e.g.][]{arn10}. 

\item $\dot{J}$, the source term on the right-hand side, describes how angular
momentum is deposited in the envelope. In the standard approximation
this is a delta function. However, models show that spiral shocks
produced by the cores are responsible for depositing much of the
angular momentum \citep{Taam00} so it seems more sensible to choose a
smoother function. 
\end{itemize}

The forms of $U$, $\mu$, $\nu$ and $\dot{J}$ are
currently unknown. More work is needed to derive reasonable
prescriptions for them based on three-dimensional results. 

We can also use three-dimensional results to refine our models for accretion of
material by the cores, the rate at which they deposit energy into the
envelope and the rate of mass loss from the system. With sensible
treatments for these effects, a one-dimensional approximation of CE
evolution could be used to predict how the ejection timescales and
post-CE properties of binary systems might vary for a wide variety of
initial conditions.

\section{Compact objects and hypercritical accretion}
\label{sec:hypercritical}

Stars spiralling into the envelope of their companion
are usually expected to be limited in the rate at which they can accrete to
the rate at which the force of the radiation released in the accretion is
equal to the inward gravitational force in a spherical model, i.e. the Eddington rate.  For a
neutron star accreting hydrogen-rich matter, this limiting rate is $\sim 1.6\times10^{-8}
{\rm M_\odot \, yr^{-1}}$.  Although a derivation based on spherical
accretion is not strictly valid when accreting material with angular
momentum, in most astrophysical phenomena, the maximum accretion rate
onto a neutron star lies within a factor of a few of this value.  

But the accretion rates in common envelope evolution can be so high that
the emitted radiation is trapped within the flow.  At these accretion rates,
the temperatures at the base of an accreting neutron star are sufficiently high
to drive neutrino emission.  These neutrinos can remove the potential energy
released from accretion without generating any significant radiation
force to prevent further accretion.  In such conditions, the
neutron star could accrete well above the Eddington rate, a process
known as hypercritical accretion.

If hypercritical accretion happens, it might prevent the formation of
some neutron-star X-ray binaries and close double-neutron-star systems 
through the canonical CE formation channel. This led to the proposal
of double-core CE evolution as an
alternative mechanism for the formation of such systems
(\citealt{Brown1995}, see also \citealt{Dewi+2006}).  Hypercritical
accretion could also prevent the formation of Thorne-\.{Z}ytkow
objects \citep{ThorneZytkow1975, ThorneZytkow1977}. 

For hypercritical accretion to occur, the photon radiation
must be trapped in the flow.  One way to estimate this trapping
is to compare the infall velocity of the accreting material
to the diffusion velocity of the radiation\citep{che93}.  The accretion
velocity ($v_{\rm acc}$) is given by the accretion rate
assuming a spherical inflow:
\begin{equation}
v_{\rm acc} = \dot{M}_{\rm acc}/(4 \pi r^2 \rho)
\end{equation}
where $\dot{M}_{\rm acc}$ is the accretion rate onto
the neutron star and $\rho$ is the density at radius $r$.
The corresponding diffusion velocity ($v_{\rm diff}$) is:
\begin{equation}
v_{\rm diff} = r/t_{\rm diff} = r \frac{\lambda_{\rm mfp} c}{r^2} = c/(\rho \kappa r)
\end{equation}
where $t_{\rm diff} = (r/\lambda_{\rm mfp})^2 \lambda_{\rm
 mfp}/c$ is the diffusion time, $\lambda_{\rm mfp} = 1/(\rho \kappa)$
is the mean free path of the photon, $\kappa$ is the photon opacity
(for ionized hydrogen, this is $0.2\,{cm^{2}}$ per g), and $c$ is the
speed of light.  For the radiation to be trapped in the flow, $v_{\rm diff}$
must be less than $v_{\rm acc}$.  Solving for the accretion rate, we
find:
\begin{eqnarray}
\dot{M}_{\rm acc} &>& 4 \pi c r/\kappa \nonumber \\
 &>& 0.003 (r/10^{11}\,{\rm cm}) (0.2\,{\rm g^{-1}\, cm^2}/\kappa) M_\odot\, {\rm y^{-1}.}
\label{eq:trapping}
\end{eqnarray}
If we assume Bondi-Hoyle accretion, the accretion rate exceeds this value for many
massive giants~\citep{fryer96}.  Actual accretion rates can be 1-2 orders of magnitude
less than the Bondi-Hoyle accretion rate because the accretion radius ($r$) is smaller than
the effective Bondi-Hoyle radius.  Even so, if the neutron star spirals deeply into
the giant envelope, the photons will be trapped, allowing the
possibilty of hypercritical accretion, i.e.\  the Eddington limit might
be beaten. 

For hypercritical accretion to work, neutrinos must effectively cool
the accreting material.  We can use equilibrium atmospheres to calculate the neutrino cooling 
timescale \citep{fryer96}. This calculation assumes that, as the material piles
onto the neutron star, it convects and forms a constant entropy atmosphere on top of the neutron star.
The neutrino cooling timescale must be shorter than the photon diffusion timescale
for it to dominate the cooling.  By comparing these timescales, \cite{fryer96} 
found that this criterion corresponds to material
entropies below 600\,$k_B$ per nucleon. Typical stellar
material has entropy below 50\,$k_B$ per nucleon, i.e.~ well below
that threshold. However, shock heating will raise the entropy ($S$) of
the accretion flow above values typical for stellar material \citep{fryer96}:

\begin{equation}
S=374 \left( \frac{M_{\rm NS}}{1.4\,M_\odot} \right)^{7/8} \left(
\frac{\dot{M}_{\rm acc}}{10^{-4}\,M_\odot {\rm yr^{-1}}} \right)^{1/4}
\left( \frac{r}{10^{10}{\rm cm}} \right)^{-3/8}
\end{equation}

\noindent where $M_{\rm NS}$ is the mass of the neutron star.
Nonetheless, this constraint
is less restrictive than equation~\ref{eq:trapping} so we can assume
that if the photons are trapped in the flow neutrino cooling will
allow hypercritical accretion.  

However, we repeat that this derivation assumed that
the unstable accretion atmosphere will efficiently convect such that
the entropy remains in instantaneous equilibrium throughout that
atmosphere.  In nature, this convection is explosive and will 
likely drive outflows that can ultimately reduce the rate of mass accretion.
These uncertainties make it difficult to determine the exact criterion
for hypercritical accretion.  To an order of magnitude, hypercritical
accretion is likely to occur if the estimated Bondi-Hoyle accretion
rate is greater than $10^{-2} {\rm M_\odot \, yr^{-1}}$.  Below
this value detailed calculations are required.  However, the
Bondi-Hoyle-Lyttleton prescription significantly overestimates the 
rate observed in simulations (see \S \ref{sec:acc_en} and the detailed
discussion in Ivanova 2011); if current simulations are producing the
correct answer for the accretion rate then it is unlikely that hypercritical
accretion will take place during CEE \cite[100 times less than estimated Bondi-Hoyle accretion rate, 
or $10^{-3} {\rm M_\odot \, yr^{-1}}$][]{Ric2011}.

\section{Linking with observations}
\label{sec:observations}

Since a CE event is short-lived, it might be argued that we are highly unlikely to catch it while it occurs
(although see \S\ref{obs_during}), in which case we could only observe
the resulting post-CE systems (including post-CE nebulae).
Our lack of full-scale simulations of all the phases does not improve
the situation, as we have few definitive physical \emph{predictions}
to offer. 

As explained in \S \ref{sec:methods}, 3D simulations currently only
help with understanding the appearance of dynamical-timescale events,
i.e.\ very short-lived phases which are unlikely to be observed. Even their
predictions for post-CE appearance are only directly applicable for CEE events
which end after the dynamical plunge-in phase.  On the other hand, the
appearance of a CE object during a long-lasting self-regulating phase is
currently provided only by 1D calculations and, since it is certain that systems
undergoing CEE (or merging) will not be spherically-symmetric, we must be cautious
about applying 1D calculations when we do see systems undergoing CEE
or during mergers. At least if a CE ends in a merger, the evolution of this merger
product can be understood by means of a regular stellar code, once the
structure of the merger product is determined.

So far, the community has mostly only been able to link models to observations
for populations of post-CE systems. Even in this case, we stress that
comparisons are usually performed within the framework of the
$\alpha$-formalism. These studies principally aim to calibrate our
existing parameterisation, fine-tuning $\alpha_{\rm CE}$-values using 
post-CE masses and periods \citep[recent examples, ][]{Zor10, Dav10,
  Dav11, Mar11}. Other parameters (e.g.\ $\alpha_{th}$) can be added,
and the parameters can be allowed to vary systematically between
systems, but even this might well miss real physical complexity. 
As discussed in \S~\ref{sec:energy},
there is no reason why the effective value of $\alpha_{\rm CE}$ cannot 
vary drastically even between systems with similar initial
conditions. Moreover, the calibration results produced by different
groups sometimes show opposite trends; see the discussions in \S~\ref{sec:onset_tides},\S~\ref{meth_comp}. 
Here we will pay attention to {\it other} characteristics of post-CE systems 
as possible keys to understanding common-envelope evolution.

\subsection{A priori expectations of appearance during CEE}
\label{sec:sim_appearance}

Whilst the plunge-in is proceeding, the envelope of the primary star expands. 
A giant donor rapidly evolves up its giant branch, though appearing  {\it  colder} than a regular giant of the same
luminosity \citep{Iva02, podsi_03_sn87a}, being closer to Hayashi line. 
The degree of the expansion of the {\it bound} envelope at the end of
the plunge-in phases -- and therefore also during  the self-regulating
spiral-in -- depends on the mass ratio, on the primary initial mass
and on its radius (or luminosity), although it is not currently
possible to specify the sensitivity with respect to these parameters.

As an example of the lack of our current understanding, we first
describe some results taken from 1D simulations (described in full detail in \citealt{Iva02} 
and in parts in \citealt{han02, iva02_mtypes,Iva03}). A $1.6  M_\odot$
giant with a pre-spiral-in radius of about 140 $R_\odot$ was found to expand
3-fold during a common envelope event with a $0.3  M_\odot$ companion, on a time-scale of  20 years.
During the plunge-in phase, a 20 $M_\odot$ giant with a pre-CE radius of $1100  R_\odot$ expanded by
a factor of about  2.5 over 100 years when the companion had a mass of
$5  M_\odot$, but when the companion had a mass of $1  M_\odot$ then the
expansion was greater (a factor of 4) and the plunge-in is more rapid 
(taking place in only $\sim  50$  years). In that second case, the
spiral-in of the $1  M_\odot$ companion never changes to become
self-regulating. The more aggressive spiral-in might partly explain why the lower-mass
secondary produced greater envelope expansion, though the difference
in spiral-in duration is less than a factor of two. 

Using a giant of 0.9 $M_\odot$ (with a radius an order of magnitude
different to the 20 $M_{\odot}$ star in the previous example), the 3D 
simulations of \citet{Passy11} found that the orbital decay of less
massive companions takes slightly \emph{longer} than for more massive
compations. It is perhaps not surprising that a very different
situation, modelled using very different methods, results in the
opposite trend. But our lack of understanding of that difference is
significant.  We note that the degree of expansion of the bound
envelope in the two cases is similar (a factor of several). 

It might be that this particular timescale comparison between codes is
invalid, i.e. that we are not comparing physically quatities with
similar meaning. We define the `fast plunge-in' to start when the
envelope begins to expand. In 1D, this fast plunge-in starts
gradually, whilst in 3D it is forced to coincide with the start of the 
simulations due to the choice of initial conditions. Also the 3D
results do not provide a single value for the radius of the envelope, 
so it is not clear when exactly fast envelope expansion started.

The stellar expansion is directly related to the increase in luminosity, by
4-16  times (by up to 3 magnitudes) for the cases described above.
At the very end of the self-regulating spiral-in
phase, if the binary is not fated to merge, the envelope experiences another fast
expansion. 

Before the envelope is ejected, 1D  simulations find that this
CE may also experience pulsations of increasing strength (see, e.g., the case
with $1.6 + 0.3  M_\odot$ from \citealt{Iva02}), 
before becoming unbound. The period of pulsations is about several years; at least
an  order  of  magnitude  longer than  star's  dynamical
timescale. There are no 3D simulations for this stage. Furthermore,
the important timescale to develop these pulsations is significantly longer than the dynamical one, 
and existing 3D hydrodynamical codes do not contain all the physics
necessary to reproduce them.

Likewise, none of the numerical methods (1D or 3D) used thus far is capable of 
obtaining the beginning of envelope ejection via 
outflows as discussed in \cite{Ivach11}.

\subsection{Observed transients as potential CE events or stellar mergers}
\label{obs_during}

Despite being a relatively short-lived event, CEE is expected to be accompanied
by a rise in luminosity which could be detected as a transient event. 
V838 Mon-type eruptions and the great eruption
of $\eta$ Car have both been argued to be potentially caused by
violent binary interactions.\footnote{We clarify here that we do not mean that 
either of V838 Mon or $\eta$ Car were definitely CEE. There are
several alternative, non-CE, scenarios which try to explain V838 Mon.
Nor is the Great Eruption of $\eta$ Car known to be a CE event;
it could perhaps have been another kind of rapid binary interaction,
e.g., a mass transfer event \citep{kashi+2010}. The fact that $\eta$
Car is currently a binary system has been used to argue against any
stellar merger models, but it is not possible to completely rule out
a CE event as it could previously have been a triple system.}
In particular, V1309 Sco  (a V838 Mon-class event)
seems to be the most promising case so far for an active CE event (or merger)
being caught in action.

The discovery of V1309 Sco was reported by \citet{2008IAUC.8972....1N}
and it was identified as  a ``red nova'' or ``V838 Mon-type eruption''
using  VLT/UVES followup observations by \citet{2010A&A...516A.108M}.
The eruption  was detected early in  September 2008 and  took place in
the    field    of    view    monitored   by    the    OGLE    project
\citep{2003AcA....53..291U}.  \citet{2011A&A...528A.114T} reported the
detection of the progenitor up to six years prior to the outburst. The
pre-outburst primary was classified photometrically as an F-type giant
\citep{Rudy+2008}. 

Prior to the outburst the  object was an eclipsing contact binary with
an orbital period  of $\sim 1.4$ days, however  the orbital period was
not constant and decreased by $1.2\%$ between 2002 and the outburst in
2008.  This orbital period is arguably too long to classify the progenitor as a
W UMa-type binary, which would be expected to merge as the primary
leaves the main sequence \citep{WebbinkContact1976,Rasio1995}. 
However the orbital period is also too
short to say that the binary contained a very evolved giant.  For such a primary, especially
considering the apparently comparatively low mass of its
companion, the theoretical prediction would be that a
common envelope event would be likely to result in a merger rather
than in envelope ejection.
 
Between 2002-2006  the light  curve showed two maxima and  two minima
during each orbital period, but  transitioned to a single  maximum and
minimum  in  2007.  During  the  same  time,  the  brightness  of  the
progenitor  increased  to $I  \simeq  15.5$  in  April 2007  and  then
decreased by $\sim 1$ magnitude  until March 2008, when the brightness
began  to rise  exponentially.  At  its  peak in  September 2008,  the
object  was $\sim  6$  magnitudes brighter  than  before the  outburst
\citep{2011A&A...528A.114T}.

During the first five months of the rise, the characteristic timescale
for the increase in luminosity  was $27$ days. During the outburst and
the subsequent decline the spectral  type changed from F9 in September
2008   to    M7   in   April    2009   and   M3   in    October   2010
\citep{2011A&A...528A.114T}.    This  is   similar  to   the  observed
evolution of V838 Mon itself \citep[see, e.g.][]{2005A&A...436.1009T}.

It is  tempting to  interpret these results  as a binary  that evolved
from  a contact  system (before  the second  peak in  the  light curve
disappears) to  a stable  common-envelope systems (after  the second  peak has
disappeared but  before the outburst)  followed by a merger  (when the
exponential increase in the  luminosity begins); see also
\citet{Stepien2011}.  
It is noticable that the behavior is qualitatively as predicted by
simulations. Quantitative comparisons are less helpful, since no
simulations for such a system have been published. Nonetheless the increases
in luminosity and radius are much larger, and the post-outburst decline in luminosity also more
rapid,  than might have been expected
based on the published simulations involving larger, more evolved
giants (as described in \S \ref{sec:sim_appearance}).

So V1309 Sco seems like an excellent candidate for an individual system which has
been observed during the CEE phase.  The fact that we have evidence for
a pre-outburst binary nature is especially compelling in that case. 
The resemblance to V838 Mon is strong enough for us to consider a link
to CEE very likely in that case too. Indeed, stellar merger models were proposed as
potential explanations very
soon after the discovery of the V838 Mon outburst 
\citep[see, e.g.,][]{Bond+2003,SokerTylenda2003,RetterMarom2003,2005A&A...436.1009T,Tylenda+2005}.

A wider class of transients with similarities to V838 Mon also invite
a possible CEE explanation: ``red novae''. Those objects are not novae
by their physical nature, despite their observational
similarities; for that reason alternative names for this class have been
suggested, including ``intermediate-luminosity red transients''
and ``intermediate luminosity optical transients''. 
These events have luminosities between novae and supernovae,
with peak absolute visual magnitudes of $-13 M_{V}$ to $-15
M_V$. During the outburst the source is cold --hence red -- unlike a
normal classical nova. 
The energy involved in producing these events is order-of-magnitude comparable
to the likely orbital energy release from CEE or the binding energy of the envelope (about $10^{47}$ erg,
\citealt{Bond+2009}; \citealt{Kulkarni+2007}). Specific examples of
this class include M85 OT 2006-1 \citep{Kulkarni+2007,Ofek+2008}, NGC300 OT 2008
(\citealt{Bond+2009}; though see \citealt{kashi+2010} for an alternative scenario which
involves rapid mass transfer from an extreme AGB star on its MS companion), PTF 10fqs
\citep{Kasliwal+2011} and M31 RV \citep[][and references
within]{Bond+2011}. The rate of similar events has been
estimated to be as much as 20\% of the core collapse SN rate
\citep{Thompson+2009}.
The observed ejecta velocities also broadly match what might
be expected from CEE (or a stellar merger). \citet{Kasliwal+2011}
detected expansion velocities in  PTF 10fqs of $\sim 1000 \rm km~s^{-1}$. 
For NGC300 OT 2008 a wide range of velocities have been published, from $\sim
75 \rm km~s^{-1}$ \citep{Bond+2009}, to $\sim 1000 \rm km~s^{-1}$ \citep{Berger+2009}. 
The low-velocity end of that range is easily compatible with CEE, or a
stellar merger model involving a large giant star.
Velocities of $\sim 1000 \rm km~s^{-1}$ suggest that the primary star would have had to be less extended, 
but this still could be compatible with an early giant, as was observed in
the example of V1309 Sco.

\subsection{{\it Post}-CE  appearance}
\label{sec:postCEobs}

If the CE event results in a merger, then the initial post-event reaction of
the star is the rapid evolution of a star out of its thermal equilibrium. In
this case the star is overluminous, and contracts towards equilibrium
as it radiates away excess energy.
This contracting sequence,  just like during the plunge-in phase, goes
along    the    giant    branch,    though    now    towards    smaller
luminosities -- during a  spiral-in and subsequent merger, the
primary star performs a loop  on the HR diagram around the giant branch. After
this fast contraction had finished, its further evolution depends on the details of
mixing  of the inner  layers, and may be similar to a normal  giant
evolution (although perhaps with abnormal surface composition).
Abnormal chemical  compositions may include enhanced abundance of 
He \citep[up to 0.4,][]{podsi_03_sn87a} or s-elements \citep{Iva03},
as well as unusual CNO ratios \citep{2003ASPC..304..339I}.  
B[e] supergiants might well be post-merger systems \citep{podsi_06_mergers}.

In some cases then post-merger massive stars are able to reach
core-collapse as a blue supergiant. This explanation for the
progenitor of SN 1987 A is now well-established, largely as a result
of the distinctive triple-ring nebula which was formed following the
merger  \citep{PhP1991,  PhP+1991, PhP1992_87A, Morris+PhP2006, Morris+PhP2007}.
Other information about the violent past of merger products
could be provided by the shape of the nebula around it \citep[e.g.,][]{2009MNRAS.399..515M}.

Post-merger giant stars could well be rapidly rotating, and
giants with unusually high surface velocities have been identified
\citep{Garcia2011KITP}.  Stars where only the \emph{surface}
layers are rapidly rotating could be especially notable:
potentially a low-mass companion is still orbiting in the outer, 
low-density, layers of the giant. 

Let us now consider in more detail cases when CE leads to survival of
the binary.

\subsubsection{Post-CE eccentricities as a constraint on time of the ejecta}

\label{obs_ecc}

One potential constraint on CEE that has received little previous
attention is the post-CE orbital eccentricity. If we
detected post-CE eccentricity then it would be a useful
indication in trying to understand the end of the preceeding CE phase. However,
eccentricity is fragile. For fixed angular momentum,
circular orbits have the lowest energy, so energy dissipation can act to circularise
orbits following the CE phase.  So the effects of tidal
circularisation \citep{zahn77} largely rule out many binaries from giving us useful information on
eccentricities (e.g.\ the large class of main-sequence + white dwarf
binaries). However, binaries in the nuclei of planetary nebulae should
still be helpful, since in their case there has been insufficient time since the ejection of the
envelope for tides to have had a significant circularising
effect. Another promising exception is a single long-period
main-sequence + white dwarf system we mention below.

Potentially useful classes of systems are white dwarf-white dwarf and white
dwarf-neutron star binaries, along with systems containing an sdB star
and a compact remnant. In these binaries post-CE
circularisation is expected to be ineffective. SdB stars are radiative, and relatively
short-lived; tidal circularisation is believed to be much less effective in radiative stars
than ones with convective envelopes \citep{zahn77}. 

What might we learn? We expect that the two stars entering a CE to be in a
near-circular orbit due to pre-CE tidal interactions. However, 3D
simulations show that the eccentricity grows rapidly during the early
spiral-in phase. Ejection immediately following the dynamical plunge
might therefore leave residual eccentricity, which could potentially be used as a
diagnostic of this phase.  Current 3D hydrodynamic simulations produce small
eccentricities at the end of this phase $\lesssim 0.1$. However, if the system continues into a slower
self-regulated spiral-in, the eccentricity built up during the
previous plunge is likely to be damped away. 
Observed eccentricities (or lack of) may then largely tell us how
effective and long-lasting this self-regulated phase is. 

Another possibility is that the post-CE eccentricity could be
increased by the presence of a dynamically significant, close
circumbinary disk -- if one exists. Tidal interactions with such a 
disk should be strongest at apastron, which tends to
amplify any existing eccentricity \citep{art91}.
Any observed eccentricities may indicate that such disks are present.

For the most part observed post-CE systems do not have significant
eccentricities. Limits are typically of the order $\epsilon < 0.05$ from
radial velocity work, although with more work upper limits on any
eccentricity present of around $0.01$
should not be hard to achieve (generally the determination of periods rather than eccentricities
has been the target of radial velocity work). In eclipsing cases, one can
reduce the errors by a further factor of 10 or so, and in pulsar binaries, one can reach
uncertainties in eccentricity of order $\epsilon \sim 10^{-6}$.
However, apparent eccentricity detections should be treated with
caution as the measurement of eccentricity is always biased to be
positive by whatever errors are present
(since the probability distribution is necessarily one-sided). 
Applying a strict $> 5\sigma$ criterion, there are two
cases of significant eccentricity amongst the sdB binaries which are
PG1232-136 ($\epsilon = 0.060 \pm 0.005$) and [CW83]~1419-09 ($\epsilon =
0.039 \pm 0.005$) \citep{ede05}. One other
interesting case is G~203-47, an M3.5V star in a 15-day orbit with a white
dwarf and having an eccentricity of $\epsilon = 0.068 \pm 0.004$ \citep{del99}. 
With only a few examples, against many
non-detections, one should be wary of Kozai-cycle driven eccentricity \citep{koz62}, 
yet perhaps there is some potential for learning
about the CE phase from eccentricities.


\subsubsection{Planetary nebulae as a constraint on ejecta velocities and timescales}

\begin{figure}
\centering
{
\includegraphics[width=0.46\textwidth]{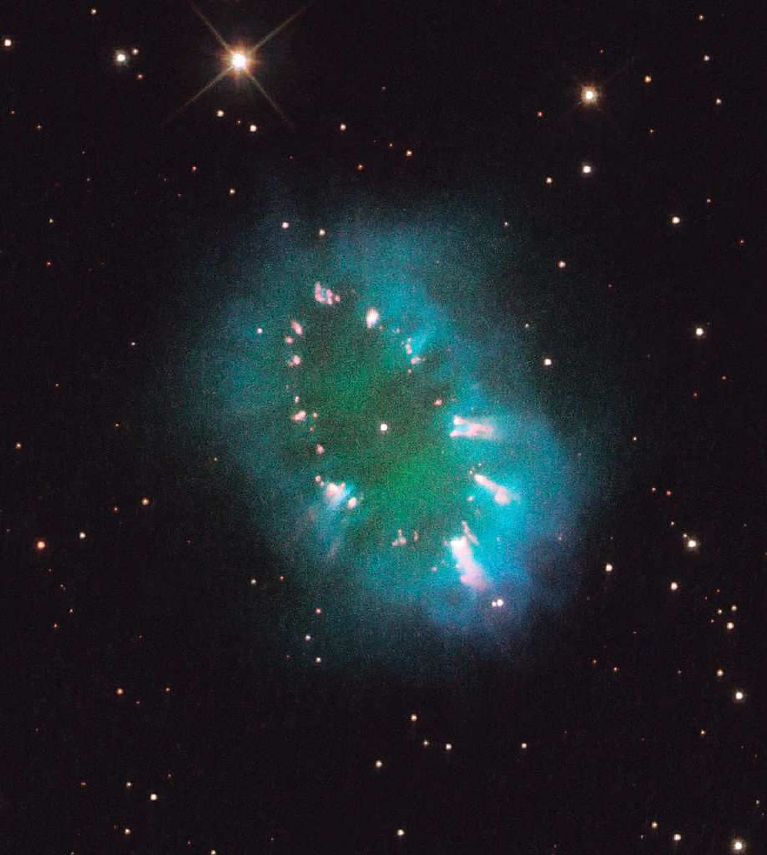}
\includegraphics[width=0.49\textwidth]{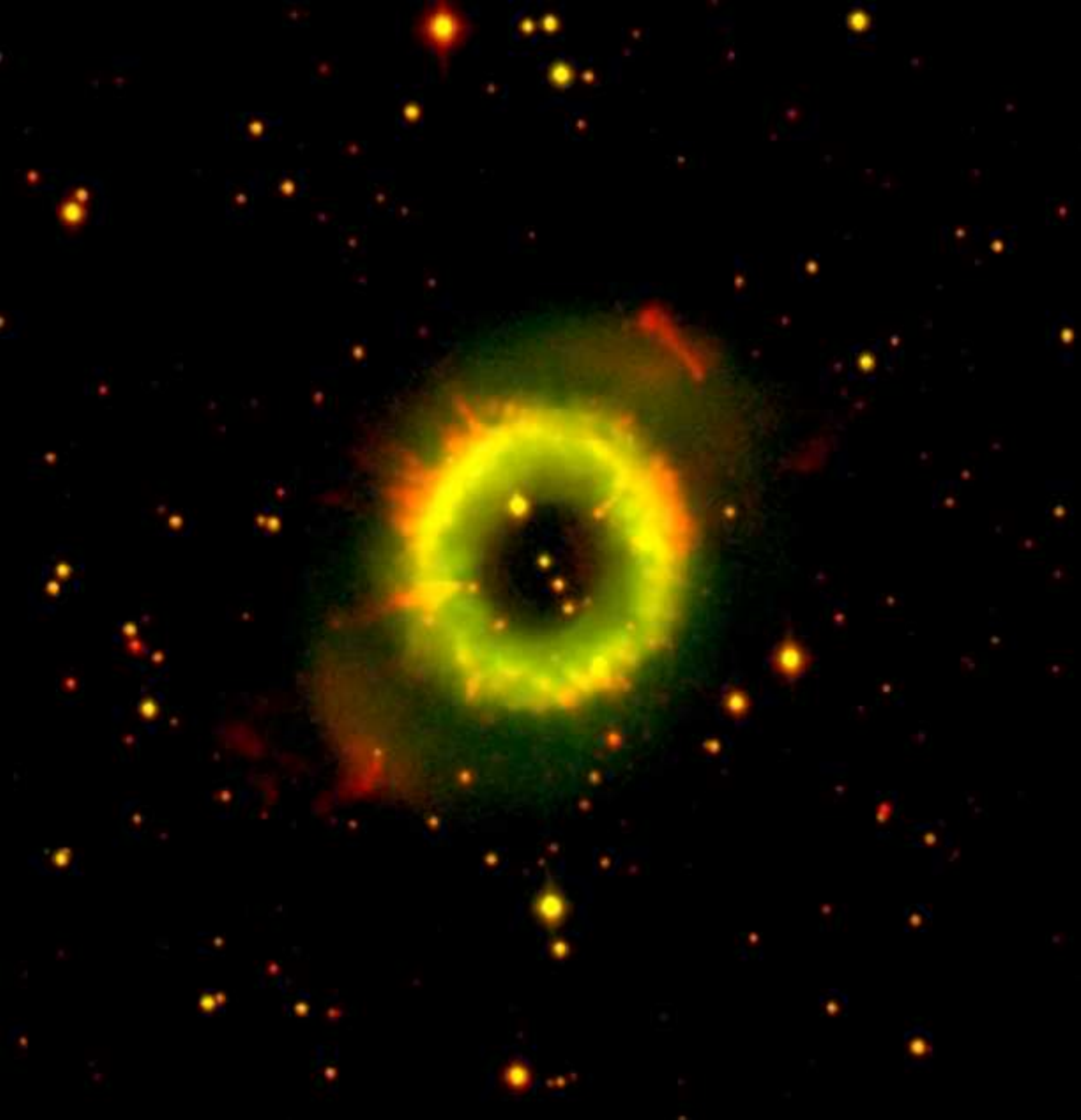}

\vskip0.1cm

\includegraphics[width=.46\textwidth]{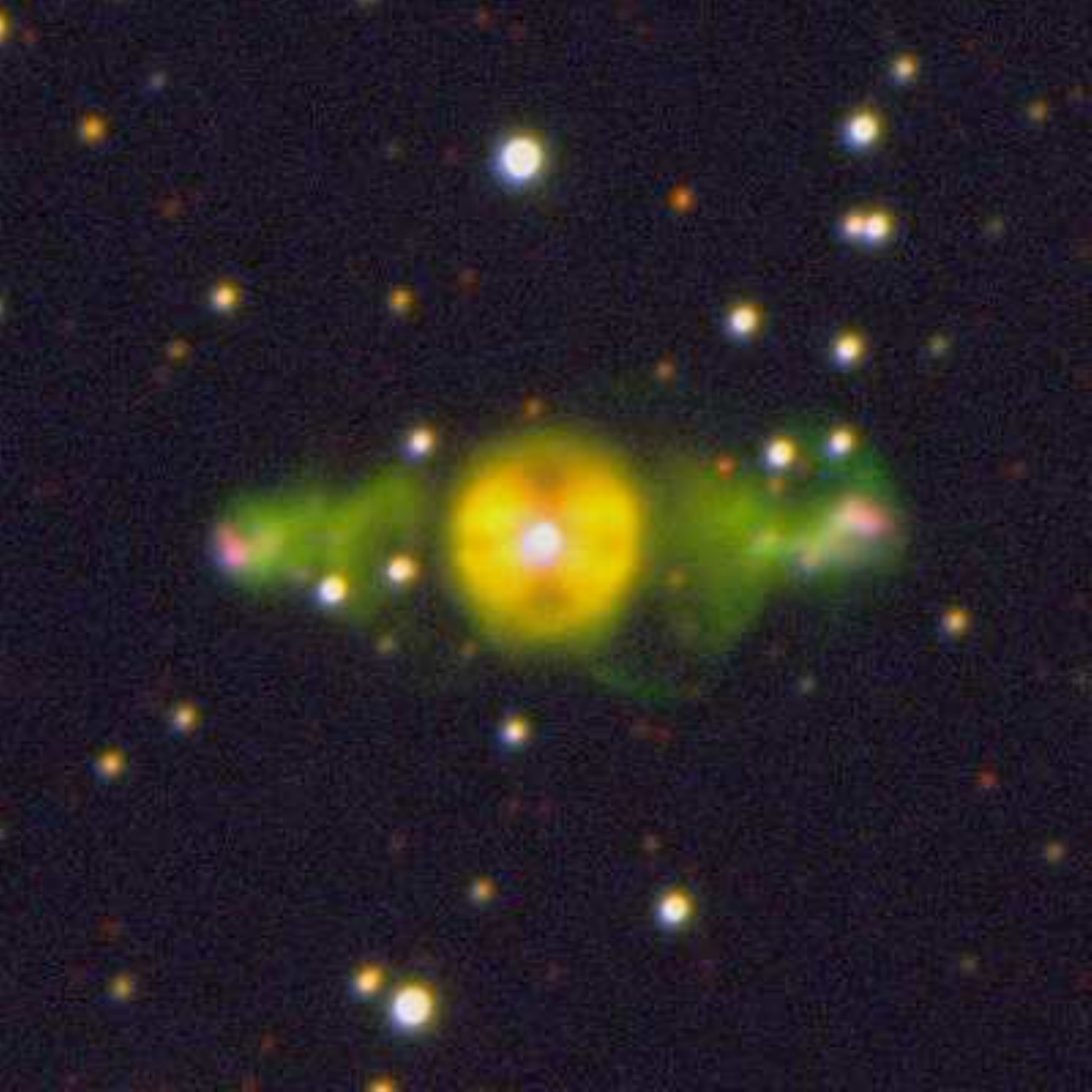}
\includegraphics[width=.49\textwidth]{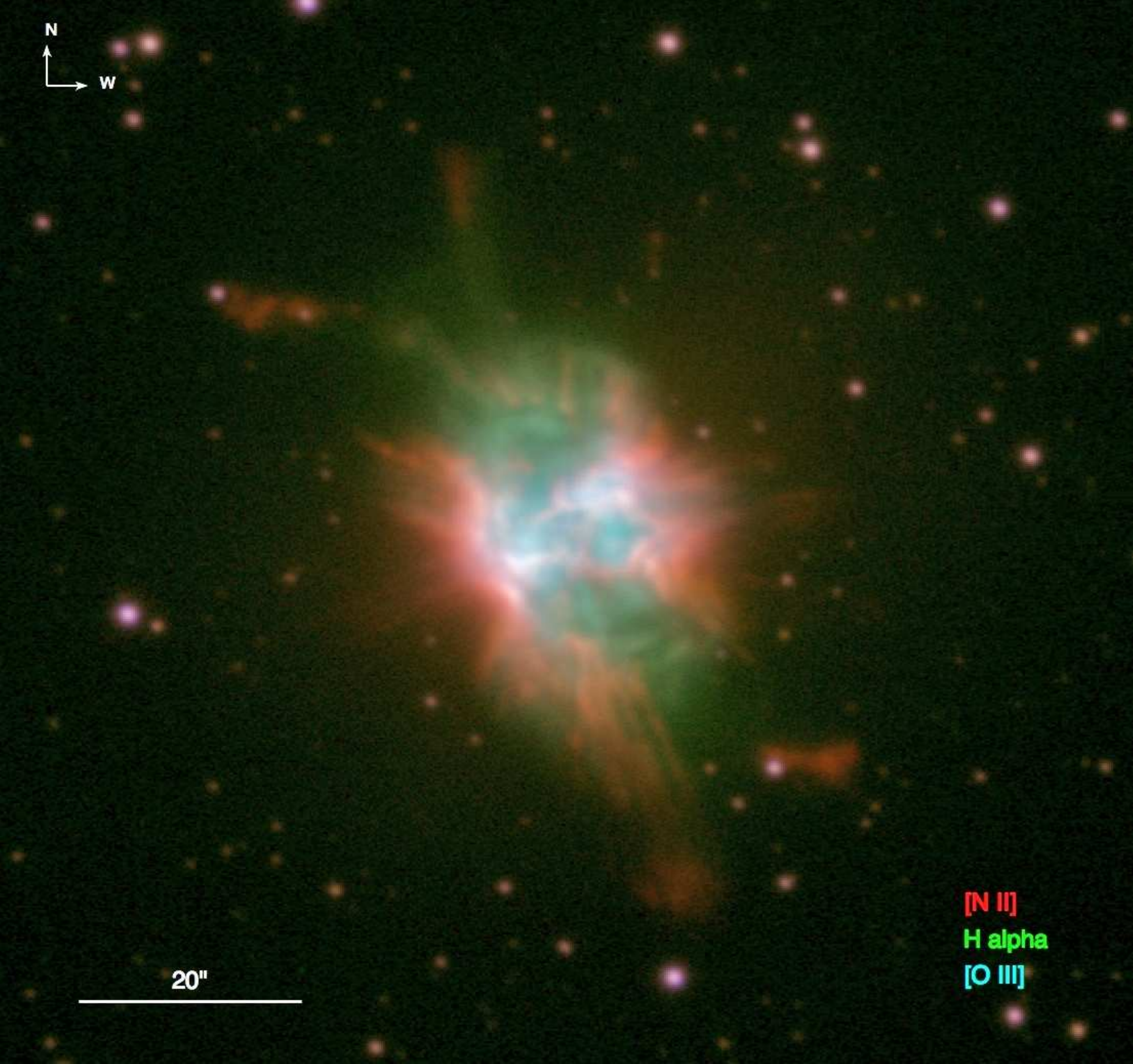}
\caption{Post-CE planetary nebulae with known compact binaries as central objects. Top left -- Necklace Nebula (image credit: NASA, ESA, and the Hubble Heritage Team (STScI/AURA), for details see \cite{2011MNRAS.410.1349C}); top right -- NGC 6337 (credit to Corradi, for more details see \cite{2000ApJ...535..823C}); bottom left -- ETHOS 1 (credit to B. Miszalski, for more details see \cite{2011MNRAS.413.1264M, 2011ASPC..447..159B}); bottom right -- NGC 6778 (credit: \cite{2012A&A...539A..47G}) }
    \label{plan_neb}
  }
\end{figure}

Approximately one  in five planetary  nebulae (PNe) are  ejected common
envelopes \citep{Han95,2000ASPC..199..115B, 2009A&A...505..249M}. The
potential role of CE ejection in shaping PNe morphology was
considered very soon after CEE was proposed \citep{webpn79}.
Hence studying the diverse shapes and velocity distributions of nebulae 
like these should give us insights into the CE ejection mechanism.
One may expect these PNe
to all  have the same  shape in virtue  of the common  phenomenon that
generated  them. 
In particular  one may  expect a  traditional bipolar
shape,  promoted by the  loss of  the AGB  envelope in  the equatorial
plane \citep{San98,2003RMxAC..18...24D}, followed  by a
spherical fast wind from the hot primary, which swept past the ejected
common envelope.
While this picture is expected, its realization can take various shapes  
-- see Fig.\ref{plan_neb} and Figs. 4 to 6 in \cite{demar2009}.

Initially, shapes around  the few known post-CE PNe  appeared not to be
systematically bipolar
\citep{1990ApJ...355..568B,2007BaltA..16...79Z}. It was noted,
however, that post-CE  PNe lack the multiple structures that
may form over several phases  of varying mass-loss, in line with their
AGB evolution having been  interrupted. \cite{1981ApJ...249..572M} and \cite{1997ApJS..112..487S} 
suggested  that   bipolarity  in  PNe  is  promoted   by  those  binary
interactions  that  avoid  a  common envelope  phase.  However,  later
studies based on  a larger number of post-CE PNe,  showed that there is
at least a  tendency for post-CE PNe to  have bipolar shapes \citep{demar2009,2009A&A...496..813M},  
 or a  shape that results  from a
faded bipolar structure.  In addition, common-envelope PNe also seem  to share a
propensity  to exhibit  low ionisation  features, knots  and filaments
embedded in larger, toroidal structures \citep{2009A&A...496..813M}.

A detailed  kinematic analysis of  post-CE PNe should be
able  to give  significant  insight on  the  common envelope  ejection
phases and timescales.  As an example, \cite{2007MNRAS.374.1404M} carried
out  a  detailed kinematical  analysis  of  the eclipsing  post-common
envelope binary  central star of PN  Abell 63. In this edge-on object, a tube-like
disk is expanding at  $17\pm1$~km~s$^{-1}$ along the orbital plane and
two  tenuous,   collimated  lobes  with  bright   caps  are  expanding
perpendicularly to the plane of the disk and the plane of the orbit at
$126\pm 23~{\rm km~s}^{-1}$.  The lobes  appear  to have  preceded the  disk
formation by a  few thousands years. Very similar  kinematics are seen
in  other post-common  envelope  PNe, such  as  ETHOS~1 \citep[][bottom left image on Fig.\ref{plan_neb}]{2011MNRAS.413.1264M}  
and the  ``Necklace" \citep[][top left image on Fig.\ref{plan_neb}]{2011MNRAS.410.1349C}.  One interpretation
of these objects is that  a collimated outflow (perhaps even a jet)
was active during  or shortly before the envelope  was ejected.
On the other hand the kinematic analysis of NGC6778 shows that the two jet pairs are 
kinematically younger than the main nebula. These two jet pairs also have different 
velocities and seem to be curved \citep{2012A&A...539A..47G}. 
Further detailed studies of PN around post-CE central stars should provide us with a great deal of insight onto the CE phase.


\subsection{Double-core common-envelope evolution}

In standard CEE, it is typically assumed that only one of the pre-CE stars has a well-developed core and
extended envelope, whilst the in-spiralling companion star is assumed to be
relatively dense.  The special case where both stars have expanded to giant-type
structures by the onset of CEE is referred to as double-core
CEE.\footnote{To help those who may be looking through early
  literature on CEE, we note that this terminology has the potential
  to be confusing, as standard CEE was itself sometimes referred to as
  ``double-core evolution''.}  Successful envelope ejection from
double-core CEE would expose \emph{both} cores, i.e.\ it would result
in a binary composed of the cores of both the pre-CEE stars. 
This possibility was briefly mentioned earlier in the context of the
onset of CEE, since double-core CEE does not normally begin
following tidal instability or dynamically unstable mass transfer (see
\S \ref{sec:onsetDynamicallyStable}). So, if double-core CEE
ever occurs,  this fact would at least increase our understanding of which
systems undergoing mass transfer are unstable to entering CEE.
Observational confirmation that double-core CEE occurs -- or does not
occur -- is not yet available, but it seems worth continuing to investigate known systems
to try to constrain the incidence of this process.
For example, the existence of double He-rich hot
subdwarf binaries might provide evidence that
double-core CEE does happen in some cases
\citep{Justham+2011}.  The formation of double neutron-star binaries
was the original motivation for suggesting double-core CEE, and some
of them may indeed be produced through this channel
\citep{Brown1995,Dewi+2006}.  
However, the different spins of the
observed double neutron star systems suggest that these known systems 
did not evolve through double-core CE.  It also seems plausible that
the apparent mild recycling of the older neutron stars in the observed
systems is due to mass transferred during the thermal core
readjustment following a normal, single, envelope ejection
\citep{Iva11}.

\section{Conclusions and Directions for future work}
\label{sec:finale}

We have attempted to reassess everything that we know to-date from the
\emph{theoretical} point of view about the \emph{physics} of CEE and
related events. This has included comparing and trying to understand
the main features of the most recent hydrodynamic simulations of CEE,
along with the relevant numerical methods.  We have also briefly
discussed some of the more direct -- and hence hopefully less
misinterpretable -- observational constraints.

Most importantly, we have tried to understand CEE from a physical
point of view with the eventual aim of replacing the existing top-down
parametrizations (such as the energy formalism) with a bottom-up
description. However, it is clear that this problem is exceptionally
complex. Any individual CEE event consists of several sub-phases
occuring on a wide range of timescales and under the influence of diverse
physical mechanisms. No existing numerical method is capable of
grasping it all. Moreover, pen-and-paper arguments still do not agree
on which are the dominant physical processes, and which physics (if
any) can be neglected.

In order to make progress, therefore, we need to determine how to
study the phases within CEE in  a self-consistent way, whereby the
outcome of one phase becomes a realistic initial condition for the
next phase. In addition to dividing CEE into separate phases in time,
each phase can be attacked from different directions: we can try to
define useful self-contained problems which are both manageable and
interesting.

It should be clear from this work that there are still many points of disagreement within the community. For example, a strong constituency believes that CEE is an intrinsically 3D problem: if that is correct, then great care would have to be taken over which 1D simultations, if any, would be worthwhile to perform. Nonetheless, there are some theoretical and modelling goals which we think are both useful and realistically tractable. These include:

\begin{itemize}
\item[1] \textbf{Understanding pre-CE evolution would help to better initialise CE simulations. We should aim to constrain:}

\item the conditions needed to start a CE phase. Even the range of systems undergoing RLOF which lead to CEE via dynamically-unstable mass transfer is not yet fully known. 

\item the angular momentum distribution of the matter at the start of hydrodynamical simulations. 

\item the pre-dynamically-unstable phase. This includes the time between the start of RLOF and CEE as well as, e.g., enhanced winds before RLOF properly begins.

\item[2] \textbf{During CEE, we should concentrate on a better physical understanding of:}

\item how one phase transforms into another phase, e.g.\ when a dynamical plunge-in becomes a self-regulated spiral-in, or when a self-regulated phase ends. 

\item whether and how recombination works in order to provide the envelope with momentum. At the very least we need to equip hydro codes such that ionization is included in their equations of state.

\item to what extent the energy from the binary orbit is transferred to the envelope through viscosity and local frictional heating, or through large-scale gravitational interactions (i.e. spiral waves). 

\item how fine-tuned envelope ejection is, i.e. how close is the ejection velocity of each element to the local escape velocity.  

\item the location of the bifurcation point that separates the ejected envelope from the bound remnant.

\item how outflowing envelopes are shaped, partly in order to allow comparison with the morphologies of PNe.

\item[3] \textbf{Developing and understanding codes and methods for CEE, by:}

\item comparing existing 3D hydrodynamic codes and results. This includes understanding the influence of the initial conditions as chosen by different groups, whether using the same or different types of code. 

\item attempting to treat the problem using coupled 1D and 3D codes, to try to take advantage of their differing strengths. For example, 3D hydro could be used to produce energy input source terms for a 1D code.  Even grafting model atmospheres onto 3D simulations could help us understand the observational signatures of systems entering CEE. 

\item thoughtfully dividing the set of possible simulations into those which can be treated with 1D codes, and which need 3D. 
\end {itemize}

Whilst it is too early to speak about a detailed comparison with observations, more observational constraints from post common-envelope binaries, and especially from observations of planetary nebulae, should be very helpful in further understanding (as well as for code verification). Specifically, observations on nebulae shapes and velocities may help to understand how the ejection proceeded. 

The eccentricity, if any, of post-CE binaries may also help to
identify and understand any CE events which resulted in envelope
ejection immediately following the dynamical plunge-in.  It is
expected that eccentricity would be lost during the slow spiral-in,
and might even be damped away during the process of envelope
ejection. The presence of eccentricity might therefore be a clue that
the envelope was ejected immediately after the the fast spiral-in
stage, not following a self-regulated spiral-in phase.  Note, however,
that observational biases tend to produce spurious apparent
eccentricities, so it is easier to give upper limits on eccentricity
than to be confident about detections.

One potential observation would give definitive information with the
detection of a single object.  If we find a single Thorne-\.{Z}ytkow
Object, we would have strong constraints on hypercritical accretion.

We note that the clarity and certainty of our understanding of CE
physics is certainly not yet good enough to predict formation rates of
many classes of system. For some post-CE systems the formation rate
inferred from observations can not currently be explained within the
mainstream energy formalism when using physically realistic
parameters (e.g., short-period black hole X-ray binaries). This
definitely strongly reduces the reliability of predictions for
formation rates, e.g.\ for systems with black holes, including close
black hole-black hole binaries which are of interest for
gravitational-wave astronomy.  

Those who study the formation rates of binary systems by means of population synthesis 
must anticipate and acknowledge that current uncertainties in
theoretically predicted rates could be about two orders of magnitude
arising from uncertain CE energetics for  systems where CEE is
involved. (In addition to this, formation channels involving CEE also
introduce other uncertainties, e.g.\ from $q_{\rm crit}$.)  In some
binary systems the major uncertainty comes from our poor understanding
of the energetics involved (and it is hence related to $\alpha_{\rm
  CE}$), whilst in others it is due to an arbitrary choice of the
remaining core's mass (hence it is related to $\lambda$).  This review
discusses several ways in which these uncertainties might be reduced
by more careful consideration of the physics involved.  Attempts to
observationally calibrate these parameters are only advisible if
performed for well-defined classes of post-CE binaries, since there is
very strong reason to expect they should not be global parameters. We
stress that trying to determine \emph{the} single effective value of
$\alpha_{\rm CE}$ is very misguided: there will be different values in
different cases, as the time-scales and energy sources and sinks
should vary from one CE event to another.

We are convinced that much work remains to be done, however we feel optimistic that the solution of the problem 
could be achieved within the next decade.

\begin{acknowledgements}
All the authors thank KIAA, the National Natural Science Foundation of
China (NSFC) and the Beijing Astronomical Society for providing
support and hospitality. The authors also thank Gijs Nelemans
  for very helpful constructive criticism, and James Lombardi
  for Fig.~3.
NI acknowledges support from NSERC Discovery and Canada Research
Chairs Program; this research was supported in part by the National Science Foundation under Grant No. NSF PHY05-51164.
 SJ thanks the Kavli Foundation, NSFC (through grants
10903001 and 11250110055) and the Chinese Academy of Sciences for support. 
XC, HG and ZH thank the NSFC (Nos.\ 10973036, 11173055,
11033008, 11203065), Chinese Academy of Sciences (No.\ KJCX2-YW-T24 and the Talent Project of Western Light) 
and Yunnan National Science Foundation (No.\ 2008CD155) for support.
The work by CF was carried out in part under the auspices of the 
National Nuclear Security Administration of the U.S. Department
of Energy at Los Alamos National Laboratory and supported by Contract No. DE-AC52-06NA25396.
XL acknowledges support by the NSFC through grant 10873008 and by the National Basic Research
Program of China (973 Program 2009CB824800). 
TRM acknowledges support from the STFC.
ATP thanks the STFC for his studentship. 
RT acknowledges support by the NSF through grant  AST-0703950.
TMT acknowledges support from Norbert Langer and the Argelander-Insitut
f\"ur Astronomie, Universit\"at Bonn.
EvdH gratefully  acknowledges support by the Leids Kerkhoven-Bosscha 
Fonds that enabled him to participate in this program. RW acknowledges
support from the Department of 
Astronomy, University of Illinois at Urbana-Champaign, and from NSFC grant 11033008.
\end{acknowledgements}

\bibliographystyle{apj}       
\bibliography{comenv}{}   

\appendix

\section{On a spatially-discrete formalism for mesh-less finite-volume methods}
\label{app:method}

Numerical finite-volume methods are usually divided into Eulerian and
Lagrangian schemes. In the former one, the discretization points are
fixed in space, whereas in the latter the geometry they are moving with the fluid
velocity, in which case it is convenient to think of them as physically
associated with a  fluid element. Usually, Eulerian scheme use a geometrical mesh,
either structured (e.g. Cartesian) or unstructured, whereas Lagrangian
methods are mostly considered to be related to particle methods. These two
seemingly different approaches actually have a lot in common. It has
recently been demonstrated that fully Lagrangian methods can be implemented on
unstructured Voronoi meshes \citep{1988CoPhC..48...39T, 2010MNRAS.406.2289H},
while Eulerian scheme can be successfully formulated in entirely mesh-less
form \citep{VILA:1999,Hietel_afinite-volume,springerlink:10.1007,1404789,
2011MNRAS.414..129G}.

Here we present a generic formalism that leads to spatially
discrete mesh-less finite-volume equations. The form of the these
equations is the same as in the case of spatial discretization on a mesh. The
geometrical quantities that are obtained from the mesh, such as volumes or
areas, are translated into spatial integrals in the mesh-less schemes. Finally,
we show that this generic formalism leads, via a few approximations, to standard
SPH equations of motions. This explains why SPH, among other Lagrangian
particle methods \citep{NME:NME547, NME:NME832, 2011MNRAS.413..271A}, remains
the most robust particle method in astrophysical computational fluid dynamics.

\subsection{ Spatially discretized finite-volume equations}

\subsubsection{Mesh-based discretized equations}
The conservative formulation of Euler equation of ideal hydrodynamics can be
compactly written in the following integral form
\begin{equation}
	\frac{d}{dt}\int_{ T}q\,dV + \oint_{\partial{T}}{\bf F}\cdot	d{\bf \Sigma} =
	\int_{T} S\,dV,
	\label{eq:q_integral}
\end{equation}
where $T$ is volume of a fluid element, $\partial T$ is its boundary with
outward pointing normal, and $q$ and ${\bf F}$ are defined as follows
\begin{equation}
	q = \left(  
	\begin{array}{c}
		\rho \\
		\rho {\bf v} \\
		e_{\rm tot}
	\end{array}
	\right)\, \quad	
	{\bf F} = \left(  
	\begin{array}{c}
		\rho ({\bf v - w}) \\
		\rho {\bf v}\otimes({\bf v - w}) + P\hat{\bf I} \\
		e_{\rm tot}({\bf v -w}) + P{\bf v}
	\end{array}
	\right).
        \label{eq:q_flux}
\end{equation}	
Here, ${\bf w}$ is the mesh velocity, $\otimes$ is a tensor product, $\hat{\bf I}$
is a unit tensor, and the rest of the symbols carry their usual
meaning. 
The source term $S$ is identically zero for an isolated system
with no external forces, i.e.:
\begin{equation}
	S_{\rm isolated} = \left(  
	\begin{array}{c}
		0 \\
		{\bf 0} \\
		0
	\end{array}
	\right).
\end{equation}
In the presence of a gravitational potential $\phi$ then $S$
would take the following form:
\begin{equation}
  	S_{\phi} = \left(  
	\begin{array}{c}
		0 \\
		-\del\phi \\
		-v\dot\del\phi 
	\end{array}
	\right).
\end{equation}

For a Eulerian scheme, the location of the mesh is time independent, thus ${\bf w} = 0$. In
Lagrangian schemes the mesh moves with fluid velocity ${\bf w}={\bf
v}$, and so that volume integral of density over the fluid element,
namely its mass, is exactly constant with time.  The physical meaning of
these equations is rather simple: the equations state that the volume integrated
amount of a quantity $q$ inside a given volume element is equal to the net
negative flux of this quantity outside this volume.

The Euler equation (Eq.~\ref{eq:q_integral}) has a simple formal spatial 
discretization on an arbitrary mesh
\begin{equation}
	\frac{d}{dt}(\bar{q}_i V_i) + \sum_{j\in{\partial T}_i} ({\bf F}\cdot{\bf
	A})_{ij} = \bar{S}_i V_i.
	\label{eq:q_integral_descrete}
\end{equation}
Here, $\bar{q}_i$ and $\bar{S}_i$ are the mean values of $q$ and $S$ 
inside a mesh cell $T_i$ which has volume $V_i$. The sum is carried out over
each of the cell boundaries, and here we introduce a compact notation
for the surface integral of a single face between mesh points $i$ and $j$:
 
\begin{equation}
\label{eq:surfaceintegral}
({\bf F}\cdot{\bf A})_{ij} = \int_{\partial T_{i,j}}{\bf F}\cdot d{\Sigma}. 
\end{equation}

\noindent  To achieve a second order approximation of this
surface integral (Eq. \ref{eq:surfaceintegral}) using the 
one-point quadrature rule, the flux of $q$ between two particles, ${\bf F}_{ij}$
is evaluated at the face centroid. This equation states that for numerical
purposes only a projection of the flux to the surface normal is required, which
we write as $F^n_{ij} A_{ij} = {\bf F}_{ij}\cdot{\bf A}_{ij}$.  It is simple
to check that $F^n_{ij} = - F^n_{ji}$. Finally, the geometrical
quantities (e.g.\ cell volumes and the areas of and normals to cell
boundaries) can be directly computed from the mesh.

This discretized form can be straightforwardly applied to an equidistant
three-dimensional Eulerian Cartesian grid. In this case, we have $V_i = V =
l^3$, $A_{ij} = A = l^2$, where $l$ is spacing between grid points along
the coordinate axes. We note that in the case
of a Eulerian scheme, $dq/dt = \partial q/\partial t$ becase ${\bf w}= 0$, thus we
have:
\begin{equation}
	V\frac{\partial\bar{q}_{i,j,k}}{\partial t} + A(F^x_{i+\half,j,k} - F^x_{i-\half,j,k}) -
	A(F^y_{i,j+\half,k} - F^y_{i,j-\half,k})  +
	A(F^z_{i,j,k+\half}-F^z_{i,j,k-\half}) = \bar{S}_{i,j,k}V,
	\label{eq:q_cartesian_integral_descrete}
\end{equation}
where $\bar{q}_{i,j,k}$ is average value of $q$ in a cell $(i,j,k)$ and
$F^x_{i+\half,j,k}, F^y_{i, j+\half,k}$, and $F^z_{i,j,k+\half}$ are projections
of fluxes of $q$ between neighbouring cell onto $x-$, $y-$ and $z-$direction
respectively. Now dividing both sides of this equation by $V$, we obtain a
conventional form of second-order spatial discretization that is used on most
Eulerian schemes on equidistant Cartesian grid
\begin{equation}
	\frac{\partial\bar{q}_{i,j,k}}{\partial t} + \frac{F^x_{i+\half,j,k} -
	F^x_{i-\half,j,k}}{l} + \frac{F^y_{i,j+\half,k} -
	F^y_{i,j-\half,k}}{l} + \frac{F^z_{i,j,k+\half}-F^z_{i,j,k-\half}}{l} =
	\bar{S}_{i,j,k}
	\label{eq:q_cartesian_diff_discrete}
\end{equation}

\subsubsection{Mesh-less discretized equations}

Any straightforward application of Eq.\,\ref{eq:q_integral_descrete} to
mesh-less scheme runs into a wall: a set of particles without underlying mesh
lack a well defined definition of volume and areas. Therefore, the simple
approach that works with mesh-based scheme fails in mesh-less case. Here, we
seek a discrete formulation of mesh-less equations by starting from a weak
form of conservation equations (e.g. \citealt{VILA:1999})
\begin{equation}
	\int (\dot\varphi q + \nabla\varphi\cdot{\bf F} + \varphi S)\,dV\,dt = 0,
	\label{eq:q_weak}
\end{equation}
where the volume integral is taken over all the space-time domain, $\varphi$ is a
differentiable function, and $\dot\varphi = \partial \varphi/\partial t + {\bf
w}\cdot\nabla\varphi$ is an advective derivative in the particle velocity field
${\bf w}$. As with the mesh-based scheme, when ${\bf w} = 0$ and ${\bf
w}={\bf v}$ the final equations will describe Eulerian and Lagrangian scheme
respectively.  One of the reasons to use this integral form of conservative
equations instead of Eq.\,\ref{eq:q_integral} is lack of explicit references to
surface integrals, which proves to be important in the subsequent derivation.

In order to spatially discretize Eq.\,\ref{eq:q_weak}, we introduce a partition
of unity $\psi_i({\bf x})$, which is a function defined such that
$\sum_i\psi({\bf x}) = 1$. Here, index ``$i$'' refers to the $i$-th particle,
and the sum is carried out over all particles. If data is defined on a
set of interpolation points, $q_i$, the interpolated value at an arbitrary
location $\bx$ is $q(\bx) = \sum_i q_i \psi_i({\bf x})$, and its gradient is $\nabla
q(\bx) = \sum_i q_i \nabla\psi_i({\bf x})$. Using these definitions we obtain
\begin{equation}
	V\bar{f} = \int\,f(\bx)\,dV = \int
	\,\sum_i f_i\psi_i({\bf	x})\,dV = \sum_i f_i \int\psi_i({\bf x})\,dV =
	\sum_i f_i V_i,
	\label{eq:tot_integral}
\end{equation}
where the integral is taken over all the computational volume. A corollary to this is
the definition of the volume. Substituting $f(\bx)=1$ into above integral we
find
\begin{equation}
	V = \sum_i V_i,\qquad {\rm where}\qquad V_i = \int\psi_i(\bx)\,dV,
	\label{eq:tot_volume}
\end{equation}
where $V_i$ is the effective volume of particle $i$. By construction, the total
volume of particles is conserved in a closed system.

In the next step, we apply similar procedure to the second term of the integral
in Eq.\,\ref{eq:q_weak}. Leaving the time integral temporarily out of the
derivation and focusing on spatial discretization, we write 
\begin{equation}
	\int\,\nabla\varphi\cdot{\bf F}(\bx)\,dV =  \int \sum_{i,j}\psi_i({\bf
	x})\varphi_j\nabla\psi_j({\bf x})\cdot{\bf F}(\bx)\,dV = \int \sum_{i,j}
	\psi_i({\bf x})(\varphi_j - \varphi_i)\del\psi_j({\bf x})\cdot{\bf F}(\bx)\,dV,
	\label{eq:int1}
\end{equation}
here, we insert $1=\sum_i\psi_i(\bx)$ into the integrand and the definition of a
gradient inside the integrand. In the last term we also made use of
the following:
\begin{eqnarray}
 &~& \sum_{i,j}\psi_i({\bf x})(-\varphi_i)  \nabla \psi_j({\bf x})\cdot{\bf
    F}({\bf x})  \\ \nonumber
 &=& \sum_i    \psi_i({\bf x})(-\varphi_i){\bf F}({\bf x})\cdot \nabla
 \sum_j\psi_j({\bf x}) \\ \nonumber
 &=& \sum_i    \psi_i({\bf x})(-\varphi_i){\bf F}({\bf x})\cdot \nabla 1
 \\ \nonumber
 &\equiv& 0
\end{eqnarray}
in order to add $\sum_{i,j}\psi_i({\bf x})(-\varphi_i)\del\psi_j({\bf x})\cdot{\bf F}(\bx)
\equiv 0$, which might appear to be introducing an unnessecary complication. 
However later -- in \S \ref{sec:AppendixConservation} -- it will become clear
that the form of that last term, which seems unnecessary in the
current formulation, is crucial for the preservation of certain vital conservative
properties when particular approximations are invoked.
We use this identity to add $\sum_{i,j} = 0$ into Eq.~\ref{eq:int1}. 
It may appear to be an unnecessary complication, but as can be seen in the next step, 
it allows us to rewrite the Eq.~\ref{eq:int1} in a manifestly symmetric form, 
from which the local conservation follows. 
This is a crucial step that provide us with the leeway to approximate 
spatial integrals with some form of a numerical quadrature without 
being worried that this fundamental symmetry is broken. An example of this approximation is shown 
in \S \ref{sec:AppendixConservation}.

To proceed further, we rewrite the double sum in the following form
\begin{equation}
	\sum_{i,j}\psi_i({\bf x})(\varphi_j - \varphi_i)\del\psi_j({\bf x})\cdot{\bf
	F}(\bx)=
	\sum_{i,j}\varphi_i(\psi_j({\bf x})\del\psi_i({\bf x}) - \psi_i({\bf
	x})\del\psi_j({\bf x}))\cdot{\bf F}(\bx),
	\label{eq:int2}
\end{equation}
where to obtain this expression the indices $i$ and $j$ have exchanged places.
Substituting this expression into the integral above, we obtain
\begin{equation}
	\int\del\varphi\cdot{\bf F}(\bx)\,dV =
	\int\sum_{i,j}\varphi_i\left[\psi_j({\bf x})\del\psi_i({\bf x}) - \psi_i({\bf
	x})\del\psi_j({\bf x})\right]\cdot{\bf F}(\bx)\,dV = \sum_{i,j} \varphi_i \int
	{\bf F}(\bx)\cdot d{\bf
	\Sigma}_{ij}.
	\label{eq:int3}
\end{equation}
The integral on the right-hand side can be interpreted as a surface integral
over a virtual boundary between particle $i$ and $j$. As a result, we can define
an effective inter-particle surface as:
\begin{equation}
	{\bf A}_{ij} = \int\,d{\bf\Sigma}_{ij} = \int\,\left[\psi_j({\bf
	x})\del\psi_i({\bf x}) - \psi_i({\bf x})\del\psi_j({\bf x})\right]\,dV.
	\label{eq:tot_area}
\end{equation}
Therefore, by analogy with Eq.\,\ref{eq:q_integral} and
Eq.\,\ref{eq:q_integral_descrete} we write
\begin{equation}
	\int\del\varphi\cdot{\bf F}(\bx)\,dV = \sum_{i,j}\varphi_i \int{\bf
	F}(\bx)\cdot d{\bf\Sigma}_{ij} = \sum_{i,j}\varphi_i \left({\bf
	F}\cdot{\bf A} \right)_{ij}.
	\label{eq:int4}
\end{equation}
Here, we can also use a second-order approximation to the last integral by
writing $({\bf F}\cdot{\bf A})_{ij} \approx {\bf F}_{ij}\cdot{\bf A}_{ij} =
F^n_{ij} A_{ij}$, where $F^n_{ij}$ is a projection of flux onto surface normal
evaluated at the centroid of the overlapping regions. For Lagrangian fluid
dynamics, this step is usually not used because we can compute fluxes directly
at each of the particle's location. However, when a mesh-less method is used
in context of a Godunov scheme, the available information is usually an
inter-particle (or inter-cell) flux projected onto the face normal. This flux is
the evaluated at the centroid of the overlapping region and then multiplied by
the effective area of two particles to achieve second order accuracy (e.g.
\citet{VILA:1999,2011MNRAS.414..129G}).

To complete the derivation, we apply spatial discretization to the first term
inside integral in Eq.\,\ref{eq:q_weak}. Here, we insert $1=\sum_i\psi_i(\bx)$
inside the integrand to obtain 
\begin{eqnarray}
	\int \left( \dot\varphi q\,dV\right)\,dt &=& \int \left(\int\dot\varphi q
	\sum_i \psi_i(\bx)\,dV\right)\,dt \approx
	\int \left(\sum_i\dot\varphi_i q_i \left[\int \psi_i(\bx)\,dV
	\right]\right)\,dt \nonumber \\
&=& \int\left( \sum_i \dot\varphi_i q_i V_i \right)\,dt,
	\label{eq:q_weak1_first}
\end{eqnarray}
where in the approximation step we applied one-point quadrature to the spatial
integral, with $q_i$ and $\varphi_i$ located at the particle's location.
Next, we apply integration by parts to the right hand side to obtain
\begin{equation}
	\int\left(\sum_i\frac{d\varphi_i}{dt}\,q_i\,V_i\right)\,dt = \sum_i \int \left(
	\frac{d}{dt}(\varphi_i\,q_i\,V_i) \right)\,dt - \int \left(\sum_i
	\frac{d}{dt}(q_i V_i)\varphi_i \right) \,dt
	\label{eq:q_weak1_second}
\end{equation}
The first term is an integral that depends only on the time-domain boundaries, which we set equal to zero.
Therefore we are left with only the second term. Combining this 
together with Eq.\,\ref{eq:int4}, we obtain the following spatially discrete
form of Eq.\,\ref{eq:q_weak}
\begin{equation}
	\int\,dt\,\sum_i\varphi_i\left(-\frac{d}{dt}(q_i V_i) - \sum_j({\bf
	F}\cdot{\bf A})_{ij} + S_i V_i\right) = 0.	
	\label{eq:space_discrete_weak}
\end{equation}
This integral must be zero for any arbitrary set of $\varphi_i$ and all times.
Hence the term inside the brackets must also always be equal to
zero. This leads to a spatially discrete form of finite-volume equations on a set of particles
\begin{equation} 
	\frac{d}{dt}(q_i V_i) + \sum_j({\bf F}\cdot{\bf A})_{ij} = S_i V_i,
	\label{eq:space_discrete_conservation}
\end{equation}
This equation has the same form as Eq.\,\ref{eq:q_flux}, except that here a surface
element between two particles is defined as a volume integral on the overlapping
regions of the corresponding partition of unity.

\subsubsection{Derivation of SPH equations of motion}

As an example of this new formalism, we use it to derive equations of Lagrangian
fluid dynamics which have the same form as the SPH equations of motion. In SPH,
the partition of unity is defined by the following expression
\begin{equation}
	\psi_i({\bf x}) = \frac{m_i}{\rho({\bf x})}W({\bf x} - {\bf x}_i, h({\bf x})),
	\label{eq:SPH_partition_of_unity}
\end{equation}
It is a simple task to check that this leads to the SPH summation identity,
$$\rho(\bx) = \sum_i m_i W(\bx - \bx_i, h(\bx)).$$ 
\noindent In the next step we evaluate
effective volume of the particle using Eq.\,\ref{eq:tot_volume}
\begin{equation}
	V_i = \int \psi_i(\bx)\,dV = \int \frac{m_i}{\rho(\bx)}W(\bx-\bx_i,
	h(\bx))\,dV
	\label{eq:SPHvolume}
\end{equation}
where $W(\bx - \bx_i, h(\bx))$ is an SPH kernel, and $h(\bx)$ is each SPH particle's
smoothing length.  In order to obtain this integral in a closed form, we
approximate the kernel inside this integrand as a delta function, namely $W(\bx
- \bx_i,h(\bx)) \approx \delta(\bx - \bx_i)$. This permits analytical evaluation 
of an SPH particle volume $V_i \approx m_i/\rho_i$, which is consistent with the
usual definition of the SPH particle number density $n_i = 1/V_i = \rho_i/m_i$.

To compute SPH forces, we apply a similar approximation to compute $({\bf
F}\cdot{\bf A})_{ij}$, in order to evaluate the flux of momentum. Because SPH is a Lagrangian method,
the flux of the momentum is a diagonal tensor ${\bf F} = P\hat{\bf I}$. Hence
we can write
\begin{eqnarray}
	 & &({\bf F}\cdot{\bf A})_{ij} = \int {\bf F}(\bx)\cdot d{\bf\Sigma}_{ij} = \int
	\left[\psi_i(\bx)\,({\bf F}(\bx):\del\psi_j(\bx))-
	\psi_j(\bx)\,({\bf F}(\bx):\del\psi_i(\bx))\,\right]\,dV \nonumber \\
	&=& \int \left(\frac{m_i}{\rho(\bx)}W(\bx-\bx_i,h(\bx))P(\bx)\del\psi_j(\bx) -
	\frac{m_j}{\rho(\bx)}W(\bx-\bx_j,h(\bx))P(\bx)\del\psi_i(\bx)\right)\,dV,
	\label{eq:SPHarea-mid}
\end{eqnarray}
where the``$:$'' symbol denotes contraction of a tensor with a vector to produce
another vector. In order to be able to analytically evaluate this integral, we
approximate the kernel (but not its gradient) using Dirac's delta function
$W(\bx-\bx_i, h(\bx))\del\psi_j(\bx) \approx \delta(\bx-\bx_i)\del\psi_j(\bx)$.
Ignoring any spatial variation in density and $h$ inside the volume
defined by the smoothing kernel for any given particle, we obtain:
\begin{equation}
	({\bf F}\cdot{\bf A})_{ij} \approx \frac{m_i m_j P_i}{\rho_i^2}\del_i W(\bx_i-\bx_j,h_i) -
	\frac{m_i m_j P_j}{\rho_j^2}\del_j W(\bx_j-\bx_i, h_j).
	\label{eq:SPHarea}
\end{equation}
To obtain this equation we used an approximation to the effective volume derived
above, $V_i = m_i/\rho_i$. Finally, if we substitute this approximation to $({\bf F}\cdot{\bf A})_{ij}$ into
Eq.\,\ref{eq:space_discrete_conservation} (in which we write $q =
\rho{\bf v}$ for the momentum), 
 then we obtain the following result
\begin{equation}
	\frac{d}{dt}(m_i{\bf v}_i) = -m_i \sum_j m_j\left(
	\frac{P_{i}}{\rho_i^2}\del_iW_{ij}(h_i) +
	\frac{P_{j}}{\rho_j^2}\del_jW_{ij}(h_j) \right) + \frac{m_i S_i}{\rho_i}.
	\label{eq:SPH-EOM-m}
\end{equation}
Here, we used standard SPH notation $W_{ij}(h_i) = W(\bx_i-\bx_j, h_i)$ and the
identity $\del_j W(\bx_j-\bx_i, h_j) = -\del_j W(\bx_i-\bx_i,h_j)$. Since 
in Lagrangian fluid dynamics the flux of mass across cell boundaries is zero, which
implies $dm_i/dt = 0$, Eq.\,\ref{eq:SPH-EOM-m} reduces to the standard
form of the SPH equations of motion
\begin{equation}
	\frac{d{\bf v}_i}{dt} = -\sum_j m_j\left(
	\frac{P_{i}}{\rho_i^2}\del_iW_{ij}(h_i) +
	\frac{P_{j}}{\rho_j^2}\del_jW_{ij}(h_j) \right) + \frac{S_i}{\rho_i}.
	\label{eq:SPH-EOM}
\end{equation}
With a more careful analysis that takes into account variations of density inside
the smoothing kernel one can derive additional correction terms due to
the varying smoothing length.

\subsection{Conservation properties}
\label{sec:AppendixConservation}

When solving finite-volume equations, certain properties ought to be
satisfied exactly by the underlying numerical scheme as a necessary condition
for obtaining the correct solution. One of the properties of utmost importance is
local conservation, which means that the amount of a quantity $q$ that leaves a
particle (mesh cell) $i$ in a particular direction (or through a given
surface) will be received by the relevant neighbour particle (mesh-cell)
$j$. A careful inspection of Eq.\,\ref{eq:space_discrete_conservation} and
Eq.\,\ref{eq:q_integral_descrete}, demonstrates that the necessary and sufficient
condition for local conservation is the antisymmetry identity $({\bf F}\cdot{\bf
A})_{ij} = -({\bf F}\cdot{\bf A})_{ji}$. In the case of the SPH equations of motion,
Eq.\,\ref{eq:SPH-EOM}, it becomes clear that local conservation guarantees that
Newton's third law is satisfied exactly for any particle distribution. The reader
can now check why we added the extra, identically zero, term to Eq.\,\ref{eq:int1} in
continuous form. Without including that term then applying the approximations
that lead to the SPH equations of motion results in the violation of
Newton's third law in the final discretized equations; in that form, 
Newton's third law would only have been satisfied to the truncation error.

Another property of the equations, which we call the closure condition, is
that the vector sum of cell boundary areas (or that of the effective areas between all
neighbours in the case of a mesh-less scheme) is identically zero, $\sum_j {\bf
A}_{ij} = 0$.  The closure condition guarantees that the time
derivatives of the relevant local quantities are always consistent
with the corresponding fluxes between the cells. For example, if the
pressure is constant for each particle or mesh cell, the net force
(which is gradient of the pressure) on each particle or mesh cell is
zero for any mesh or particle distribution. While the sum $\sum_j{\bf
  A}_{ij} = 0$ can be geometrically proven when referring to a mesh, 
this identity is not intuitive in case of particles; however, it can
be easily proven. Starting with
Eq.\,\ref{eq:tot_area}, we write
\begin{eqnarray}
	\sum_j{\bf A}_{ij} &=& \sum_j \int\left[\psi_j({\bf x})\del\psi_i({\bf x}) -
	\psi_i\del\psi_j({\bf x})\right]\,dV \nonumber \\ &=& \sum_j \int\left[\del(\psi_j({\bf
	x})\psi_i({\bf x})) - 2\psi_i\del\psi_j({\bf x})\right]\,dV = 0.
	\label{eq:prp1_proof}
\end{eqnarray}
Here we applied integration by parts to the first term of the integral.
Applying the divergence theorem to the first integral on the right-hand side, we
can rewrite it as a surface integral over the boundary between overlapping regions of the
corresponding partition of unity. Therefore, by construction this integral
vanishes, because the value of the partition of unity at its boundary is zero. In the
second term, we include the sum inside the integral, $\sum_j \int 2\psi_i({\bf
x})\del\psi_j({\bf x}) = 2\int\psi_i({\bf x})\sum_j\del\psi_j({\bf x})$.
Applying the identity $\sum_j\del\psi_j(\bx) \equiv 0$, we can demonstrate that this
term vanishes as well, therefore completing the proof that $\sum_j {\bf A}_{ij}
= 0$ in mesh-less schemes.

The approximations that lead to Eq.\,\ref{eq:SPH-EOM} may result in the
loss of some properties. Indeed, while the equations maintain their local
conservative character (Newton's third law is still exactly satisfied), the
closure condition does not hold anymore. Indeed, setting $P_{ij} = P_0={\rm
const}$, we obtain
\begin{equation}
	\frac{d}{dt}(m_i{\bf v}_i) = -m_i P_0 \sum_j m_j 
	\left(\frac{\del_iW_{ij}(h_i)}{\rho_i^2} + \frac{\del_j
	W_{ij}(h_j)}{\rho_j^2}\right) = -m_i P_0 C_i,
	\label{eq:SPHclosure}
\end{equation}
where $C_i = 0$ only when particles are distributed on a regular lattice. It is
well known in SPH that for irregular particle distributions there is a net
force proportional to the mean pressure of the system. This effect, called 
`pressure leak', tends to regularize the particle distribution (which
in turn tends to
decrease the strength of the effect). This pressure leak can be
attributed to the violation of the exact
closure condition, and it becomes more damaging with increasing degrees of
particle irregularity.  One manifestation of this pressure leak is the
presence of
``surface tension'' forces at a discontinuity in the particle distribution. It
is possible to reduce this effect by more accurate evaluation of the volumes and
areas in Eq.\,\ref{eq:SPHvolume} and Eq.\,\ref{eq:SPHarea-mid} respectively
either by numerical quadratures (e.g.
\citealt{springerlink:10.1007}) or using an inter-particle model for the
density distribution \citep{2002JCoPh.179..238I}.

\subsubsection{Angular momentum conservation}

Spatially discrete Euler equations exactly conserve the total mass, linear momentum
and energy of the system. However, conservation of these quantities
does not automatically guarantee conservation of angular momentum,
even though it is guaranteed in a continuous approximation. 
Let consider a system of particles or mesh points
with coordinates ${\bf x}_i$. To check whether a discretized form of the
equations conserves angular momentum, we compute
\begin{equation}
	\sum_i \dot{\bf L}_i = \sum_i\frac{d}{dt}\left({\bf x}_i\times m_i{\bf v}_i\right)
	= \sum_i\dot{\bf x}_i\times m_i{\bf v}_i + \sum_i{\bf
	x}_i\times\frac{d}{dt}(m_i{\bf v}_i),
	\label{eq:angular_moment}
\end{equation}
where $\dot{\bf x}_i = {\bf w}_i$ is the velocity of a particle or a mesh point.
Using Eq.\,\ref{eq:q_integral_descrete} with $q = \rho_i{\bf v}_i$ together with
momentum flux form Eq\,\ref{eq:q_flux}, we obtain
\begin{eqnarray}
	\frac{d}{dt}(m_i v^\alpha_i) &=& -\sum_j \left[(
	P\delta^{\alpha\beta}A^\beta)_{ij} + (\rho v^\alpha (v-w)^\beta
	A^\beta)_{ij} \right] \nonumber \\ &=& -\sum_j \left[ (PA^\alpha)_{ij} + (\rho
	v^\alpha ({\bf v}-{\bf w})\cdot{\bf A})_{ij}\right],
	\label{eq:momentum_eq}
\end{eqnarray}
where we use Einstein summation convention over Greek indices (which
here denote the 
three-dimensional components of a vector). Substituting this into
Eq.\,\ref{eq:angular_moment}, we obtain an equation for the time
evolution of the 
angular momentum in component form:
\begin{equation}
	\sum_i \dot{L}_i^\alpha = \eabc \sum_i m_i w^\beta v^\gamma - \eabc \sum_i x_i^\beta
	\sum_j \left[ (P A^\gamma)_{ij} + (\rho v^\gamma ({\bf v}-{\bf
	w})\cdot{\bf A})_{ij}\right]
	\label{eq:momentum_eq1}
\end{equation}
where we write a cross product of two vectors as $({\bf v}\times{\bf w})^\alpha =
\eabc v^\beta w^\gamma$, and $\eabc$ is the three-dimensional Levi-Civita symbol.
The second term in this equation can be rewritten in the following form
\begin{equation}
	\eabc\frac{1}{2}\sum_{i,j} (x_i - x_j)^\beta (PA^\gamma)_{ij}+
	\eabc\frac{1}{2}\sum_{i,j} (x_i - x_j)^\beta (\rho v^\gamma ({\bf v}-{\bf
	w})\cdot{\bf A})_{ij}.
	\label{eq:moment_eq2}
\end{equation}

\noindent Several properties can be derived from this equation. The first term vanishes
if the normal to the surface between mesh-cell (or particles) $i$ and $j$,
${\bf A}_{ij}/A_{ij}$, is parallel to the their separation vector, $\bx_i -
\bx_j$.  The second term, however, only automatically and generally vanishes if the mesh
points move with the fluid velocity, i.e. ${\bf w} = {\bf v}$. So, in
general the necessary condition for angular momentum conservation to
be guaranteed is a Lagrangian formulation of
hydrodynamics,\footnote{In the special case of a cylindrical
  coordinate system, only the z-component of angular momentum is conserved.} in which case we have
\begin{equation}
	\sum_i \dot{\bf L}_i = -\frac{1}{2}\sum_{i,j} ({\bf x}_i-{\bf x}_j)\times(P{\bf
	A})_{ij}.
	\label{eq:Lconser_lagrang}
\end{equation}
It becomes clear that, in order to conserve angular momentum exactly, the normal
vector to the area between particle $i$ and $j$ must be directed along the line
connecting these two particles. In the mesh-based Lagrangian method
that uses a
Voronoi mesh \citep{1988CoPhC..48...39T, 2010MNRAS.406.2289H}, these
normal vectors are always parallel to the separation vector between
two particles, as required.  In the
case of mesh-less schemes, however, it cannot be proven from
Eq.\,\ref{eq:SPHarea-mid} that $(P{\bf A})_{ij}$ is, in general, parallel to the
separation vector.  Nevertheless, the nature of the approximation that
led to the SPH
equations of motion, although sacrificing the closure condition, does
guarantee exact conservation of angular momentum.

\subsection{ Discussion}

We have presented a generic formalism for the mesh-less
discretization of finite-volume equations. Formally, the spatially discrete
equations (Eq.\,\ref{eq:space_discrete_conservation}) have the same form as
mesh-based equations (Eq.\,\ref{eq:q_integral_descrete}), and therefore posses
the same properties, such as local conservation and the closure condition. The former
property is of the utmost importance, as the violation of it will most
likely produce incorrect results in problems involving strong shocks. 

We have shown that, with few approximation, Eq.\,\ref{eq:space_discrete_conservation}
reduces to the SPH equations of motion. While the resulting equations conserve
angular momentum exactly, the closure condition is no longer satisfied for
generic particle distribution. This will result in a pressure leak
effect. In a positive 
pressure system with a random particle
distribution, this pressure leak will tend to regularize the particle
distribution such as to
minimise the leak, whilst it will cause tensile instability in the case of negative
stresses.  Amongst other issues, this pressure leak causes the inability of
SPH to resolve certain fluid instabilities, such as the Kelvin-Helmholtz and
Rayleigh-Taylor instabilities. A more accurate approximation
Eq.\,\ref{eq:SPHarea-mid} can substantially reduce the damaging effects of pressure leak
and the associated surface tension forces \citep{2002JCoPh.179..238I,
2010MNRAS.403.1165C}.  Alternatively, one may apply a high order numerical
quadrature to integrate Eq.\,\ref{eq:SPHarea-mid}, and therefore enforce
satisfaction of the closure conditions to high order
\citep{springerlink:10.1007}, but this may still result in
violation of angular momentum conservation. 

Particle-based scheme require various approximations to derive equations
that can be numerically integrated. On the other hand, the mesh-based schemes require a rather
complex process of construction of the unstructured mesh in order to
obtain the necessary geometrical quantities. The advantage of the latter is that the geometry is exact
and well-defined, which means that local conservation and the closure condition are
satisfied by construction. On the contrary, to achieve these properties with
mesh-less schemes, one needs to invoke high-order numerical quadrature over complex spatial
domains, such as the overlapping regions of two partitions of unity;  
these are conceptually simple but computationally expensive.
Therefore both methods can reach comparable degrees of accuracy, and the eventual computational
cost and overall complexity of each of the methods might also be
comparable. Furthermore, as we have shown, there is no well-defined
line separating the underlying principles of mesh-based and mesh-less
methods. Hence we expect that the generalisation outlined here will help
future simulations to take advantage of the best properties of both
kinds of scheme.

\section{Analysis of energy- and
  angular-momentum based parametrizations of CEE using the $E$--$J$ plane}
\label{app:gamma}

Here we \emph{illustrate} the energy and angular-momentum balance at
the end of CEE resulting from two  
parametrizations common in the literature (the energy formalism and
the $\gamma$-formalism).  
Clearly both energy conservation and angular momentum conservation
should physically take place during CEE; here we examine the
relationship between these parametrizations. 

To compare the two prescriptions and their outcomes, it is necessary
to adopt a relationship between orbital energy ($E$) and angular
momentum ($J$). 
We choose to assume that the post-CE binaries have circular, Keplerian
orbits. This seems reasonable (we do not expect high post-CE
eccentricities) and relatively robust (since, at fixed $E$, a non-zero
eccentricity [$e$] would lead to a correction in $J$ by only a factor
of $\sqrt{1-e^{2}}$).  So for the $\gamma$-formalism, we take the
post-CE orbital energy as it would be for a Keplerian binary (and for
the energy formalism we fix $J$ in the same way).

We will use the following to indicate the possible
post-CE states from the different formalisms (see Figs. \ref{alpha}, \ref{gamma1}
and \ref{gamma2}): 
\begin{itemize}
\item Black curves represent sets of possible outcomes produced by the
  $\gamma$-formalism, for given fixed values of $\gamma$. 
\item Blue lines are the sets of outcomes produced by the energy formalism,
  assuming $\alpha_{\rm CE}=1$.
\end{itemize}

We emphasize that the above curves are \emph{not} intended to represent the evolution during CEE, but
only possible {\bf final states}.  They could \emph{only} represent
the evolution during CEE if the instantaneous value of $\gamma$ or $\alpha_{\rm CE}$ is
constant throughout the CE phase, which would be \emph{extremely} unexpected. 
Each curve represents a collection of possible final states for fixed
CE parameter ($\gamma$ or $\alpha_{\rm CE}$), where different points
on these curves represent different final remnant masses.

Note that final states higher in $|E|$ tend to represent
\emph{tighter} binaries; orbital energy increases as the period decreases.

We also use the following conventions:
\begin{itemize}
\item The thick red line separates the merger region from the
  non-merger region. This represents the condition for the remnant
  core to not to overfill its Roche lobe, assuming that the remnant
  core does not expand upon mass loss. Since we know that a remnant
  core will almost always expand \citep[see the discussion
  in][]{del10,Iva11}, this definition usually represents the 
  closest possible post-CE orbits. Realistically, the
  final position of the binary should be below this line -- the actual
  state depends on how close the post-CE system is to being Roche-lobe
  filling.
\item  The dashed-dotted green line separates regions where the
  post-CE binary has more or less orbital energy than at the start of the CE. 
\item  The dotted green line separates the region where the post-CE
  system is wider ($a>a_\mathrm{i}$) than at the start of the CE from
  the region where the separation has decreased ($a<a_\mathrm{i}$).
\item Thin red lines show all Keplerian solutions for the minimum and maximum possible core masses (see below).
For any given remnant mass these are straight lines on the $E$--$J$ plane.
\end{itemize}

We mark the initial state of a binary with a star. We take into
account the orbital energy $E_\mathrm{orb, i}$ and the rotational energy of the giant's envelope
(assumed to be synchronized with the initial orbit), 
as well as the orbital angular momentum $J_\mathrm{orb, i}$ and angular momentum of
the giant's envelope. 

For this analysis, our choice of the \textit{lower} bound on the possible core mass $m_\mathrm{c, min}$
is the hydrogen-exhausted core $m_\mathrm{d, X}$ (the region where $X < 10^{-10}$) and is rather standard.
For the \textit{upper} bound on the possible core mass $m_\mathrm{c, max}$ we choose the minimum between
the central mass which contains less than 10\% hydrogen and the bottom of the outer convective envelope ($m_\mathrm{BCE}$). 
The solutions/outcomes for the post-CE core masses above the maximum possible core mass (to the right of the right thin red line) 
are provided only to show the behavior throughout the $E$--$J$ plane, but should not be considered as physically likely final states.
(Note also that it is for those masses where the effect of remnant
expansion should be the greatest.)

Based on our assumptions, \emph{only the solutions bounded from above by the thick solid merger line and lying within 
the region between minimum and maximum core masses are expected to be
permitted for a self-consistent post-CE binary unless another
  -- currently unidentified -- energy source is available.}
This condition is independent of whether the energy formalism or the $\gamma$-formalism is used.

\subsection{Examining outcomes of the $\alpha$-formalism in the
    $E$--$J$ plane}
\label{sec:example1}

\begin{figure}
\centering
\includegraphics[height=.3\textheight]{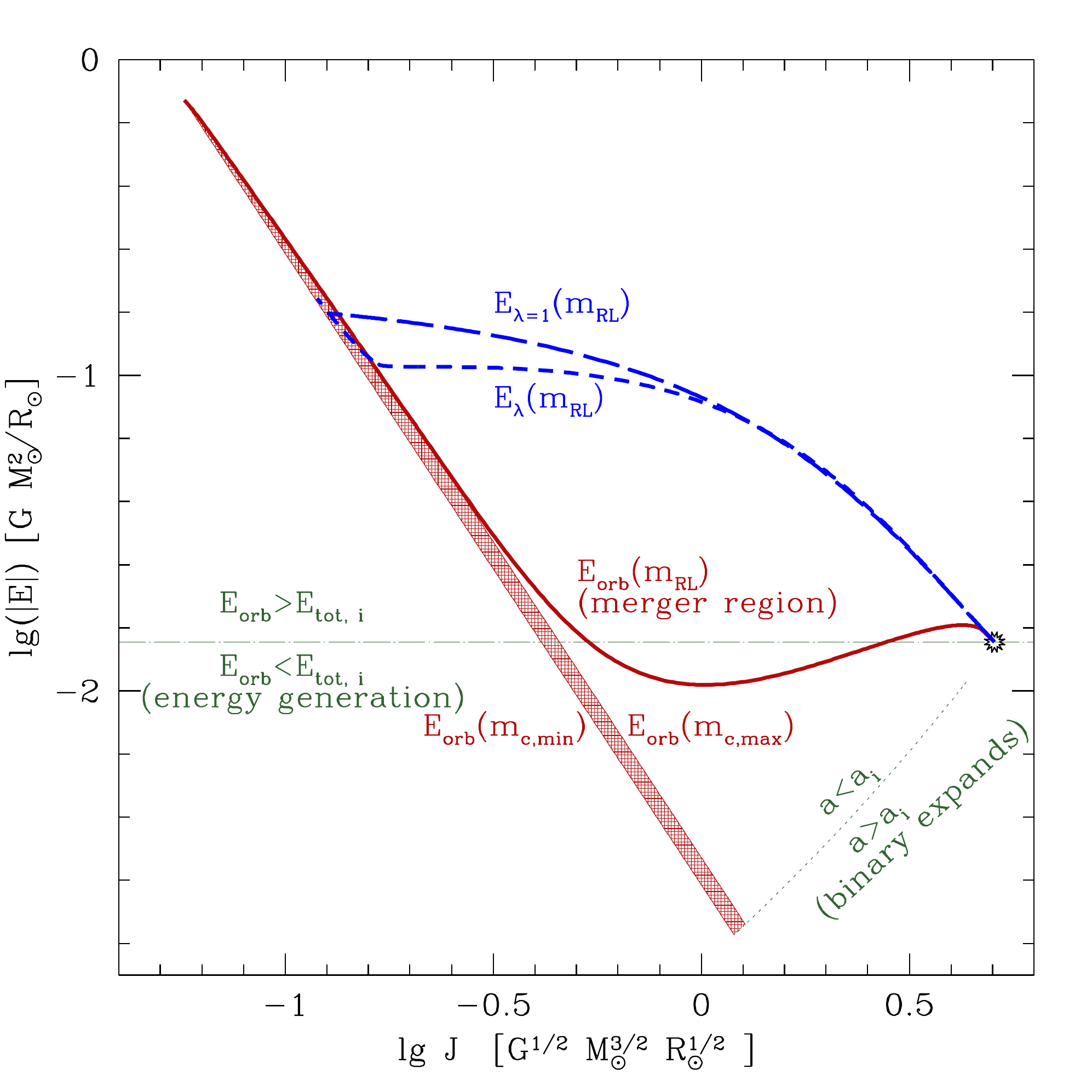}
  \caption{  Orbital angular momentum $J$ and energy $E$ for a CE in a $2\,M_\odot + 0.5\,M_\odot$ binary
    ($r_\mathrm{d}=23\,R_\odot$ and the hydrogen-exhausted core $m_\mathrm{d,X}=0.317\,M_\odot$).
    The black star indicates the state of the binary at the onset of the CE.
    The solid line (red) shows the Keplerian $E_\mathrm{orb}- J_\mathrm{orb}$ relation for the final binary
    assuming that the mass of the post-CE remnant consists of all the RG mass that was originally 
    contained within the final Roche lobe ($m_\mathrm{RL}$), in other words, the maximum possible
    remnant mass for this orbit.
    The shaded region contains Keplerian orbits for core masses
    bounded by $m_\mathrm{c,min} < m_\mathrm{d, c} < m_\mathrm{c, max}$ (see the text).  
    A final state of the  post-CE binary must lie within this area. 
    The dashed lines (blue) represent the set of all possible final states as per the energy formalism (being a function of all possible core masses), 
    for $\lambda$ calculated using the stellar model and for $\lambda=1$.
    The dashed-dotted green line separates regions where the post-CE binary has more or less energy than at the start of the CE.
    The dotted green  line separates regions where the post-CE is wider ($a>a_\mathrm{i}$) than at the start of the CE and where it shrunk ($a<a_\mathrm{i}$). 
    The stellar model was calculated using the evolutionary code  described in \cite{Iva02, IvanovaTaam2004}.
    \label{alpha}
  }
\end{figure}

As an example, we consider a binary with a giant of $2\,M_\odot$ and a WD of $0.5\,M_\odot$,
where the CE is initiated when the giant is  evolved to a hydrogen-exhausted
core mass $m_\mathrm{d, X}\sim 0.317\,M_\odot$ and has a radius of $23\,R_\odot$ (see Fig.\,\ref{alpha}).
In this figure, we shade the region containing the expected range of potential core masses. 
As we said above,  only the strip bounded from above by the thick solid line and lying within the shaded region 
is permitted for a self-consistent final post-CE binary. 
As was discussed in \S4, even for a properly computed $\lambda$, the final orbital separations can vary by over a factor of 10 in the region of likely core masses.

In the energy formalism, as expected, the post-CE binary can not have more energy than the initial binary and still satisfy 
conservation of energy (see the dashed-dotted green line on Fig.~\ref{alpha}), 
since some energy must be used to expel the common envelope to infinity.
We can also find the position that a final binary would have if it kept the same orbital separation, for any valid post-CE mass
of the donor $m_\mathrm{c, min}<m_\mathrm{d,c}<m_\mathrm{d}$. In the
absence of another energy source, a widened orbit would violate energy
conservation and so should not be produced by this energy formalism.

The whole range of possible states for a post-CE binary described by the energy formalism 
is bounded by the blue dashed line at the bottom,
the solid red line at the top and the shaded red area.
When the CE is initialised at a different giant radius (i.e.\ at a different orbital period), 
and accordingly a different $m_\mathrm{d,X}$, the picture is qualitatively similar, although the 
uncertainty that is introduced by the core-mass definition could vary.

\subsection{Examining outcomes of the $\gamma$-formalism in the
    $E$--$J$ plane }
\label{sec:example2}

\begin{figure}
\centering
\includegraphics[height=.3\textheight]{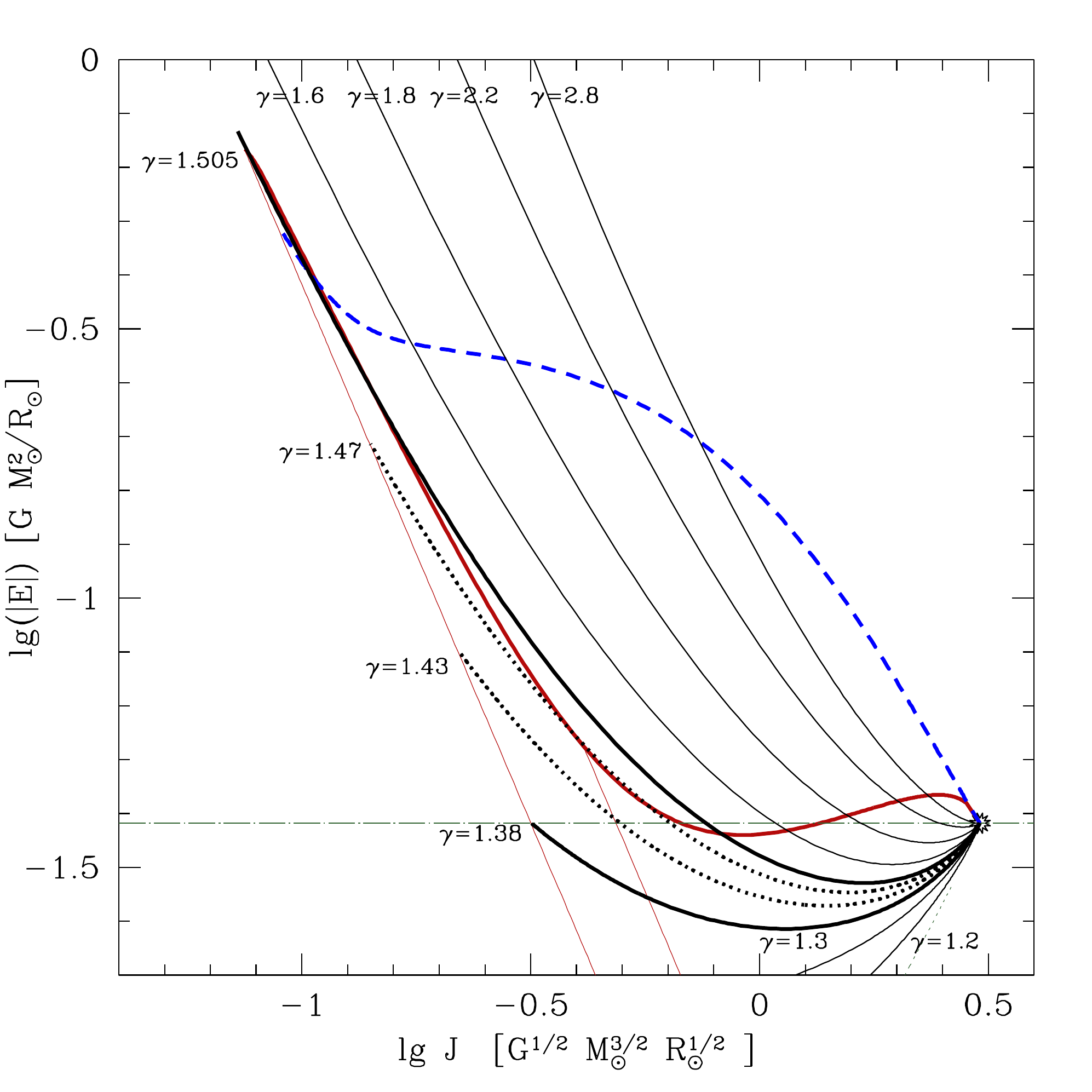}
\caption{Orbital angular momentum and energy for a $2\,M_\odot + 0.5\,M_\odot$ binary
    ($r_\mathrm{d}=86\,R_\odot$ and $m_\mathrm{d,X}=0.38\,M_\odot$).
    Thick and thin solid red lines show the only possible final binary configurations for various
    adopted core masses. For comparison we also show the minimum energy expenditures to release the envelope 
    (blue dashed line, see also the caption of Fig.\,\ref{alpha}).
    Black solid and dotted lines indicate possible final binary
    configurations, assuming angular momentum is lost
    in accordance with the $\gamma$-formalism, where the thick black
    solid lines show $\gamma_{\rm E}$ and $\gamma_{\rm M}$, i.e the minimum and maximum $\gamma$ 
    that make a binary which avoids the need for energy input or
    merger. Dotted black lines show values of $\gamma$ that lead to formation
    of binaries which satisfy those constraints for the full range of
    core mass definitions. Other line-styles are as in Fig.\,\ref{alpha}.
    \label{gamma1}
  }
\end{figure}

Here we choose a $2\,M_\odot$ giant,   evolved to a hydrogen-exhausted
core mass $m_\mathrm{d, X}\sim 0.38\,M_\odot$ and has a radius of $86\,R_\odot$. This represents
a common case in the study of \citet{Nelemans05}, where many double WD binaries
have an older companion of $0.5\,M_\odot$ and a younger WD of $0.4\,M_\odot$ (see Fig.\,\ref{gamma1}).
From Fig.\,\ref{gamma1}, we see that the set of solutions for $\gamma=1.5$ roughly coincides with a possible 
final binary configuration for these particular companion
masses.\footnote{For clarity, we repeat that the curves represent
    set of possible solutions -- end points -- for the outcome of CEE,
    \textbf{not} the paths taken to reach those states.} 
The set of $\gamma\approx 1.5$ 
solutions crosses the final binary configuration at approximately the
location mandated by energy consideration.  This may therefore help to
explain why $\gamma = 1.5$ is successful in fitting the observed
systems which are thought to be similar to this one. 

We can also study the restricted set of outcomes available when CEE is limited by the available
orbital energy reservoir. We repeat this is \emph{not} an assumption in the
$\gamma$-formalism, but it is important for understanding how the
$\gamma$-formalism related to canonical CEE, with a significant spiral-in, as described by the
longstanding energy prescription. Here we make use of the limiting cases
defined in \S \ref{sec:restrictedgamma}, which represent the $\gamma$-values which
lead to merger ($\gamma_{\rm M}$) or require an additional energy
source ($\gamma_{\rm E}$). 

For $\gamma > \gamma_{\rm M}=1.505$ this binary would merge 
as Eq.~(\ref{gammaform}) predicts negative post-CE angular momentum.
For  $\gamma<\gamma_{\rm E}=1.38$, Eq.~(\ref{gammaform}) produces an
orbital separation such that a Keplerian post-CE binary could only be explained 
if some form of extra energy input was provided to the system during a CEE. 
Even within this narrow range of $\gamma_{\rm E}<\gamma<\gamma_{\rm M}$ 
some outcomes would require either energy generation or would be
mergers: the details depend on the size of the core and on the 
exact definition of the ejected envelope. For smaller ejected masses,
the range also becomes smaller.
For example, in the case when CEE removes only the convective envelope
(rather than removing the entire hydrogen burning shell to leave only
the naked core), then $\gamma>1.43$ is required to produce a binary
without apparent energy generation and  $\gamma<1.47$ is required for the binary not to merge.

\begin{figure}
  \includegraphics[height=.3\textheight]{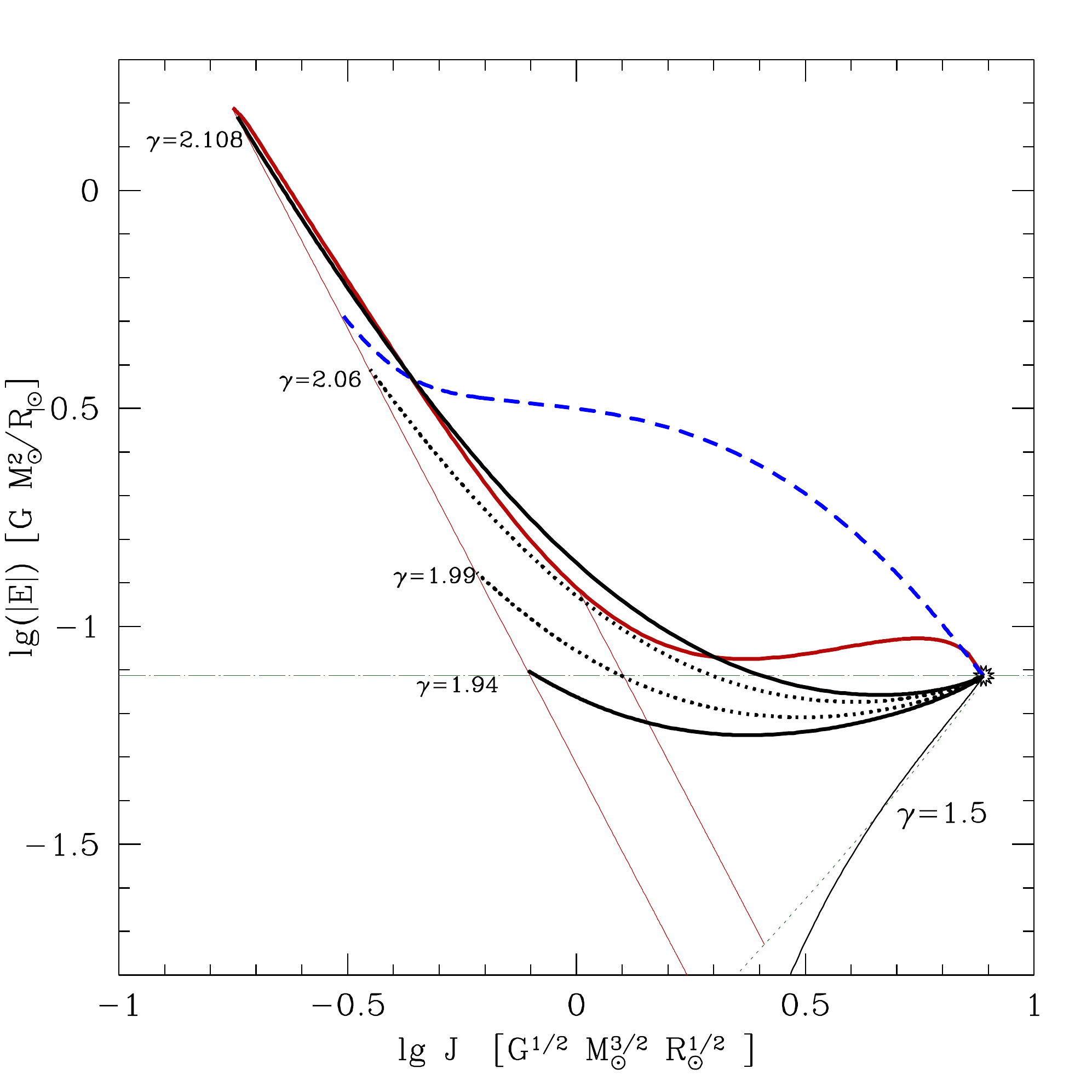}
  \includegraphics[height=.3\textheight]{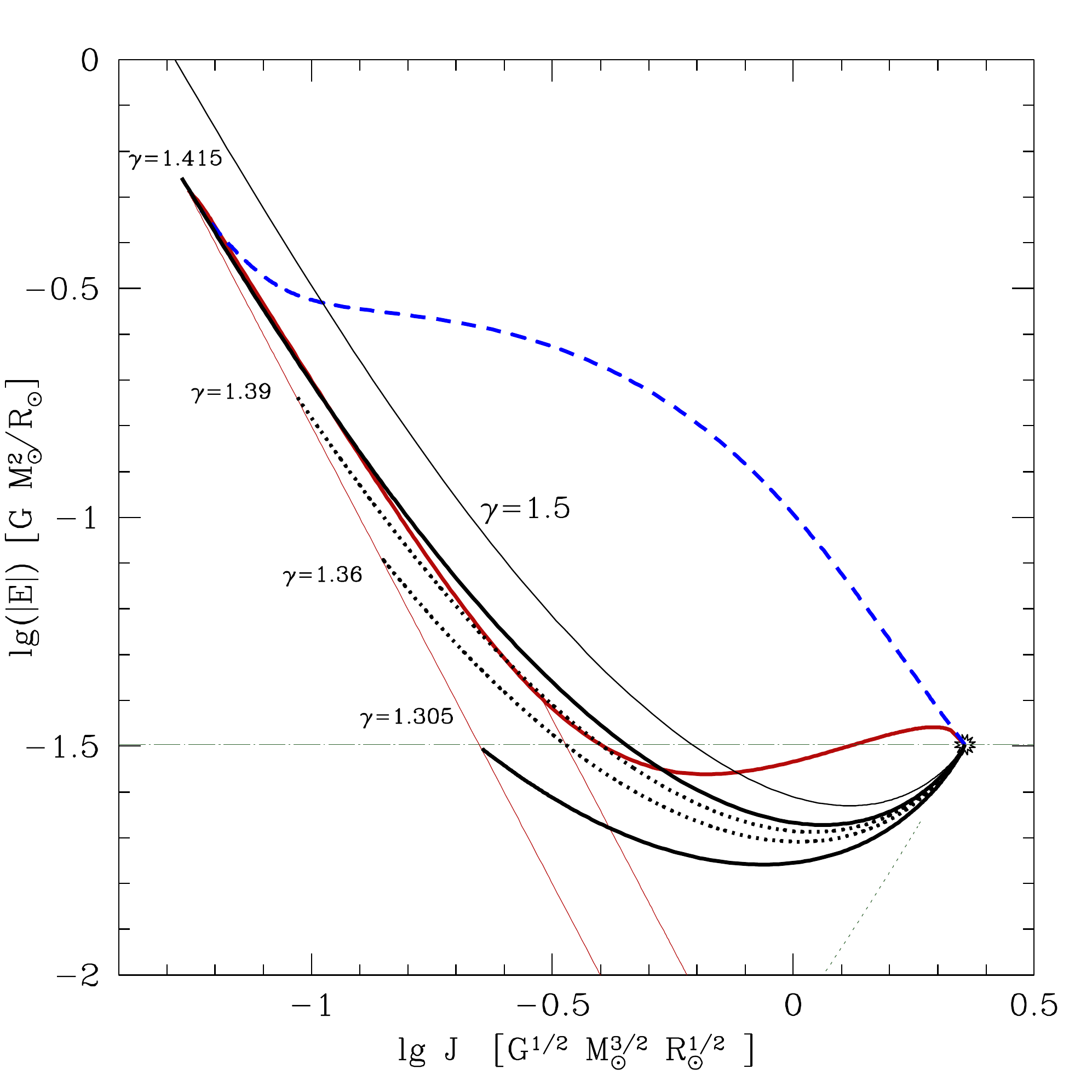}
  \caption{Left: Orbital angular momentum and energy for a $2\,M_\odot + 1.5\,M_\odot$ binary
    ($r_\mathrm{d}=86\,R_\odot$ and $m_\mathrm{d,X}=0.38\,M_\odot$).
 Right: Orbital angular momentum and energy for a $2\,M_\odot + 0.35\,M_\odot$ binary
    ($r_\mathrm{d}=86\,R_\odot$ and $m_\mathrm{d,X}=0.38\,M_\odot$).
    Line styles are as in Fig.\,\ref{gamma1}.
    \label{gamma2}
  }
\end{figure}

This behaviour is related to the sensitivity to the value of $\gamma$ of
the post-CE separations predicted using Eq.\ref{gammaform}, as discussed in \S
\ref{sec:gammasensitivity}.  For this particular binary,  
the change in the input $\gamma$ values by as little 
as $\delta \gamma=0.125$, from 1.38 to 1.505, 
provides the difference in post-CE binary separations for spiral-ins   
by a factor of 20 for the same adopted core mass; any larger change would modify the outcome
qualitatively into either a merger or an outcome with apparent energy generation.

Now consider the same $2\,M_\odot$ giant, with the same $m_\mathrm{c, min}\sim 0.4\,M_\odot$,
but with a $1.5\,M_\odot$ companion (Fig.\,\ref{gamma2}, left). This is similar to some initial binaries 
considered to be progenitors for DWDs in \cite{Sluys2006}.
Assuming a spiral-in limited by the available orbital energy, this
binary could survive without merging only between
$\gamma_{\rm M}=2.108$ and  $\gamma_{\rm E}=1.94$ (we note that this is
close to values of $\gamma$ used
in some similar systems in \cite{Sluys2006}, see their Table 6); orbital expansion happens for $\gamma<1.5$. 
Taking into account that not all of the hydrogen shell might be expelled and that the core has finite size,
this range is reduced to $\gamma\approx1.99-2.06$. 
With $\gamma=1.5$, the binary is \textit{even wider} than it was at the beginning --- 
the same effect as having negative $\alpha$ or a stellar wind. 
As a consequence, the binary becomes wider during mass loss (of course
this is both legal and natural for the $\gamma$-formalism, since it was 
created to model exactly such widening for the first episode of MT in
DWDs formation and includes no restrictions on the overall energy balance).

Another example involves a binary with a less massive companion
($0.35\,M_\odot$, Fig.\,\ref{gamma2}, right).  Keplerian 
solutions can easily be found for $\gamma \approx 1.36-1.39$, where the
extreme cases are $\gamma_{\rm M} = 1.415$ and  $\gamma_{\rm
  E} = 1.305$.
With $\gamma=1.5$, the post-CE binary should merge if realistic core sizes are taken into account.

We stress that the analysis in this subsection assumes constraints which are not
contained within the original $\gamma$-formalism, although those
limits are natural ones in the absence of an additional, currently
unidentified, energy source. It also takes the standard position
that the post-CE systems have circular, Keplerian orbits.
Nonetheless,
we suggest that the overall results above indicate the
$\gamma$-formalism, as currently expressed,
is not ideal for making predictions about CE phases which are limited by the
available orbital energy, or (similarly) those which involve a
significant spiral-in and mass loss.

\end{document}